\documentclass[epj]{svjour}
\usepackage{amsmath, amssymb,  dcolumn, bm, epsfig, youngtab, bbm}
\usepackage{hyperref}
\usepackage{slashed}
\usepackage{verbatim}
\usepackage{mathtools}
\usepackage{longtable}
\usepackage{bbold}
\usepackage{soul}   
\usepackage{mywidetext}
\newcommand{\Slash}[1]{\ooalign{\hfil/\hfil\crcr$#1$}}   
\newcommand{\tr}{\textbf{tr}}

\long\def\Omit#1{}

\allowdisplaybreaks[4]

\renewcommand*{\eqref}[1]{(\ref{eq:#1})}
\newcommand*{\eqlab}[1]{\label{eq:#1}}

\newcommand{\SP}[3]{\left<{#1}\right|{#2}\left|{#3}\right>}

\newcommand{\beq}{\begin{equation}}
\newcommand{\eeq}{\end{equation}}

\newcommand{\ghat}{\hat g}

\newcommand{\rbot}{{r_{^{\tiny \rfloor \! \!\lfloor} }\!}}
\newcommand{\rbarbot}{{\bar r_{^{\tiny \rfloor \! \!\lfloor} }\!}}
\newcommand{\wRbot}{{w_{^{\tiny \lfloor} }\! }}
\newcommand{\wLbot}{{w_{^{\tiny \rfloor} }\!}}

\newcommand{\wRbarbot}{{\bar w_{^{\tiny \lfloor} }\! }}
\newcommand{\wLbarbot}{{\bar w_{^{\tiny \rfloor} }\!}}

\newcommand{\etext}[1]{\mbox{$#1$}}

\newcommand{\tsp}[2]{\left<\mbox{\small{${#1}$}}\right|T \left|\mbox{\small{${#2}$}}\right>}

\def\rescale{\fontsize{7}{1.9}}

\begin{document}
\title{On kinematical constraints in the hadrogenesis conjecture for the baryon resonance spectrum}
\author{
  Yonggoo\ Heo\thanks{\email{y.heo@gsi.de}} \and Matthias F.M.\ Lutz\thanks{\email{m.lutz@gsi.de}}
}                     
%
%
\institute{
  GSI Helmholtzzentrum f\"ur Schwerionenforschung GmbH, Planck Str. 1, 64291 Darmstadt, Germany
}
\date{Received: date / Revised version: date}
%
\abstract{
  We consider the reaction dynamics of bosons with negative parity and spin $0$ or $1$ and fermions with positive parity
  and spin $\frac{1}{2}$ or $\frac{3}{2}$. Such systems are of central importance for the computation of the
  baryon resonance spectrum in the hadrogenesis conjecture. Based on a chiral Lagrangian the coupled-channel partial-wave
  scattering amplitudes have to be computed. We study the generic properties of such amplitudes. A decomposition of the
  various scattering amplitudes into suitable sets of invariant functions expected to satisfy Mandelstam's
  dispersion-integral representation is presented. Sets are identified that are free from kinematical constraints and that can
  be computed efficiently in terms of a novel projection algebra. From such a representation one can deduce the analytic
  structure of the partial-wave amplitudes. The helicity and the conventional angular-momentum
  partial-wave amplitudes are kinematically constrained at the Kibble conditions. Therefore an application of
  a dispersion-integral representation is prohibitively cumbersome. We derive covariant partial-wave amplitudes that are free
  from kinematical constraints at the Kibble conditions. They correspond to specific polynomials in the 4-momenta and Dirac
  matrices that solve the various Bethe-Salpeter equations in the presence of short-range interactions analytically.
 } 
\maketitle
%
\section{Introduction}
\label{sec:intro}

The light vector mesons play a crucial role in the hadrogenesis
conjecture \cite{Lutz:2001dr,Lutz:2001yb,Lutz:2001mi,Lutz:2003fm,Kolomeitsev:2003kt,Lutz:2004dz,Lutz:2005ip,Lutz:2007sk,Lutz:2008km,Terschlusen:2012xw}.
Together with the Goldstone bosons they are identified to be the quasi-fundamental hadronic degrees of freedom
that are expected to be responsible for the formation of the meson spectrum. If supplemented with the baryon-octet and decuplet ground states
the baryon resonance spectrum is conjectured to be generated by coupled-channel dynamics.
For instance it was shown that the leading chiral interaction of Goldstone bosons with the light vector mesons generates
an axial-vector meson spectrum that is quite close to the empirical spectrum \cite{Lutz:2003fm,Lutz:2008km}.
Similarly s- and d-wave baryon resonances were generated by the leading chiral interaction of the Goldstone bosons with
the baryon octet and decuplet states \cite{Kaiser:1995cy,GarciaRecio:2003ks,Kolomeitsev:2003kt,Lutz:2004tb}

Extensions of such computations to systems involving intermediate states of a vector meson and
a baryon, as suggested by the hadrogenesis conjecture, are formidable challenges. Though it is well known how to
incorporate more massive degrees of freedom into the chiral Lagrangian, it is not so clear how to organize systematic
applications. For instance, the low-energy interaction of vector mesons with the baryons is characterized by various
unknown two-body counter terms. This is analogous to the low-energy interaction of two nucleons \cite{Ordonez:1995rz,Lutz:1999yr,Epelbaum:2008ga,Gasparyan:2012km}.
Here chiral dynamics does not predict the dominant structure of the coupled-channel interaction. This motivated a
phenomenological study that parameterized a set of quasi-local coupled-channel interactions and adjusted their st\-rength
to empirical data from pion- and photon-nucleon scattering \cite{Lutz:2001mi}. By construction all long-range forces
from t- and u-channel exchange processes are assumed to be integrated out and therefore the approach is applicable in the
resonance region only. The number of adjusted parameter, about 50, were used to describe about 2000 data points. Various
baryon resonances were shown to be dynamically generated by s-wave channels with a vector meson. Clearly, a more systematic
and predictive approach would be highly desirable \cite{Romanets:2012hm}. The challenge is the consistent treatment of both long- and short-range
forces from t- and u-channel exchange processes. Only then it may be possible to establish a reliable link of a chiral
Lagrangian to the resonance spectrum.

The key issue is the identification of an optimal set of degrees of freedom in combination with the construction of
power counting rules. A novel counting scheme for the chiral Lagrangian which
includes the nonet of light vector mesons in the tensor field representation was
explored in \cite{Lutz:2008km,Terschlusen:2012xw}. It is based on the hadrogenesis conjecture and large-$N_c$
considerations \cite{Lutz:2001dr,Lutz:2003fm,Lutz:2004dz,Lutz:2005ip,Lutz:2007sk}. The counting scheme would be
a consequence of an additional mass gap of QCD in the chiral limit, that may arise if the
number of colors increases. How to systematically include the baryon octet and decuplet states into this
Lagrangian remains an open issue.

In this work we prepare the ground for systematic coupled-channel computations involving active vector meson degrees
of freedom. Based on a chiral Lagrangian coupl\-ed-channel partial-wave amplitudes need to be established in a
controlled approximation. Any conceivable scheme should obey the coupled-channel unitarity condition and
generate amplitudes, which analytic structures are consistent with the constraints set by micro-causality.
Following \cite{Gasparyan:2010xz,Danilkin:2010xd,Gasparyan:2011yw,Danilkin:2011fz,Danilkin:2012ua,Gasparyan:2012km}
we use the concept of a generalized potential. A partial-wave scattering amplitude $T^J(\sqrt{s}\,)$
is decomposed into two contributions
\begin{eqnarray}
T^J(\sqrt{s}\,) =U^J(\sqrt{s}\,) + \int \frac{\mathrm{d} w}{\pi}\,\frac{T^{J}(w)\,\rho^J(w)\,T^{J \dagger}(w)}{w- \sqrt{s}-i\,\epsilon}\,.
\label{def-U}
\end{eqnarray}
By definition the generalized potential $U^J(\sqrt{s\,})$ receives contributions with left-hand cuts only. All right-hand
cuts from the s-channel unitarity condition are generated by the second term in (\ref{def-U}). Given an approximated
generalized potential one can solve for the partial-wave scattering amplitude $T^J(\sqrt{s}\,)$. A systematic
unitarization scheme based on (\ref{def-U}) was developed recently in
\cite{Gasparyan:2010xz,Danilkin:2010xd,Gasparyan:2011yw,Danilkin:2011fz,Danilkin:2012ua,Gasparyan:2012km}.
It is applicable in the presence of long- and short-range forces and therefore suitable for coupled-channel studies
with active vector meson degrees of freedom.

Once intermediate states with a vector meson are considered there are almost always long-range forces implied by
t- or u-channel exchange processes that lead to non-trivial left-hand branch points in the partial-wave amplitudes.
The left- and right-hand branch cuts almost always overlap and therefore any algebraic or separable
approach has to be rejected. Let us be specific and exemplify our statement: consider the $\omega$-meson nucleon
scattering process. The partial-wave amplitudes have a left-hand branch point at 1351 MeV that is caused by the
u-channel nucleon exchange process according to the Landau condition \cite{Landau:1959fi}.
The branch point is right from the pion-nucleon threshold and therefore constitutes
an example for an overlap of left- and right-hand cuts. If the on-shell reduction scheme of \cite{Lutz:2001yb} or any
other separable scheme would be applied to the system with two channels $\pi N$ and $\omega N$, the partial-wave scattering
amplitudes develop necessarily unphysical left-hand branch points \cite{Lutz:2003fm,Gasparyan:2010xz,Danilkin:2011fz}. This holds at any
finite truncation of the interaction kernel: all pion-nucleon partial-wave amplitudes would have an unphysical branch point
at 1351 MeV. The resulting partial-wave amplitudes are analytic functions, however, they have
a cut structure that is inconsistent with the constraints set by micro causality.

Though it is straight forward to introduce partial-wave scattering amplitudes in the helicity formalism of Jacob and
Wick \cite{Jacob:1959at}, it is a nontrivial task to derive transformations that lead to amplitudes that are kinematically
unconstrained \cite{Chew:1957tf,Nakanishi:1962,Berends:1967vi,Lutz:1999yr,Lutz:2001mi,Lutz:2001yb,Lutz:2001dq,Lutz:2007bh,Korpa:2008ut,Gasparyan:2010xz,Gasparyan:2012km}.
A kinematical constraint of a partial-wave amplitude requires a boundary condition on the
non-linear integral equations (\ref{def-U}) that complicates the search of its solutions significantly. Therefore it is useful
to find transformations of the helicity partial-wave amplitudes to covariant partial-wave amplitudes that do
not require boundary conditions in (\ref{def-U}). In a previous work one of the authors studied
the scattering of $0^-$ and $1^-$ particles \cite{Lutz:2003fm,Lutz:2011xc} and fermion-antifermion annihilation processes
with $\frac{1}{2}^+$ and $\frac{3}{2}^+$ particles \cite{Stoica:2011cy}.

So far most reactions involving two-body states with $0^-$ or $1^-$ particles and $\frac{1}{2}^+$ or $\frac{3}{2}^+$ particles
have not been dealt with. It is the purpose of the present work to derive the covariant partial-wave amplitudes for the latter
reactions. The technique applied in this work has been used previously in studies of two-body scattering systems with photons,
pions and nucleons \cite{Chew:1957tf,Ball:1961zza,Barut:1963zzb,Hara:1964zza,PhysRev.169.1248,PhysRev.142.1187,PhysRev.170.1606,Scadron:1969rw,CohenTannoudji1968239,Bardeen:1969aw,Stoica:2011cy,Lutz:2011xc}.
Since the applications of the unitarization scheme \cite{Gasparyan:2010xz,Danilkin:2010xd,Gasparyan:2011yw,Danilkin:2011fz,Danilkin:2012ua,Gasparyan:2012km}
requires a detailed study of the analytic structure of any contribution to the generalized potential, it is instrumental
to generate from a given chiral Lagrangian analytic expressions for such driving terms. In an initial step we decompose
the scattering amplitude into invariant functions that are free of kinematical constraints
\cite{Chew:1957tf,MacDowell:1959zza,Ball:1961zza,Barut:1963zzb,Hara:1964zza,PhysRev.169.1248,Scadron:1969rw,CohenTannoudji1968239,Bardeen:1969aw,Cheung:1972tt}.
Such amplitudes are expected to satisfy a Mandelstam's
dispersion-integral representation~\cite{Mandelstam:1958xc,Ball:1961zza}. Like in the previous work \cite{Lutz:2011xc} we will
derive a projection algebra that allows the analytic derivation of contributions to the invariant amplitudes
from a given chiral Lagrangian by means of a computer algebra program. In a second step the non-trivial transformation
from the helicity to the covariant partial-wave amplitudes is derived. As a side product of such an analysis we generate
specific polynomials in the 4-momenta and Dirac matrices that solve the various Bethe-Salpeter equations in the presence
of short range interactions analytically. This generalizes and systematizes the results of previous
works \cite{Lutz:2001mi,Lutz:2001yb,Nieves:2001wt,Bruns:2010sv}.

The work is organized as follows. Section~\ref{sec:on-shell-scattering-amp} introduces the conventions used for the kinematics of the various
two-body reactions. The scattering amplitudes are decomposed into sets of invariant amplitudes free of kinematical
constraints. In the following section  the helicity partial-wave amplitudes are constructed within the given convention.
The transformation to partial-wave amplitudes free of kinematical constraints are derived, discussed and presented.
Section~\ref{sec:summary} offers a short summary.

\section{On-shell scattering amplitudes}
\label{sec:on-shell-scattering-amp}
We consider two-body reactions of a boson with $J^P=0^-,1^-$ and a fermion with $J^P=\frac{1}{2}^+,\frac{3}{2}^+$ where we
use the conventions introduced in \cite{Stoica:2011cy,Lutz:2011xc} for the kinematics and wave functions. All derivations
will be completely generic. A two-body reaction is characterized by the three Mandelstam variables $s, t$ and $u$ with
\begin{eqnarray}
  s+t+u = m^2+M^2+\bar m^2+\bar M^2
  \,,
  \label{def-Mandelstam}
\end{eqnarray}
with  the initial and final masses $m, M$ and $\bar m, \bar M$ respectively. In the center-of-momentum frame
the 4-momenta $q$ and $\bar{q}$ of the incoming and outgoing boson and those of the fermion, $p$ and $\bar{p}$ are
determined by the scattering angle $\theta$ and the magnitudes of the initial and final three-momenta $q_{\rm cm}$
and $\bar{q}_{\rm cm}$. From \cite{Stoica:2011cy,Lutz:2011xc} we recall the further useful notations
\begin{eqnarray}
  &&
  w^\mu = q^\mu+p^\mu = \bar{q}^\mu + \bar{p}^\mu
  \,,\nonumber\\ &&
  k^\mu=\frac{1}{2}\,(q^\mu-p^\mu)\,, \quad \;\;\,  r_\mu = k_\mu - \frac{1}{2}\, \frac{q^2-p^2}{s}\,w_\mu
  \,,\nonumber\\ &&
  \bar k^\mu =\frac{1}{2}\,(\bar q^\mu-\bar p^\mu)
  \,, \quad \;\; \,
  \bar r_\mu = \bar k_\mu-\frac{1}{2}\, \frac{\bar q^2-\bar p^2}{s}\,w_\mu
  \,,\nonumber\\ &&
  \bar r \cdot \bar r = -\bar q_{\rm cm}^2 \,, \quad\, \bar r \cdot r = -\bar q_{\rm cm}\,q_{\rm cm}\,\cos \theta
  \,,\nonumber\\ &&
  r \cdot r = - q_{\rm cm}^2 \,,\quad \,\bar r \cdot w = 0 = w \cdot r \,, \quad \,w \cdot w = s
  \,,
  \label{def-w}
\end{eqnarray}
where the two 4-vectors $r_\mu$ and $\bar r_\mu $ are orthogonal to $w_\mu$.

The on-shell production and scattering amplitudes are defined in terms of plane-wave matrix elements of the scattering
operator where we do not make explicit internal degrees of freedom like isospin or strangeness quantum numbers for simplicity.
We decompose the scattering amplitudes into sets of invariant functions. The merit of the decomposition lies in the
transparent analytic properties of the functions $F_n(s,t)$, which are expected to satisfy Mandelstam's dispersion-integral representation~\cite{Mandelstam:1958xc,Ball:1961zza}. For reactions involving non-zero spin particles it
is not straight forward to identify such amplitudes.

We begin with the elastic scattering of a pseudoscalar boson off a spin-one-half fermion
\begin{eqnarray}
  &&
  T_{0\,\frac{1}{2} \to \,0\,\frac{1}{2}}(\bar k,\, k,\, w)
  =
  F^+_1 (\sqrt{s},t) \,\langle P_+\rangle_{0\,\frac{1}{2} \to \,0\,\frac{1}{2}}
  \nonumber\\ && \qquad \qquad \qquad \qquad
  +\, F^-_1 (\sqrt{s},t) \,\langle P_-\rangle_{0\,\frac{1}{2} \to \,0\,\frac{1}{2}}
  \,,\nonumber\\ \nonumber\\ &&
  \langle P_\pm \rangle_{0\,\frac{1}{2} \to \,0\,\frac{1}{2}}
  =
  \bar u (\bar p,\lambda_{\bar p})\,P_\pm\,u (p,\lambda_{p})
  \,,
  \label{def-01/2to01/2}
\end{eqnarray}
where the convention for the baryon wave functions $u(p, \lambda_{p})$ and $ \bar u(\bar p, \lambda_{\bar p})$ with the
helicity projections $\lambda_{p}$ and $\lambda_{\bar p}$ is taken from \cite{Stoica:2011cy}. In (\ref{def-01/2to01/2})
we use the projection matrices $P_\pm $ with  $P_\pm \,P_\mp =0$ and
\begin{eqnarray}
  &&
  P_\pm = \frac{1}{2\,\sqrt{s}}\,\Big(\sqrt{s} \pm  \Slash{w}\Big)
  \,, \qquad
  P_{\pm }\,P_{\pm}  =P_\pm
  \,.
  \label{def-Ppm}
\end{eqnarray}
The reaction is characterized by two scalar function. It is well studied in the literature (see e.g. \cite{MacDowell:1959zza,Frazer:1960zz,Liu:1971qt}).
The number of invariant amplitudes follows readily from the number of on-shell independent Dirac matrices.
Further Dirac structures like $\Slash{p}$ or $\Slash{\bar p}$  are redundant
\begin{eqnarray}
  &&
  \langle \Slash{p}\,P_\pm \rangle_{0\,\frac{1}{2} \to \,0\,\frac{1}{2}}
  =
  M \,\langle P_\pm \rangle_{0\,\frac{1}{2} \to \,0\,\frac{1}{2}}
  \,,\nonumber\\ &&
  \langle P_\pm\,\Slash{\bar p} \rangle_{0\,\frac{1}{2} \to \,0\,\frac{1}{2}}
  =
  \bar M \,\langle P_\pm \rangle_{0\,\frac{1}{2} \to \,0\,\frac{1}{2}}
  \,.
\end{eqnarray}
Owing to the MacDowell symmetry \cite{MacDowell:1959zza} it holds
\begin{eqnarray}
  F^-_1(+\sqrt{s}, t) = F^+_1(-\sqrt{s},t) \,.
  \label{def-MacDowell}
\end{eqnarray}
While the functions $F^\pm_1$ depend on $\sqrt{s}$ and $t$ the particular combinations
\begin{eqnarray}
&& F_1(s,t) = F^+_1(\sqrt{s},t) +F^-_1(\sqrt{s},t)\,,
\nonumber\\
&& F_2(s,t) = \sqrt{s}\,\big(F^+_1(\sqrt{s},t) -F^-_1(\sqrt{s},t)\big)\,,
\end{eqnarray}
depend on $s$ and $t$ and do satisfy Mandelstam's dispersion-integral
representation~\cite{Mandelstam:1958xc,Ball:1961zza}. This implies that the functions $F_n(s,t)$,
unlike the functions $F^\pm_1(\sqrt{s},t)$, do not have a square-root branch point at $s=0$.

The invariant amplitudes $F^\pm_1$ can be derived by means of the following projection algebra
\begin{eqnarray}
  &&
  \frac{1}{2}\,\tr \big(P_\pm\, \Lambda\;Q_\pm\, \bar \Lambda\big) = 1
  \,, \qquad
  \Lambda = \Slash{p}+M
  \,,\nonumber\\ &&
  \frac{1}{2}\,\tr \big(P_\pm\, \Lambda\;Q_\mp\, \bar \Lambda\big) = 0
  \,, \qquad
  \bar \Lambda =  \Slash{\bar p}+\bar M
  \,,
  \label{def-Q-01/2to01/2}
\end{eqnarray}
with
\begin{eqnarray}
  &&
  Q_\pm = \frac{s}{v^2}\,\Big( (\bar r \cdot r)\,P_\mp - \bar E_\mp \,E_\mp\,P_\pm \Big)
  \,,
  \nonumber\\ &&
  E_\pm = \frac{\sqrt{s}}{2}\,\big( 1- \delta \big)  \pm M
  \,, \qquad
  \delta = \frac{m^2-M^2}{s}
  \,,\nonumber\\ &&
  \bar E_\pm = \frac{\sqrt{s}}{2}\,\big( 1- \bar \delta \big)  \pm \bar M
  \,, \qquad
  \bar \delta = \frac{\bar m^2-\bar M^2}{s}
  \,,
  \nonumber\\ &&
  v^2 = s\,\big(( \bar r \cdot r)^2 - \bar r^2 \,r^2\big) =-s\,\bar q^2_{\rm cm}\,q^2_{\rm cm} \,\sin^2\theta \,.
  \label{def-Qpm}
\end{eqnarray}

A slightly more complicated process involves one vector particle in the final state
\begin{eqnarray}
  &&
  T_{0\,\frac{1}{2} \to \,1\,\frac{1}{2}}(\bar k,\, k,\, w)
  =
  \sum_{\pm, n} F^\pm_n (\sqrt{s},t) \,\langle  T^{(n)}_{\pm,\bar \mu} \rangle^{\bar \mu}_{0\,\frac{1}{2} \to \,1\,\frac{1}{2}}
  \,,\nonumber\\ &&
  \langle T^{(n)}_{\pm ,\bar \mu }\rangle^{\bar \mu}_{0\,\frac{1}{2} \to \,1\,\frac{1}{2}}
  =
  \epsilon^{\dagger \bar \mu }(\bar q, \lambda_{\bar q})\,\bar u (\bar p,\lambda_{\bar p})\,T^{(n)}_{\pm , \bar \mu}\,u (p,\lambda_{p})
  \,,\nonumber\\
  \nonumber\\ &&
  \begin{array}{ll}
    T^{(1)}_{\pm, \bar \mu } = \hat \gamma_{\bar \mu}\,P_\pm \,i\,\gamma_5\,, \qquad \qquad  &
    T^{(2)}_{\pm, \bar \mu } = w_{\bar \mu}\,P_\pm \,i\,\gamma_5\,,  \\
    T^{(3)}_{\pm, \bar \mu } = r_{\bar \mu}\,P_\pm \,i\,\gamma_5\,,  &
  \end{array}
  \label{def-01/2to11/2}
\end{eqnarray}
where we use a notation analogous to the one introduced in \cite{Stoica:2011cy}.
The sign convention for the vector wave functions $\epsilon_\mu(q, \lambda_q)$ and
$\epsilon_\mu(\bar q, \lambda_{\bar q})$ are given in \cite{Lutz:2011xc}.
In (\ref{def-01/2to11/2}) we use (\ref{def-Ppm}) and
\begin{eqnarray}
  &&
  \hat\gamma_\mu =  \gamma_{ \mu} - \frac{1}{s}\,\Slash{w}\,w_{ \mu}
  \,, \qquad
  P_\pm \,\hat \gamma_\mu =\hat \gamma_\mu\,P_\mp
  \,.
  \label{def-hatgamma}
\end{eqnarray}
For notational simplicity we do not introduce different notations for the invariant amplitudes $F^\pm_n(\sqrt{s},t)$ in the two
reactions (\ref{def-01/2to01/2}, \ref{def-01/2to11/2}).

The number of invariant on-shell amplitudes $F_n^\pm$ with $n=1,2,3$ follows from the number of helicity amplitudes. Since we are assuming parity conservation the total number of independent helicity amplitudes is
\begin{eqnarray}
  \etext{\frac{1}{2}} \,(2 \,S_{q}+1)\,(2 \,S_{p}+1)\,(2 \,S_{\bar q}+1)\,(2 \,S_{\bar p}+1)
  \,,
  \label{def-number-of-helicity-amplitudes}
\end{eqnarray}
where $S_{q}, S_{p}$ and $S_{\bar q}, S_{\bar p}$ are the spins of the initial and final particles.

Since the invariant amplitudes are supposed to be free of kinematical constraints the tensor structures should involve the minimal number of  momenta. There is one structure, $\gamma_{\bar \mu}\,i\,\gamma_5$ which does not involve any momentum. Owing to the transversality of the spin-one wave functions with
\begin{eqnarray}
  &&
  \epsilon_{\bar \mu}(\bar q, \lambda_{\bar q})\,k^{\bar \mu}=  2\,\epsilon_{\bar \mu}(\bar q, \lambda_{\bar q})\,w^{\bar \mu}
  \,,
\end{eqnarray}
there are three structures with one momentum involved, $\gamma_{\bar \mu}\,\Slash{w}\,i\,\gamma_5$, $i\,\gamma_5\,r_{\bar \mu}$ and $i\,\gamma_5\,w_{\bar \mu}$. Those 4 structures are part of our basis. It is left to identify the remaining two tensors, which involve necessarily two momenta. Our basis suggests the two structures $\Slash{w}\,i\,\gamma_5\,r_{\bar \mu}$ and $\Slash{w}\,i\,\gamma_5\,w_{\bar \mu}$, but there are three more candidates built from the tensor $\epsilon_{\bar \mu \mu \alpha \beta}\,\gamma^\mu$ contracted with two momenta. It is left to show that for on-shell conditions the latter three tensors can be decomposed into our basis without generating kinematical singularities. Indeed, it holds
\begin{eqnarray}
  &&
  \sqrt{s}\, \langle
  \epsilon_{\bar \mu \mu \alpha \beta}\,\gamma^\mu\,P_{\pm}\,\bar r^\alpha \,\,r^\beta
  \rangle^{\bar \mu}_{0\,\frac{1}{2} \to \,1\,\frac{1}{2}}
  =
  \langle
  \pm\,\sqrt{s}\, \bar E_\pm \, T^{(3)}_{\mp, \bar \mu }
  \nonumber\\ && \qquad
  -\,\sqrt{s}\,\bar E_\pm\, E_\pm\,T^{(1)}_{\pm, \bar \mu } + \sqrt{s}\, (\bar r \cdot r)\,T^{(1)}_{\mp, \bar \mu }
  \nonumber\\ && \qquad
  -\, (\bar M \mp \sqrt{s}\,)\,\,E_\pm\,T^{(2)}_{\pm, \bar \mu }  \mp (\bar r \cdot r)\,T^{(2)}_{\mp, \bar \mu }
  \rangle^{\bar \mu}_{0\,\frac{1}{2} \to \,1\,\frac{1}{2}}
  \,,\nonumber\\
  &&
  \frac{1}{\sqrt{s}}\, \langle
  \epsilon_{\bar \mu \mu \alpha \beta}\,\gamma^\mu\,P_{\pm}\,w^\alpha \,r^\beta
  \rangle^{\bar \mu}_{0\,\frac{1}{2} \to \,1\,\frac{1}{2}}
  =
  \langle
  -\, E_\pm\, T^{(1)}_{\pm, \bar \mu }
  \nonumber\\ && \hspace{4.1cm}
  \pm\, T^{(3)}_{\mp, \bar \mu }
  \rangle^{\bar \mu}_{0\,\frac{1}{2} \to \,1\,\frac{1}{2}}
  \,,\nonumber\\
  &&
  \frac{1}{\sqrt{s}}\, \langle
  \epsilon_{\bar \mu \mu \alpha \beta}\,\gamma^\mu\,P_{\pm }\,\bar r^\alpha \,w^\beta
  \rangle^{\bar \mu}_{0\,\frac{1}{2} \to \,1\,\frac{1}{2}}
  =
  \langle
  -\, \bar E_\mp\, T^{(1)}_{\mp, \bar \mu }
  \nonumber\\ && \hspace{3.1cm}
  \mp\,{\textstyle{1\over 2}}\,(\bar \delta +1)\,T^{(2)}_{\mp, \bar \mu }
  \rangle^{\bar \mu}_{0\,\frac{1}{2} \to \,1\,\frac{1}{2}}\,.
  \label{eps-gamma-redundance}
\end{eqnarray}
It is almost obvious that any tensor involving three or more momenta has a regular on-shell decomposition into our basis. For instance, consider the two additional structures $v_\mu \,P_\pm$ with
\begin{eqnarray}
  v_{\mu} = \epsilon_{\mu \alpha \tau \beta}\,\bar k^\alpha \,w^{\tau}\, k^\beta\,.
  \label{def-eps-vector}
\end{eqnarray}
The on-shell identity
\begin{eqnarray}
  &&
  \frac{1}{\sqrt{s}}\, \langle v_{\bar \mu} \,P_\pm \rangle^{\bar \mu}_{0\,\frac{1}{2} \to \,1\,\frac{1}{2}}
  =
  \langle
  \pm \,\bar E_\pm \,E_\pm \,T^{(1)}_{\pm, \bar \mu }
  \mp  (\bar r \cdot r)\, T^{(1)}_{\mp, \bar \mu }
  \nonumber\\ && \qquad
  -\, {\textstyle{1\over 2}}\,(\bar \delta +1)\,E_\pm \, T^{(2)}_{\pm, \bar \mu }
  - \bar E_\pm\, T^{(3)}_{\mp, \bar \mu }
  \rangle^{\bar \mu}_{0\,\frac{1}{2} \to \,1\,\frac{1}{2}}\,,
\label{v-redundance}
\end{eqnarray}
shows that such structures are linear dependent of the six tensors $T^{(n)}_{\pm, \bar \mu}$ introduced in
(\ref{def-01/2to11/2}). We conclude that the particular choice (\ref{def-01/2to11/2})
leads to invariant amplitudes $F^\pm_n(\sqrt{s}, t)$ that are free of kinematical constraints and manifest the
MacDowell relations  with
\begin{eqnarray}
  F^-_n(+\sqrt{s}, t) = F^+_n(-\sqrt{s},t)
  \,.
  \label{def-MacDowell-general}
\end{eqnarray}

It is useful to derive a suitable projection algebra analogous to the one displayed in (\ref{def-Q-01/2to01/2}). We find
\begin{eqnarray}
  &&
  \frac{1}{2}\,\tr \big(T^{(n)}_{a, \bar \mu}\, \Lambda\;Q^{\bar \mu}_{b, k}\, \bar \Lambda\big)
  =
  \delta_{nk}\,\delta_{ab}
  \,, \qquad
  \bar q_{\bar \mu}\, Q^{\bar \mu}_{\pm, n} =0
  \,,\nonumber\\ &&
  Q^{\bar \mu}_{\pm, 1}
  =
  \pm \frac{\sqrt{s}}{v^2}\,P_\pm \,v^{\bar \mu }
  \,, \qquad\nonumber\\ &&
  Q^{\bar \mu}_{\pm, 2}
  =
  -\, i\,\gamma_5\,R_\pm\, \wLbot^{\bar \mu}
  - {\textstyle{1\over 2}}\, (1+\bar \delta \,)\,\frac{\sqrt{s}}{v^2}\,E_\pm \,Q_\pm\,v^{\bar \mu}
  \,, \qquad\nonumber\\ &&
  Q^{\bar \mu}_{\pm, 3}
  =
  -\, i\, \gamma_5\, R_\pm \,\rbot^{\bar \mu}
  -  \frac{\sqrt{s}}{v^2}\,\bar E_\mp\, Q_\mp \,v^{\bar \mu}
  \,,
  \label{def-Q-01/2to11/2}
\end{eqnarray}
with
\begin{eqnarray}
  &&
  R_{\pm} = \frac{s}{v^2}\,\Big( \bar E_\mp \,E_\pm\,P_\pm -(\bar r \cdot r)\,P_\mp   \Big)
  \,.
  \label{def-Rpm}
\end{eqnarray}
Following \cite{Lutz:2011xc} the 4-vectors $\rbot, \wLbot$ and $\wRbot$ are suitable linear combinations of $\bar r, r$ and $w$ as to have the convenient properties
\begin{eqnarray}
  &&
  \,\rbot \cdot r \,= 1\,,\qquad \,\,\rbot \cdot \bar r = 0 = \,\rbot \cdot w
  \,,\nonumber\\ &&
  \wLbot  \cdot w = 1\,,\qquad \wLbot  \cdot r = 0 = \wLbot  \cdot \bar q
  \,,\nonumber\\ &&
  \wRbot  \cdot w = 1\,,\qquad \wRbot  \cdot r = 0 = \wRbot  \cdot \bar p
  \,.
\label{bot-properties}
\end{eqnarray}
The index ${\small \rfloor}$ and ${\small \lfloor}$ of a vector indicates whether it is orthogonal to the 4 momentum of the first or second particle respectively. The patched symbol ${\small \rfloor \! \!\lfloor}$ implies the orthogonality to both 4 momenta. We recall the explicit form of the auxiliary vectors $\rbot, \wLbot$ and $\wRbot$ from \cite{Lutz:2011xc}. Given three 4-vectors $a_\mu, b_\mu$ and $c_\mu$ we introduce a vector, $a^\mu_{b\, c}= a^\mu_{c\, b}$, as follows
\begin{eqnarray}
  &&
  \frac{a^\mu_{b \,c}}{a_{b \,c} \cdot a_{b\, c}} = a^\mu  - \frac{a \cdot c}{c \cdot c}\,c^\mu
  \nonumber\\ && \qquad \qquad
  \,-\, \frac{a \cdot (b- \frac{c \cdot b }{c \cdot c}\,c )}{(b- \frac{c \cdot b }{c \cdot c}\,c )^2}
  \left( b^\mu - \frac{c \cdot b }{c \cdot c}\,c^\mu  \right)
  \,,\nonumber\\ &&
  a^\mu_{b\, c} \,a_\mu =1 \,, \qquad a^\mu_{b\, c} \,b_\mu =0 \,,\qquad a^\mu_{b\, c} \,c_\mu =0
  \,.
  \label{def-abc}
\end{eqnarray}
In the notation of (\ref{def-abc}) the desired vectors are identified with
\begin{eqnarray}
  &&
  \rbot^\mu = r^\mu_{\bar r \,w} \,, \qquad \wLbot^\mu = w^\mu_{r \,\bar q}\,, \qquad  \wRbot^\mu = w^\mu_{r \,\bar p}
  \,,\nonumber\\ &&
  \rbarbot^\mu = \bar r^\mu_{r\, w} \,,\qquad \wLbarbot^\mu = w^\mu_{\bar r\, q}\,, \qquad  \wRbarbot^\mu = w^\mu_{\bar r \,p}
  \,,
  \label{def-rbot}
\end{eqnarray}
where we introduced the additional vectors $\rbarbot, \wLbarbot$ and $\wRbarbot$ that will turn useful below. In the
derivation of (\ref{def-Q-01/2to11/2})  the following expressions are useful
\begin{eqnarray}
  &&
  v^2 = s\,\big(( \bar r \cdot r)^2 - \bar r^2 \,r^2\big)
  \,,\nonumber\\ &&
  v^2\,(\,\rbot \cdot \,\rbarbot ) = s\, (\bar r\cdot r)
  \,,\quad
  v^2\,(\,\rbot \cdot \,\rbot ) = - s\, \bar r^2
  \,,\nonumber\\ &&
  s\,(\wRbot \cdot \wLbot) =  1 + {\textstyle {1 \over 4}}\,(\bar \delta +1) \,(\bar \delta -1)\,s\,(\rbarbot \cdot \rbarbot)
  \,,\nonumber\\ &&
  s\,(\wRbot \cdot \wLbarbot) =  1 + {\textstyle {1 \over 4}}\,(\bar \delta -1) \,(\delta +1)\,s\,(\rbot \cdot \rbarbot)
  \,,\nonumber\\ &&
  v^2\,(\,\rbot \cdot \wLbot ) = - {\textstyle {1 \over 2}}\,(\bar \delta +1) \,s\,(\bar r \cdot r)
  \,,\nonumber\\ &&
  v^2\,(\,\rbot \cdot \wRbot) = - {\textstyle {1 \over 2}}\, (\bar \delta -1) \,s\, (\bar r \cdot r)
  \,,\nonumber\\ &&
  (\wLbot \cdot \bar r) = - {\textstyle {1 \over 2}}\, (\bar \delta +1)
  \,,\quad
  (\wRbot \cdot \bar r) = - {\textstyle {1 \over 2}}\, (\bar \delta -1)
  \,,\nonumber\\ &&
  (\wLbarbot \cdot r) = - {\textstyle {1 \over 2}}\, (\delta +1)
  \,,\quad
  (\wRbarbot \cdot r) = - {\textstyle {1 \over 2}}\, (\delta -1)
  \,.
\label{res-rbot-wbot}
\end{eqnarray}

We turn to the scattering of spin-one bosons off spin-one-half fermions. The scattering amplitude takes the generic form
\begin{eqnarray}
  &&
  T_{1\,\frac{1}{2} \to \,1\,\frac{1}{2}}(\bar k,\, k,\, w)
  =
  \sum_{\pm, n} F^\pm_n (\sqrt{s},t) \,\langle T^{(n)}_{\pm,\bar \mu \mu} \rangle^{\bar \mu \mu}_{1\,\frac{1}{2} \to \,1\,\frac{1}{2}}
  \,,\nonumber\\ &&
  \langle T^{(n)}_{\pm ,\bar \mu \mu}\rangle^{\bar \mu \mu}_{1\,\frac{1}{2} \to \,1\,\frac{1}{2}}
  =
  \epsilon^{\dagger \bar \mu }(\bar q, \lambda_{\bar q})\,\bar u (\bar p,\lambda_{\bar p})
  \nonumber\\ && \qquad \qquad \qquad  \;\;
  \,\times\, T^{(n)}_{\pm , \bar \mu \mu}\, u (p,\lambda_{p})\,\epsilon^\mu(q, \lambda_{q} )
  \,,\nonumber\\
  \nonumber\\ &&
  \begin{array}{ll}
    T^{(1)}_{\pm, \bar \mu \mu} = \hat g_{\bar \mu \mu} \,P_\pm \,,\qquad \qquad \qquad &
    T^{(2)}_{\pm, \bar \mu \mu} = \hat \gamma_{\bar \mu}\,P_\pm \,\hat \gamma_{ \mu}\,, \\
    T^{(3)}_{\pm, \bar \mu \mu} = \hat \gamma_{\bar \mu}\,P_\pm \,w_{ \mu}\,,  &
    T^{(4)}_{\pm, \bar \mu \mu} = w_{\bar \mu}\,P_\pm \,\hat \gamma_{ \mu} \,, \\
    T^{(5)}_{\pm, \bar \mu \mu} = \hat \gamma_{\bar \mu}\,P_\pm \,\bar r_{ \mu}\,,  &
    T^{(6)}_{\pm, \bar \mu \mu} = r_{\bar \mu}\,P_\pm \,\hat \gamma_{ \mu} \,,\\
    T^{(7)}_{\pm, \bar \mu \mu} = w_{\bar \mu}\,P_\pm \,\bar r_{ \mu}\,,&
    T^{(8)}_{\pm, \bar \mu \mu} = r_{\bar \mu}\,P_\pm \,w_\mu\,, \\
    T^{(9)}_{\pm, \bar \mu \mu} = w_{\bar \mu}\,P_\pm \,w_\mu\,,  &
  \end{array}
  \label{def-11/2to11/2}
\end{eqnarray}
where we use
\begin{eqnarray}
  \ghat_{ \mu  \nu} = g_{ \mu  \nu}- \frac{w_{\mu}\,w_{\nu}}{s}
  \,.
\end{eqnarray}
The invariant amplitudes $F^\pm_n(\sqrt{s}, t)$ satisfy the MacDowell relations (\ref{def-MacDowell-general}).
At first there appear many possible tensor structures.  Besides the 18 structures introduced in (\ref{def-11/2to11/2})
there are for instance the tensors
$v_{\bar \mu}\, P_\pm\,v_{\mu} , r_{\bar \mu}\,P_\pm \,\bar r_\mu$ and $\epsilon_{\bar \mu \mu \alpha \beta}\,\bar r^\alpha\,r^\beta \,P_\pm \,\gamma_5$.
They can be decomposed into our basis tensors with regular coefficients. As an example we display the identity
\begin{eqnarray}
  &&
  v_{\bar \mu}\, v_{\mu}
  =
  v^2\,\Big[ g_{\bar \mu \mu} - (\wLbot \cdot \wLbarbot) \,w_{\bar \mu}\, w_{\mu} -(\wLbot \cdot \rbarbot )\, w_{\bar \mu}\, \bar r_{\mu}
  \nonumber\\ && \qquad \qquad
  -\, (\rbot \cdot \wLbarbot)\,r_{\bar \mu}\, w_{\mu}- (\rbot \cdot \rbarbot )\,r_{\bar \mu}\,  \bar r_{\mu} \Big]
  \,.
  \label{res-vv}
\end{eqnarray}

A derivation of the invariant amplitudes $F^\pm_n(\sqrt{s},t)$ for a given physical system turns more and more tedious
as the spins of the involved particles and therewith the number of invariant amplitudes increases. For the considered
case there are 18 amplitudes to be determined and without a systematized approach this appears prohibitively cumbersome.
Fortunately it is possible to establish a projection algebra analogous to the one displayed
in (\ref{def-Q-01/2to01/2}, \ref{def-Q-01/2to11/2}) also for the more complicated cases.  By means of computer
algebra programs it is then straight forward to determine the amplitudes. We find
\begin{eqnarray}
  &&
  \frac{1}{2}\,\tr \big(T^{(n)}_{a, \bar \mu \mu }\, \Lambda\;Q^{\bar \mu \mu}_{b, k}\, \bar \Lambda\big)
  =
  \delta_{nk}\,\delta_{ab}
  \,,
  \label{def-Q-11/2to11/2}
  \\ &&
  {\rm with} \qquad
  \bar q_{\bar \mu}\, Q^{\bar \mu \mu }_{\pm, k} =0  \,,\qquad q_{ \mu}\, Q^{\bar \mu \mu }_{\pm, k} =0
  \,,\nonumber\\ &&
  Q^{\bar \mu \mu}_{\pm, 1} = \frac{1}{v^2}\,v^{\bar \mu }\,Q_\pm \,v^\mu -  Q^{\bar \mu \mu}_{\mp, 2}
  \,,\nonumber\\ &&
  Q^{\bar \mu  \mu}_{\pm, 2} =  \rbot^{\bar \mu}\,\Big[P_\pm -2\,(\bar r \cdot r) \,Q_\mp \Big]\, \rbarbot^{\mu}
  \nonumber\\ &&\quad
  - \,(\rbarbot \cdot \rbot)\,\frac{1}{v^2}\,v^{\bar \mu}\,\Big[P_\pm -2\,(\bar r \cdot r) \,Q_\mp \Big]\,v^\mu
  \nonumber\\ &&\quad
  + \,\bar E_\pm\,\frac{\sqrt{s}}{v^2}\,v^{\bar \mu}\,i\,\gamma_5\,\Big[P_\mp +2\,(\bar r \cdot r) \,R_\pm  \Big]\,\rbarbot^\mu
  \nonumber\\ &&\quad
  - \,E_\pm\,\frac{\sqrt{s}}{v^2}\,\rbot^{\bar \mu}\,i\,\gamma_5\,\Big[P_\pm +2\,(\bar r \cdot r) \,R_\mp \Big]\,v^{\mu}
  \,,\nonumber\\ &&
  Q^{\bar \mu  \mu}_{\pm, 3} =  \mp \,\frac{\sqrt{s}}{v^2}\,v^{\bar \mu}\,i\,\gamma_5\,P_\pm \,\wLbarbot^{\mu}
  \nonumber\\ &&\quad
  \pm \,{\textstyle{1\over 2}}\,(\delta+1)\,\frac{s}{v^2} \,\Big[  (\bar r \cdot r)\,\bar E_\pm\,Q^{\bar \mu  \mu}_{\mp, 2}
  - (\bar r \cdot \bar r)\,E_\mp\,Q^{\bar \mu  \mu}_{\pm, 2} \Big]
  \,,\nonumber\\ &&
  Q^{\bar \mu  \mu}_{\pm, 4} =  \pm \,\frac{\sqrt{s}}{v^2}\,\wLbot^{\bar \mu}\,P_\pm \,i\,\gamma_5\,v^{\mu}
  \nonumber\\ && \quad
  \pm \,{\textstyle{1\over 2}}\,(\bar \delta+1)\,\frac{s}{v^2} \,\Big[  (\bar r \cdot r)\,E_\pm\,Q^{\bar \mu  \mu}_{\mp, 2}
  - (r \cdot r)\,\bar E_\mp\,Q^{\bar \mu  \mu}_{\pm, 2} \Big]
  \,,\nonumber\\ &&
  Q^{\bar \mu  \mu}_{\pm, 5} = \mp \,\frac{\sqrt{s}}{v^2}\,v^{\bar \mu}\,i\,\gamma_5\,P_\pm \,\rbarbot^{\mu}
  \nonumber\\ && \quad
  \pm\,\frac{s}{v^2}\,\Big[  ( r \cdot r)\,\bar E_\pm\,Q^{\bar \mu  \mu}_{\mp, 2}
  - (\bar r \cdot r)\,E_\mp\,Q^{\bar \mu  \mu}_{\pm, 2} \Big]
  \,,\nonumber\\ &&
  Q^{\bar \mu  \mu}_{\pm, 6} = \pm \,\frac{\sqrt{s}}{v^2}\,\rbot^{\bar \mu}\,P_\pm \,i\,\gamma_5\,v^{\mu}
  \nonumber\\ && \quad
  \pm\,\frac{s}{v^2}\,\Big[  ( \bar r \cdot \bar r)\,E_\pm\,Q^{\bar \mu  \mu}_{\mp, 2}
  - (\bar r \cdot r)\,\bar E_\mp\,Q^{\bar \mu  \mu}_{\pm, 2} \Big]
  \,,\nonumber\\ &&
  Q^{\bar \mu  \mu}_{\pm, 7} =Q_\pm \,\Big( \wLbot^{\bar \mu} \,\rbarbot^{ \mu} - (\wLbot \cdot \rbarbot)\,\frac{1}{v^2}\,v^{\bar \mu }\,v^\mu\Big)
  \nonumber\\ && \quad
  \pm\, (r \cdot r) \,\bar E_\mp\,\frac{s}{v^2}\,\Big[Q^{\bar \mu  \mu}_{\mp, 4} - {\textstyle{1\over 2}}\,(\bar \delta + 1)\,Q^{\bar \mu  \mu}_{\pm, 5}\Big]
  \nonumber\\ && \quad
  \mp\, (\bar r \cdot r) \,E_\mp\,\frac{s}{v^2}\,\Big[Q^{\bar \mu  \mu}_{\pm, 4} - {\textstyle{1\over 2}}\,(\bar \delta + 1)\,Q^{\bar \mu  \mu}_{\mp, 5}\Big]
  \,,\nonumber\\ &&
  Q^{\bar \mu  \mu}_{\pm, 8} = Q_\pm \,\Big( \rbot^{\bar \mu} \,\wLbarbot^{ \mu} - (\rbot \cdot \wLbarbot)\,\frac{1}{v^2}\,v^{\bar \mu }\,v^\mu \Big)
  \nonumber\\ && \quad
  \pm\, (\bar r \cdot \bar r) \,E_\mp\,\frac{s}{v^2}\,\Big[Q^{\bar \mu  \mu}_{\mp, 3} - {\textstyle{1\over 2}}\,(\delta + 1)\,Q^{\bar \mu  \mu}_{\pm, 6}\Big]
  \nonumber\\ && \quad
  \mp\, (\bar r \cdot r) \,\bar E_\mp\,\frac{s}{v^2}\,\Big[Q^{\bar \mu  \mu}_{\pm, 3} - {\textstyle{1\over 2}}\,(\delta + 1)\,Q^{\bar \mu  \mu}_{\mp, 6}\Big]
  \,,\nonumber\\ &&
  Q^{\bar \mu  \mu}_{\pm, 9} =   Q_\pm \,\Big( \wLbot^{\bar \mu} \,\wLbarbot^\mu - \Big(\wLbot \cdot \wLbarbot - \frac{1}{s} \Big)\,\frac{1}{v^2}\,v^{\bar \mu} \,v^\mu \Big)
  \nonumber\\ && \quad
  -\,{\textstyle{1\over 4}}\, (\bar \delta + 1)\,(\delta + 1)\,\frac{s}{v^2}\, \Big[ (\bar r \cdot r)\,Q^{\bar \mu  \mu}_{\mp, 2}
  - \bar E_\mp \,E_\mp\,Q^{\bar \mu  \mu}_{\pm, 2}    \Big]
  \nonumber\\ && \quad
  \pm\,{\textstyle{1\over 2}}\, (\bar \delta + 1)\,\frac{s}{v^2}\,\Big[ (\bar r \cdot r)\,E_\mp\,Q^{\bar \mu  \mu}_{\mp, 3}
  - (r \cdot r) \,\bar E_\mp\,Q^{\bar \mu  \mu}_{\pm, 3}   \Big]
  \nonumber\\ && \quad
  \pm\,{\textstyle{1\over 2}}\, (\delta + 1)\,\frac{s}{v^2}\,\Big[ (\bar r \cdot r)\,\bar E_\mp\,Q^{\bar \mu  \mu}_{\mp, 4}
  - (\bar r \cdot \bar r) \,E_\mp\,Q^{\bar \mu  \mu}_{\pm, 4}   \Big]
  \,.
  \nonumber
\end{eqnarray}

We turn to the reactions involving the spin-three-half fermions. The simplest reaction is the production process
\begin{eqnarray}
  &&
  T_{0\,\frac{1}{2} \to \,0\,\frac{3}{2}}(\bar k,\, k,\, w)
  =
  \sum_{\pm, n} F^\pm_n (\sqrt{s},t) \,\langle T^{(n)}_{\pm,\bar \nu } \rangle^{\bar \nu}_{0\,\frac{1}{2} \to \,0\,\frac{3}{2}}
  \,,\nonumber\\ &&
  \langle T^{(n)}_{\pm ,\bar \nu  }\rangle^{\bar \nu}_{0\,\frac{1}{2} \to \,0\,\frac{3}{2}}
  =
  \bar u^{\bar \nu} (\bar p, \lambda_{\bar p})\,T^{(n)}_{\pm , \bar \nu }\,u (p,\lambda_p)
  \,,\nonumber\\
  \nonumber\\ &&
  \begin{array}{ll}
    T^{(1)}_{\pm, \bar \nu } = w_{\bar \nu}\,P_\pm \,i\,\gamma_5\,, \quad  &
    \qquad T^{(2)}_{\pm, \bar \nu } = r_{\bar \nu}\,P_\pm \,i\,\gamma_5\,,
  \end{array}
  \label{def-01/2to03/2}
\end{eqnarray}
where we refer to \cite{Stoica:2011cy} for the convention used for the spin-three-half wave function
$\bar u_{\bar \nu} (\bar p, \lambda_{\bar p})$. For the associated projection algebra we derive
\begin{eqnarray}
  &&
  \frac{1}{2}\,\tr \big(T^{(n)}_{a, \bar \nu}\, \Lambda\;Q^{\bar \nu}_{b, k}\, \bar \Lambda \big)
  =
  \delta_{nk}\,\delta_{ab}
  \,, \qquad\nonumber\\ &&
  {\rm with} \qquad
  \Lambda\;Q^{\bar \nu}_{\pm, k}\, \bar \Lambda \,\gamma_{\bar \nu} = 0 \,, \qquad \bar p_{\bar \nu}\,Q^{\bar \nu}_{\pm,k}=0
  \,,\nonumber\\ &&
  Q^{\bar \nu}_{\pm, 1} = \frac{s}{v^2}\,\Big[ (\bar r\cdot r)\,P^{\bar \nu}_{\mp, 1} - \bar E_\mp\,E_\pm\,P^{\bar \nu}_{\pm, 1}\Big]
  \,,\nonumber\\ &&
  Q^{\bar \nu}_{\pm, 2} = \frac{s}{v^2}\,\Big[ (\bar r\cdot r)\,P^{\bar \nu}_{\mp, 2} - \bar E_\mp\,E_\pm\,P^{\bar \nu}_{\pm, 2}\Big]
  \,,
  \label{def-Q-01/2to03/2}
\end{eqnarray}
where $P^{\bar \nu}_{\pm, k}$ are designed to satisfy
on-shell conditions  as follows
\begin{eqnarray}
  && \Lambda\,P^{\bar \nu}_{\pm, k}\,\bar\Lambda \,\gamma_{\bar \nu} = 0\,, \qquad
  \bar p_{\bar \nu}\,P^{\bar \nu}_{\pm,k}=0 \,,
  \nonumber\\ \nonumber\\
  &&
  P^{\bar \nu}_{\pm, 1}
  =
  \wRbot^{\bar\nu}\,i\,\gamma_5\,P_\pm - v^{\bar\nu}\,((\bar r\cdot r)\,P_\mp \pm\,\bar M\,E_\mp\,P_\pm)/v^2
  \,,\nonumber\\ &&
  P^{\bar \nu}_{\pm, 2}
  =
  \rbot^{\bar\nu}\,i\,\gamma_5\,P_\pm - v^{\bar\nu}\,(\sqrt{s}\,\bar E_\pm\,P_\mp)/v^2
  \,.\nonumber
\end{eqnarray}
There are left further reactions involving at least one spin-three-half particle.
We derived complete lists of regular tensors and their associated projection algebras
\begin{eqnarray}
  &&
  T_{1\,\frac{1}{2} \to \,0\,\frac{3}{2}}(\bar k,\, k,\, w)
  =
  \sum_{\pm, n} F^\pm_n (\sqrt{s},t) \,\langle T^{(n)}_{\pm,\bar \nu \mu} \rangle^{\bar \nu \mu}_{1\,\frac{1}{2} \to \,0\,\frac{3}{2}}
  \,,\nonumber\\
  &&
  \langle T^{(n)}_{\pm ,\bar \nu \mu}\rangle^{\bar \nu \mu}_{1\,\frac{1}{2} \to \,0\,\frac{3}{2}}
  =
  \bar u^{\bar \nu} (\bar p,\lambda_{\bar p})\,T^{(n)}_{\pm , \bar \nu \mu}\, u (p,\lambda_{p})\,\epsilon^\mu(q, \lambda_{q} )
  \,,\nonumber\\ \nonumber\\
  &&
  \begin{array}{ll}
    T^{(1)}_{\pm, \bar \nu \mu} = \hat g_{\bar \nu \mu} \,P_\pm \,,\qquad \qquad \qquad &
    T^{(2)}_{\pm, \bar \nu \mu} = w_{\bar \nu}\,P_\pm \,\hat \gamma_{ \mu} \,, \\
    T^{(3)}_{\pm, \bar \nu \mu} = r_{\bar \nu}\,P_\pm \,\hat \gamma_{ \mu} \,,&
    T^{(4)}_{\pm, \bar \nu \mu} = w_{\bar \nu}\,P_\pm \,\bar r_{ \mu}\,,\\
    T^{(5)}_{\pm, \bar \nu \mu} = r_{\bar \nu}\,P_\pm \,w_\mu\,, &
    T^{(6)}_{\pm, \bar \nu \mu} = w_{\bar \nu}\,P_\pm \,w_\mu\,,
  \end{array}
   \nonumber\\ \nonumber\\
 && \frac{1}{2}\,\tr \big(T^{(n)}_{a, \bar \nu \mu }\, \Lambda\;Q^{\bar \nu \mu}_{b, k}\, \bar \Lambda\big)
  =
  \delta_{nk}\,\delta_{ab}
  \label{def-11/2to03/2}
  \,,\\
  && {\rm with} \qquad
  \bar p_{\bar \nu}\, Q^{\bar \nu \mu }_{\pm, k} =0 = \Lambda\,Q^{\bar \nu \mu }_{\pm, k}\,\bar\Lambda\,\gamma_{\bar \nu}
  \,,\quad
  q_{ \mu}\, Q^{\bar \nu \mu }_{\pm, k} =0
  \,,
\nonumber\\  \nonumber\\
&& T_{0\,\frac{1}{2} \to \,1\,\frac{3}{2}}(\bar k,\, k,\, w)
  =
  \sum_{\pm, n} F^\pm_n (\sqrt{s},t) \,\langle T^{(n)}_{\pm,\bar \mu \bar \nu} \rangle^{\bar \mu \bar \nu}_{0\,\frac{1}{2} \to \,1\,\frac{3}{2}}
  \,,\nonumber\\
  &&
  \langle T^{(n)}_{\pm ,\bar \mu \bar \nu}\rangle^{\bar \mu \bar \nu}_{0\,\frac{1}{2} \to \,1\,\frac{3}{2}}
  =
  \epsilon^{\dagger \bar \mu }(\bar q, \lambda_{\bar q})\,\bar u^{\bar \nu} (\bar p,\lambda_{\bar p})\,
  T^{(n)}_{\pm , \bar \mu \bar \nu}\,u(p,\lambda_{p})
  \,,\nonumber\\ \nonumber\\
  &&
  \begin{array}{ll}
    T^{(1)}_{\pm, \bar \mu \bar \nu} = \hat g_{\bar \mu \bar \nu} \,P_\pm \,,\qquad \qquad \qquad &
    T^{(2)}_{\pm, \bar \mu \bar \nu} = \hat \gamma_{\bar \mu}\,w_{\bar \nu}\, P_\pm \,,  \\
    T^{(3)}_{\pm, \bar \mu \bar \nu} = \hat \gamma_{\bar \mu}\,r_{\bar \nu}\, P_\pm \,,  &
    T^{(4)}_{\pm, \bar \mu \bar \nu} = w_{\bar \mu}\,w_{\bar  \nu}\, P_\pm \,,\\
    T^{(5)}_{\pm, \bar \mu \bar \nu} = w_{\bar \mu}\,r_{\bar \nu}\, P_\pm \,, &
    T^{(6)}_{\pm, \bar \mu \bar \nu} = r_{\bar \mu}\,r_{\bar \nu}\, P_\pm \,,
  \end{array}
  \nonumber\\ \nonumber\\
 && \frac{1}{2}\,\tr \big(T^{(n)}_{a, \bar \mu \bar \nu }\, \Lambda\;Q^{\bar \mu \bar \nu}_{b, k}\, \bar \Lambda\big)
  =
  \delta_{nk}\,\delta_{ab}
  \label{def-01/2to13/2}
  \,,\\
  && {\rm with} \quad
  \bar q_{\bar \mu}\, Q^{\bar \mu \bar \nu }_{\pm, k} =0
  \,,\quad
  \bar p_{ \bar\nu}\, Q^{\bar \mu \bar \nu }_{\pm, k} =0= \Lambda\,Q^{\bar \mu \bar \nu }_{\pm, k}\,\bar\Lambda\,\gamma_{\bar \nu}
  \,,\nonumber\\ \nonumber\\
&&
  T_{0\,\frac{3}{2} \to \,0\,\frac{3}{2}}(\bar k,\, k,\, w)
  =
  \sum_{\pm, n} F^\pm_n (\sqrt{s},t) \,\langle T^{(n)}_{\pm,\bar \nu \nu} \rangle^{\bar \nu \nu}_{0\,\frac{3}{2} \to \,0\,\frac{3}{2}}
  \,,\nonumber\\
  &&
  \langle T^{(n)}_{\pm ,\bar \nu \nu}\rangle^{\bar \nu \nu}_{0\,\frac{3}{2} \to \,0\,\frac{3}{2}}
  =
  \bar u^{\bar \nu} (\bar p,\lambda_{\bar p})\,T^{(n)}_{\pm , \bar \nu \nu}\, u^\nu (p,\lambda_{p})
  \,,\nonumber\\ \nonumber\\
  &&
  \begin{array}{ll}
    T^{(1)}_{\pm, \bar \nu \nu} = \hat g_{\bar \nu \nu} \,P_\pm \,,\qquad \qquad \qquad &
    T^{(2)}_{\pm, \bar \nu \nu} = w_{\bar \nu}\,P_\pm \,\bar r_{ \nu}\,,\\
    T^{(3)}_{\pm, \bar \nu \nu} = r_{\bar \nu}\,P_\pm \,w_\nu\,, &
    T^{(4)}_{\pm, \bar \nu \nu} = w_{\bar \nu}\,P_\pm \,w_\nu\,,
  \end{array}
   \nonumber\\ \nonumber\\
 && \frac{1}{2}\,\tr \big(T^{(n)}_{a, \bar \nu \nu }\, \Lambda\;Q^{\bar \nu \nu}_{b, k}\, \bar \Lambda\big) = \delta_{nk}\,\delta_{ab}
  \label{def-03/2to03/2}
  \,,\\
  && {\rm with} \qquad
  \bar p_{\bar \nu}\, Q^{\bar \nu \nu }_{\pm, k} =0 = \Lambda\,Q^{\bar \nu \nu }_{\pm, k}\,\bar\Lambda\,\gamma_{\bar \nu}
  \,,\nonumber\\
  && \qquad  \qquad
  p_{ \nu}\, Q^{\bar \nu \nu }_{\pm, k} =0=\gamma_\nu\,\Lambda\,Q^{\bar \nu \nu }_{\pm, k}\,\bar\Lambda
  \,,
  \nonumber\\ \nonumber\\
  && T_{1\,\frac{1}{2} \to \,1\,\frac{3}{2}}(\bar k,\, k,\, w)
  =
  \sum_{\pm, n} F^\pm_n (\sqrt{s},t) \,\langle T^{(n)}_{\pm,\bar \mu \bar \nu \mu} \rangle^{\bar \mu \bar \nu \mu}_{1\,\frac{1}{2} \to \,1\,\frac{3}{2}}
  \,,\nonumber\\
  &&
  \langle T^{(n)}_{\pm ,\bar \mu \bar \nu \mu}\rangle^{\bar \mu \bar \nu \mu}_{1\,\frac{1}{2} \to \,1\,\frac{3}{2}}
  =
  \epsilon^{\dagger \bar \mu }(\bar q, \lambda_{\bar q})\,\bar u^{\bar \nu} (\bar p,\lambda_{\bar p})
  \,\nonumber\\
  &&\qquad\qquad\qquad \qquad\qquad \times
  T^{(n)}_{\pm , \bar \mu \bar \nu \mu}\,u(p,\lambda_{p})\, \epsilon^{\mu}(q,\lambda_{q})
  \,,\nonumber\\ \nonumber\\
  &&
  \begin{array}{ll}
    T^{(1)}_{\pm, \bar \mu \bar \nu \mu} = \hat g_{\bar \mu \bar \nu} \,P_\pm \,\hat\gamma_{\mu}\,  i\,\gamma_5\,,  \qquad &
    T^{(2)}_{\pm, \bar \mu \bar \nu \mu} = \hat\gamma_{\bar\mu} \,\hat g_{\bar \nu \mu}\,P_\pm \,  i\,\gamma_5\,,\\
    T^{(3)}_{\pm, \bar \mu \bar \nu \mu} = \hat g_{\bar \mu \bar \nu} \,P_\pm \,w_{\mu} \,  i\,\gamma_5\,, &
    T^{(4)}_{\pm, \bar \mu \bar \nu \mu} = w_{\bar \mu}\,\hat g_{\bar \nu \mu} \,P_\pm \,  i\,\gamma_5\,,\\
    T^{(5)}_{\pm, \bar \mu \bar \nu \mu} = \hat g_{\bar \mu \bar \nu} \,P_\pm \,\bar r_{\mu} \,  i\,\gamma_5\,, &
    T^{(6)}_{\pm, \bar \mu \bar \nu \mu} = r_{\bar \mu}\,\hat g_{\bar \nu \mu} \,P_\pm \,  i\,\gamma_5\,,\\
    T^{(7)}_{\pm, \bar \mu \bar \nu \mu} = \hat \gamma_{\bar \mu}\,w_{\bar \nu}\,  P_\pm \,\hat \gamma_{\mu}\,i\,\gamma_5\,, &
    T^{(8)}_{\pm, \bar \mu \bar \nu \mu} = \hat \gamma_{\bar \mu}\,r_{\bar \nu}\,  P_\pm \,\hat \gamma_{\mu}\,i\,\gamma_5\,, \\
    T^{(9)}_{\pm, \bar \mu \bar \nu \mu} = \hat \gamma_{\bar \mu}\,w_{\bar \nu}\,  P_\pm \,\bar r_{\mu}\,i\,\gamma_5\,,&
    T^{(10)}_{\pm, \bar \mu \bar \nu \mu} = w_{\bar \mu}\,w_{\bar \nu}\, P_\pm \,\hat \gamma_{\mu} \,i\,\gamma_5\,, \qquad \\
    T^{(11)}_{\pm, \bar \mu \bar \nu \mu} = \hat \gamma_{\bar \mu}\,r_{\bar \nu}\,  P_\pm \,w_{\mu}\,i\,\gamma_5\,, &
    T^{(12)}_{\pm, \bar \mu \bar \nu \mu} = w_{\bar \mu}\,r_{\bar \nu}\, P_\pm \,\hat \gamma_{\mu} \,i\,\gamma_5\,,  \\
    T^{(13)}_{\pm, \bar \mu \bar \nu \mu} = \hat \gamma_{\bar \mu}\,w_{\bar \nu}\,  P_\pm \,w_{\mu}\,i\,\gamma_5\,, &
    T^{(14)}_{\pm, \bar \mu \bar \nu \mu} = r_{\bar \mu}\,r_{\bar \nu}\, P_\pm \,\hat \gamma_{\mu} \,i\,\gamma_5\,,  \\
    T^{(15)}_{\pm, \bar \mu \bar \nu \mu} = w_{\bar \mu}\,w_{\bar \nu}\, P_\pm \,w_{ \mu}\,i\,\gamma_5\,, &
    T^{(16)}_{\pm, \bar \mu \bar \nu \mu} = w_{\bar \mu}\,w_{\bar \nu}\, P_\pm \,\bar r_\mu\,i\,\gamma_5\,, \\
    T^{(17)}_{\pm, \bar \mu \bar \nu \mu} = w_{\bar \mu}\,r_{\bar \nu}\, P_\pm \,w_\mu\,i\,\gamma_5\,, &
    T^{(18)}_{\pm, \bar \mu \bar \nu \mu} = r_{\bar \mu}\,r_{\bar \nu}\, P_\pm \,w_\mu\,i\,\gamma_5\,,
  \end{array}
  \nonumber\\ \nonumber\\
   &&
  \frac{1}{2}\,\tr \big(T^{(n)}_{a, \bar \mu \bar \nu \mu}\, \Lambda\;Q^{\bar \mu \bar \nu \mu}_{b, k}\, \bar \Lambda\big)
  =
  \delta_{nk}\,\delta_{ab}
  \label{def-11/2to13/2}
  \,,\\
  && {\rm with} \qquad
  \bar q_{\bar \mu}\, Q^{\bar \mu \bar \nu \mu}_{\pm, k} =0  \,,\qquad q_{\mu}\, Q^{\bar \mu \bar \nu \mu}_{\pm, k} =0
  \,,\nonumber\\
  && \qquad  \qquad
  \bar p_{ \bar\nu}\, Q^{\bar \mu \bar \nu \mu}_{\pm, k} =0= \Lambda\,Q^{\bar \mu \bar \nu \mu}_{\pm, k}\,\bar\Lambda\,\gamma_{\bar \nu}
  \,,\nonumber\\ \nonumber\\
  &&
  T_{0\,\frac{3}{2} \to \,1\,\frac{3}{2}}(\bar k,\, k,\, w)
  =
  \sum_{\pm, n} F^\pm_n (\sqrt{s},t) \,\langle T^{(n)}_{\pm,\bar \mu \bar \nu \nu} \rangle^{\bar \mu \bar \nu \nu}_{0\,\frac{3}{2} \to \,1\,\frac{3}{2}}
  \,,\nonumber\\ &&
  \langle T^{(n)}_{\pm ,\bar \mu \bar \nu \nu}\rangle^{\bar \mu \bar \nu \nu}_{0\,\frac{3}{2} \to \,1\,\frac{3}{2}}
  =
  \epsilon^{\dagger \bar \mu }(\bar q, \lambda_{\bar q})\,\bar u^{\bar \nu} (\bar p,\lambda_{\bar p})
  T^{(n)}_{\pm , \bar \mu \bar \nu \nu}\,u^{\nu}(p,\lambda_{p})
  \,,\nonumber\\ \nonumber\\ &&
  \begin{array}{ll}
    T^{(1)}_{\pm, \bar \mu \bar \nu \nu} = \hat g_{\bar \nu \nu} \,\hat\gamma_{\bar\mu} \,P_\pm \,  i\,\gamma_5\,,  \qquad &
    T^{(2)}_{\pm, \bar \mu \bar \nu \nu} = \hat g_{\bar \nu \nu} \,w_{\bar \mu}\,P_\pm \,  i\,\gamma_5\,,\\
    T^{(3)}_{\pm, \bar \mu \bar \nu \nu} = \hat g_{\bar \nu \nu} \,r_{\bar \mu}\,P_\pm \,  i\,\gamma_5\,, &
    T^{(4)}_{\pm, \bar \mu \bar \nu \nu} = \hat g_{\bar \nu \bar \mu} \,P_\pm \,w_{\nu} \,  i\,\gamma_5\,,\\
    T^{(5)}_{\pm, \bar \mu \bar \nu \nu} = \hat g_{\bar \nu \bar \mu} \,P_\pm \,\bar r_{\nu}\,i\,\gamma_5\,, &
    T^{(6)}_{\pm, \bar \mu \bar \nu \nu} = \hat \gamma_{\bar \mu}\,w_{\bar \nu}\, P_\pm \,w_{\nu}\,i\,\gamma_5\,, \\
    T^{(7)}_{\pm, \bar \mu \bar \nu \nu} = \hat \gamma_{\bar \mu}\,w_{\bar \nu}\, P_\pm \,\bar r_{\nu}\,i\,\gamma_5\,,&
    T^{(8)}_{\pm, \bar \mu \bar \nu \nu} = \hat \gamma_{\bar \mu}\,r_{\bar \nu}\, P_\pm \,w_{\nu}\,i\,\gamma_5\,, \qquad \\
    T^{(9)}_{\pm, \bar \mu \bar \nu \nu} = w_{\bar \mu}\,w_{\bar \nu}\, P_\pm \,w_{ \nu}\,i\,\gamma_5\,, &
    T^{(10)}_{\pm, \bar \mu \bar \nu \nu} = w_{\bar \mu}\,w_{\bar \nu}\, P_\pm \,\bar r_\nu\,i\,\gamma_5\,, \\
    T^{(11)}_{\pm, \bar \mu \bar \nu \nu} = w_{\bar \mu}\,r_{\bar \nu}\, P_\pm \,w_\nu\,i\,\gamma_5\,, &
    T^{(12)}_{\pm, \bar \mu \bar \nu \nu} = r_{\bar \mu}\,r_{\bar \nu}\, P_\pm \,w_\nu\,i\,\gamma_5\,,
  \end{array}
  \nonumber\\ \nonumber\\
  &&
  \frac{1}{2}\,\tr \big(T^{(n)}_{a, \bar \mu \bar \nu \nu}\, \Lambda\;Q^{\bar \mu \bar \nu \nu}_{b, k}\, \bar \Lambda\big)
  =
  \delta_{nk}\,\delta_{ab}
  \label{def-03/2to13/2}
  \,,\\
  && {\rm with} \qquad
  \bar q_{\bar \mu}\, Q^{\bar \mu \bar \nu \nu}_{\pm, k} =0
  \,,\nonumber\\
  && \qquad  \qquad
  \bar p_{\bar\nu}\, Q^{\bar \mu \bar \nu \nu}_{\pm, k} =0= \Lambda\,Q^{\bar \mu \bar \nu \nu}_{\pm, k}\,\bar\Lambda\,\gamma_{\bar \nu}
  \,,\nonumber\\
  && \qquad  \qquad
  p_{\nu}\, Q^{\bar \mu \bar \nu \nu}_{\pm, k} =0= \gamma_{\nu}\,\Lambda\,Q^{\bar \mu \bar \nu \nu}_{\pm, k}\,\bar\Lambda
  \,,\nonumber\\ \nonumber
   &&
  T_{1\,\frac{3}{2} \to \,1\,\frac{3}{2}}(\bar k,\, k,\, w)
  =
  \sum_{\pm, n} F^\pm_n (\sqrt{s},t) \,\langle T^{(n)}_{\pm,\bar \mu \bar \nu \mu \nu} \rangle^{\bar \mu \bar \nu \mu \nu}_{1\,\frac{3}{2} \to \,1\,\frac{3}{2}}
  \,,\nonumber\\
  &&
  \langle T^{(n)}_{\pm,\bar \mu \bar \nu \mu \nu} \rangle^{\bar \mu \bar \nu \mu \nu}_{1\,\frac{3}{2} \to \,1\,\frac{3}{2}}
  =
  \epsilon^{\dagger \bar \mu }(\bar q, \lambda_{\bar q})\,\bar u^{\bar \nu} (\bar p,\lambda_{\bar p})
  \,\nonumber\\
  &&\qquad\qquad\qquad \qquad\qquad \times
  T^{(n)}_{\pm , \bar \mu \bar \nu \mu \nu}\,u^{\nu}(p,\lambda_{p})\, \epsilon^{\mu}(q,\lambda_{q})
  \,,\nonumber\\ \nonumber\\
  &&
  \begin{array}{ll}
    T^{(1)}_{\pm, \bar \mu \bar \nu \mu \nu} = \hat g_{\bar \mu \bar \nu}\,\hat g_{\mu \nu}\,P_\pm\,,\qquad &
    T^{(2)}_{\pm, \bar \mu \bar \nu \mu \nu} = \hat g_{\bar \mu \mu}\,\hat g_{\bar \nu \nu}\, P_\pm\,,\\
    T^{(3)}_{\pm, \bar \mu \bar \nu \mu \nu} = \hat g_{\bar \nu \nu}\,\hat\gamma_{\bar \mu}\,P_\pm\,\hat\gamma_{\mu}\,,\qquad &
    T^{(4)}_{\pm, \bar \mu \bar \nu \mu \nu} = \hat g_{\bar \mu \bar \nu}\,P_\pm\,\hat\gamma_{\mu}\,w_{\nu}\,,\\
    T^{(5)}_{\pm, \bar \mu \bar \nu \mu \nu} = \hat g_{\mu \nu}\,\hat\gamma_{\bar \mu}\,w_{\bar \nu}\, P_\pm\,,\qquad &
    T^{(6)}_{\pm, \bar \mu \bar \nu \mu \nu} = \hat g_{\bar \mu \bar \nu}\,P_\pm\,\hat\gamma_{\mu}\,\bar r_{\nu}\,,\\
    T^{(7)}_{\pm, \bar \mu \bar \nu \mu \nu} = \hat g_{\mu \nu}\,\hat\gamma_{\bar \mu}\,r_{\bar \nu}\,P_\pm\,,\qquad &
    T^{(8)}_{\pm, \bar \mu \bar \nu \mu \nu} = \hat g_{\bar \nu \nu}\,\hat\gamma_{\bar \mu}\,P_\pm\,w_{\mu}\,,\\
    T^{(9)}_{\pm, \bar \mu \bar \nu \mu \nu} = \hat g_{\bar \nu \nu}\,w_{\bar \mu}\,P_\pm\,\hat\gamma_{\mu}\,,\qquad &
    T^{(10)}_{\pm, \bar \mu \bar \nu \mu \nu} = \hat g_{\bar \nu \nu}\,\hat\gamma_{\bar \mu}\,P_\pm\,\bar r_{\mu}\,,\\
    T^{(11)}_{\pm, \bar \mu \bar \nu \mu \nu} = \hat g_{\bar \nu \nu}\,r_{\bar \mu}\,P_\pm\,\hat\gamma_{\mu}\,,\qquad &
    T^{(12)}_{\pm, \bar \mu \bar \nu \mu \nu} = \hat g_{\bar \mu \bar \nu}\,P_\pm\,w_{\mu}\,w_{\nu}\,,\\
    T^{(13)}_{\pm, \bar \mu \bar \nu \mu \nu} = \hat g_{\mu \nu}\,w_{\bar \mu}\,w_{\bar \nu}\,P_\pm\,,\qquad &
    T^{(14)}_{\pm, \bar \mu \bar \nu \mu \nu} = \hat g_{\bar \mu \bar \nu}\,P_\pm\,w_{\mu}\,\bar r_{\nu}\,,\\
    T^{(15)}_{\pm, \bar \mu \bar \nu \mu \nu} = \hat g_{\mu \nu}\,w_{\bar \mu}\,r_{\bar \nu}\,P_\pm\,,\qquad &
    T^{(16)}_{\pm, \bar \mu \bar \nu \mu \nu} = \hat g_{\bar \mu \bar \nu}\,P_\pm\,\bar r_{\mu}\,\bar r_{\nu}\,,\\
    T^{(17)}_{\pm, \bar \mu \bar \nu \mu \nu} = \hat g_{\mu \nu}\,r_{\bar \mu}\,r_{\bar \nu}\,P_\pm\,,\qquad &
    T^{(18)}_{\pm, \bar \mu \bar \nu \mu \nu} = \hat g_{\bar \nu \nu}\,w_{\bar \mu}\,P_\pm\,w_{\mu}\,,\\
    T^{(19)}_{\pm, \bar \mu \bar \nu \mu \nu} = \hat g_{\bar \nu \nu}\,w_{\bar \mu}\,P_\pm\,\bar r_{\mu}\,,\qquad &
    T^{(20)}_{\pm, \bar \mu \bar \nu \mu \nu} = \hat g_{\bar \nu \nu}\,r_{\bar \mu}\,P_\pm\,w_{\mu}\,,\\
    T^{(21)}_{\pm, \bar \mu \bar \nu \mu \nu} = \hat\gamma_{\bar \mu}\,w_{\bar \nu}\, P_\pm\,\hat\gamma_{\mu}\,w_{\nu}\,,\qquad &
    T^{(22)}_{\pm, \bar \mu \bar \nu \mu \nu} = \hat\gamma_{\bar \mu}\,w_{\bar \nu}\, P_\pm\,\hat\gamma_{\mu}\,\bar r_{\nu}\,,\\
    T^{(23)}_{\pm, \bar \mu \bar \nu \mu \nu} = \hat\gamma_{\bar \mu}\,r_{\bar \nu}\, P_\pm\,\hat\gamma_{\mu}\,w_{\nu}\,,\qquad &
    T^{(24)}_{\pm, \bar \mu \bar \nu \mu \nu} = \hat\gamma_{\bar \mu}\,w_{\bar \nu}\, P_\pm\,w_{\mu}\,w_{\nu}\,,\\
    T^{(25)}_{\pm, \bar \mu \bar \nu \mu \nu} = w_{\bar \mu}\,w_{\bar \nu}\,P_\pm\,\hat\gamma_{\mu}\,w_{\nu}\,,\qquad &
    T^{(26)}_{\pm, \bar \mu \bar \nu \mu \nu} = \hat\gamma_{\bar \mu}\,w_{\bar \nu}\, P_\pm\,w_{\mu}\,\bar r_{\nu}\,,\\
    T^{(27)}_{\pm, \bar \mu \bar \nu \mu \nu} = w_{\bar \mu}\,r_{\bar \nu}\,P_\pm\,\hat\gamma_{\mu}\,w_{\nu}\,,\qquad &
    T^{(28)}_{\pm, \bar \mu \bar \nu \mu \nu} = \hat\gamma_{\bar \mu}\,w_{\bar \nu}\, P_\pm\,\bar r_{\mu}\,\bar r_{\nu}\,,\\
    T^{(29)}_{\pm, \bar \mu \bar \nu \mu \nu} = r_{\bar \mu}\,r_{\bar \nu}\,P_\pm\,\hat\gamma_{\mu}\,w_{\nu}\,,\qquad &
    T^{(30)}_{\pm, \bar \mu \bar \nu \mu \nu} = \hat\gamma_{\bar \mu}\,r_{\bar \nu}\, P_\pm\,w_{\mu}\,w_{\nu}\,,\\
    T^{(31)}_{\pm, \bar \mu \bar \nu \mu \nu} = w_{\bar \mu}\,w_{\bar \nu}\,P_\pm\,\hat\gamma_{\mu}\,\bar r_{\nu}\,,\qquad &
    T^{(32)}_{\pm, \bar \mu \bar \nu \mu \nu} = w_{\bar \mu}\,w_{\bar \nu}\,P_\pm\,w_{\mu}\,w_{\nu}\,,\\
    T^{(33)}_{\pm, \bar \mu \bar \nu \mu \nu} = w_{\bar \mu}\,w_{\bar \nu}\,P_\pm\,w_{\mu}\,\bar r_{\nu}\,,\qquad &
    T^{(34)}_{\pm, \bar \mu \bar \nu \mu \nu} = w_{\bar \mu}\,r_{\bar \nu}\,P_\pm\,w_{\mu}\,w_{\nu}\,,\\
    T^{(35)}_{\pm, \bar \mu \bar \nu \mu \nu} = w_{\bar \mu}\,w_{\bar \nu}\,P_\pm\,\bar r_{\mu}\,\bar r_{\nu}\,,\qquad &
    T^{(36)}_{\pm, \bar \mu \bar \nu \mu \nu} = r_{\bar \mu}\,r_{\bar \nu}\,P_\pm\,w_{\mu}\,w_{\nu}\,,
  \end{array}
\nonumber\\ \nonumber\\
&&
  \frac{1}{2}\,\tr \big(T^{(n)}_{a, \bar \mu \bar \nu \mu \nu}\, \Lambda\;Q^{\bar \mu \bar \nu \mu \nu}_{b, k}\, \bar \Lambda\big)
  =
  \delta_{nk}\,\delta_{ab}
  \label{def-13/2to13/2}
  \,,\\
  && {\rm with} \qquad
  \bar q_{\bar \mu}\, Q^{\bar \mu \bar \nu \mu \nu}_{\pm, k} =0
  \,,\quad
  q_{\mu}\, Q^{\bar \mu \bar \nu \mu \nu}_{\pm, k} =0
  \,,\nonumber\\
  && \qquad  \qquad
  \bar p_{\bar\nu}\, Q^{\bar \mu \bar \nu \mu \nu}_{\pm, k} =0= \Lambda\,Q^{\bar \mu \bar \nu \mu \nu}_{\pm, k}\,\bar\Lambda\,\gamma_{\bar \nu}
  \,,\nonumber\\
  && \qquad  \qquad
  p_{\nu}\, Q^{\bar \mu \bar \nu \mu \nu}_{\pm, k} =0= \gamma_{\nu}\,\Lambda\,Q^{\bar \mu \bar \nu \mu \nu}_{\pm, k}\,\bar\Lambda
  \,,\nonumber
\end{eqnarray}
Since the expressions for the projection algebras are increasingly
tedious we refrain from detailing all of them in the main text. In \nameref{app-A} explicit expressions for the $Q$'s
can be found for all reactions except for the most tedious case.

We summarize that all amplitudes satisfy the MacDowell relations $F^-_n(+\sqrt{s},t)= F^+_n(-\sqrt{s},t)$ and
the even amplitude combinations
\begin{eqnarray}
  F^+_n(\sqrt{s},t) &+& F^-_n(\sqrt{s},t) \,, \qquad
  \nonumber\\
  \sqrt{s}\,\big(F^+_n(\sqrt{s},t) &-&  F^-_n(\sqrt{s},t)\big)\,
\end{eqnarray}
as introduced in this section  are truly uncorrelated and satisfy Mandelstam's dispersion-integral representation~\cite{Mandelstam:1958xc,Ball:1961zza}.

\section{Partial-wave decomposition}
\label{sec:partial-wave-decomp}
The scattering operator, $T$, is decomposed into partial-wave amplitudes characterized by the total angular
momentum $J$. Following the seminal work of Jacob and Wick \cite{Jacob:1959at} we consider
helicity projections  $\lambda_{q},\lambda_{p}$ and $\lambda_{\bar q}, \lambda_{\bar p}$ of the
scattering matrix, where we apply the helicity wave functions in the convention as introduced
in \cite{Stoica:2011cy,Lutz:2011xc}. We write
\begin{eqnarray}
  && \SP{\lambda_{\bar q}\lambda_{\bar p}}{T}{\lambda_{q}\lambda_{p}}= \sum_{J}
  (2\, J + \!1) \,\langle \lambda_{\bar q} \lambda_{\bar p} | \,T_J | \lambda_{q} \lambda_{p} \rangle \,
  d^{(J)}_{\lambda,\bar{\lambda}} (\theta)\,,
  \nonumber\\
  && d^{(J)}_{\lambda, \bar \lambda} (\theta )=(-)^{\lambda-\bar \lambda}\, d^{(J)}_{-\lambda, -\bar \lambda} (\theta )
  \nonumber\\
  && \qquad \quad \;= (-)^{\lambda-\bar \lambda}\, d^{(J)}_{\bar \lambda, \lambda} (\theta ) =
  d^{(J)}_{-\bar \lambda, -\lambda} (\theta )\,,
  \label{def-Wigner} \\
  && \langle \lambda_{\bar q} \lambda_{\bar p} | \,T_J | \lambda_{q} \lambda_{p} \rangle
  = \int_{-1}^{1} \frac{\mathrm{d}\cos\theta}{2} \tsp{\lambda_{\bar q}\lambda_{\bar p}}{\lambda_{q} \lambda_{p}} \,
  d^{(J)}_{\lambda,\bar{\lambda}} (\theta)\,, \nonumber
\end{eqnarray}
with $\lambda=\lambda_{q}-\lambda_{p}$ and $\bar \lambda=\lambda_{\bar q}-\lambda_{\bar p}$. In (\ref{def-Wigner})
we recall some general properties of the Wigner rotation functions, $d^{(J)}_{\lambda,\bar \lambda}(\theta)$.
The phase conventions assumed in this work imply the parity relations
\begin{eqnarray}
  && \langle -\lambda_{\bar q}, -\lambda_{\bar p} | \,T\, | -\lambda_{q} ,-\lambda_{p} \rangle
  = (-)^{\Delta  } \,
  \langle \lambda_{\bar q}, \lambda_{\bar p} | \,T\, | \lambda_{q}, \lambda_{p} \rangle\,,
  \nonumber\\
  && \qquad {\rm with} \qquad \Delta = S_{q}-S_{p} -S_{\bar q} +S_{\bar p} + \lambda - \bar \lambda \,.
  \label{def-helicity-flip}
\end{eqnarray}
The two parity sectors with $P=\pm 1$ are decoupled by introducing parity eigenstates of good total angular
momentum $J$, formed in terms of the helicity states \cite{Jacob:1959at}. Following (\ref{def-Wigner}) we
introduce the angular momentum projection, $| \lambda_{q}, \lambda_{p} \,\rangle_J$, of the helicity
state $|\lambda_{q}, \lambda_{p} \,\rangle$. We write
\begin{eqnarray}
  | \lambda_{q}, \lambda_{p}  \,\rangle_J \,, \quad {\rm with}\quad
  T\,| \lambda_{q}, \lambda_{p} \,\rangle_J = T_J|\lambda_{q}, \lambda_{p} \,\rangle \,.
  \label{def-angular-projection}
\end{eqnarray}
We introduce  states, $|n_\pm ,J\, \rangle$, that are eigenstates of the total angular momentum
operator $J$ and the parity operator $P$. The following state convention
\begin{eqnarray}
&&|1_\pm ,J \,\rangle_{0\,\frac{1}{2}} = {1\over \sqrt{2}}\,\Big( | \,0\,,-{\textstyle{1\over 2 }}\, \rangle_J \pm  | \,0\,,+{\textstyle{1\over 2 }}\, \rangle_J \Big)\,,
\nonumber\label{def-states-PB}\\ \nonumber\\
&&|1_\pm ,J \,\rangle_{1\,\frac{1}{2}} = {1\over \sqrt{2}}\,\Big( | \,0\,,-{\textstyle{1\over 2 }}\, \rangle_J \mp  | \,0\,,+{\textstyle{1\over 2 }}\, \rangle_J\Big)\,,
\nonumber\\
&&|2_\pm ,J \,\rangle_{1\,\frac{1}{2}} = {1\over \sqrt{2}}\,\Big( | +1\,,+{\textstyle{1\over 2 }}\, \rangle_J \mp  | -1\,,-{\textstyle{1\over 2 }}\, \rangle_J\Big)\,,
\nonumber\\
&&|3_\pm ,J \,\rangle_{1\,\frac{1}{2}} = {1\over \sqrt{2}}\,\Big( | +1\,,-{\textstyle{1\over 2 }}\, \rangle_J \mp  | -1\,,+{\textstyle{1\over 2 }}\, \rangle_J\Big)\,,
\nonumber\label{def-states-VB}\\ \nonumber\\
&&|1_\pm ,J \,\rangle_{0\,\frac{3}{2}} = {1\over \sqrt{2}}\,\Big( | \,0\,,-{\textstyle{1\over 2 }}\, \rangle_J \mp  | \,0\,,+{\textstyle{1\over 2 }}\, \rangle_J\Big)\,,
\nonumber\\
&&|2_\pm ,J \,\rangle_{0\,\frac{3}{2}} = {1\over \sqrt{2}}\,\Big( | \,0\,,-{\textstyle{3\over 2 }}\, \rangle_J \mp  | \,0\,,+{\textstyle{3\over 2 }}\, \rangle_J\Big)\,,
\nonumber\label{def-states-PD}\\ \nonumber\\
&&|1_\pm ,J \,\rangle_{1\,\frac{3}{2}} = {1\over \sqrt{2}}\,\Big( | \,0\,,-{\textstyle{1\over 2 }}\, \rangle_J \pm  | \,0\,,+{\textstyle{1\over 2 }}\, \rangle_J \Big)\,,
\nonumber\\
&&|2_\pm ,J \,\rangle_{1\,\frac{3}{2}} = {1\over \sqrt{2}}\,\Big( | +1\,,+{\textstyle{1\over 2 }}\, \rangle_J \pm  | -1\,,-{\textstyle{1\over 2 }}\, \rangle_J\Big)\,,
\nonumber\\
&&|3_\pm ,J \,\rangle_{1\,\frac{3}{2}} = {1\over \sqrt{2}}\,\Big( | -1\,,-{\textstyle{3\over 2 }}\, \rangle_J \pm  | +1\,,+{\textstyle{3\over 2 }}\, \rangle_J\Big)\,,
\nonumber\\
&&|4_\pm ,J \,\rangle_{1\,\frac{3}{2}} = {1\over \sqrt{2}}\,\Big( | +1\,,-{\textstyle{1\over 2 }}\, \rangle_J \pm  | -1\,,+{\textstyle{1\over 2 }}\, \rangle_J\Big)\,,
\nonumber\\
&&|5_\pm ,J \,\rangle_{1\,\frac{3}{2}} = {1\over \sqrt{2}}\,\Big( | \,0\,,-{\textstyle{3\over 2 }}\, \rangle_J \pm  | \,0\,,+{\textstyle{3\over 2 }}\, \rangle_J\Big)\,,
\nonumber\\
&&|6_\pm ,J \,\rangle_{1\,\frac{3}{2}} = {1\over \sqrt{2}}\,\Big( | +1\,,-{\textstyle{3\over 2 }}\, \rangle_J \pm  | -1\,,+{\textstyle{3\over 2 }}\, \rangle_J\Big)\,,
\label{def-helicity-states}
\end{eqnarray}
will be used, where we omit the sector index $S_q\,S_p$ on the right-hand sides for notational convenience.
It holds
\begin{eqnarray}
P\,|n_\pm ,J \,\rangle_{S_q S_p} = \pm \,(-1)^{J+ \frac{1}{2}}\,|n_\pm ,J \,\rangle_{S_q S_p}  \,,
\label{res-parity}
\end{eqnarray}
for all considered systems of this work. The helicity partial-wave  amplitudes, $t^{J}_{\pm,ab}$, that carry good
angular momentum $J$ and good parity $P$ are defined by
\begin{eqnarray}
  t^{J}_{\pm,ab} =  \SP{a_{\pm},J\,}{T}{b_{\pm},J}\,,
  \eqlab{def-tij}
\end{eqnarray}
where $a$ and $b$ label the states in the convention (\ref{def-helicity-states}). For sufficiently large
$s$ the unitarity condition takes the simple form
\begin{eqnarray}
  \Im \big[ t^{J}_\pm\big]^{-1}_{ab} =
  -\frac{q^{\rm cm}_a\,M_a}{4\pi\,\sqrt{s}}\,\delta_{ab} \,,
  \label{unitarity-appendix}
\end{eqnarray}
where $M_a$  is the baryon mass in the channel $a$. According to (\ref{def-w})
the momentum $q_a^{\rm cm}$ is the on-shell value of $\sqrt{-r^2}$ in
the channel $a$.

\begin{table*}[t]
  \renewcommand{\arraystretch}{2.5} 
  \setlength{\tabcolsep}{13pt} 
  \rescale
  \begin{center}
    \begin{tabular}{|l@{\;\;\;:\quad}c@{}c@{}c|l@{\;\;\;:\quad}c@{}c@{}c|}
      \hline
      $\displaystyle
      \phantom{+}k\;\;n$ & &
      $\displaystyle
      \Big[a^{J+k}_{\pm\,n}\Big]_{11}$
      & &
      $\displaystyle
      \phantom{+}k\;\;n$ & &
      $\displaystyle
      \Big[b^{J+k}_{\pm\,n}\Big]_{11}$
      & \\[5pt] \cline{3-3}\cline{7-7}
      $\displaystyle
      +\frac{3}{2}\;\;2$ & &
      $\displaystyle
      \sqrt{\frac{2}{3}}\,\frac{q_{\rm cm}^2\,\bar E_\pm\,E_\mp}{2\,(2\,J+2)\,s}\left(
        \begin{subarray}{l}\displaystyle
          \mp(2\,J+3)\,\bar E + (2\,J-3)\,\bar M
          \\ \\ \displaystyle
          \mp (2\,J-1)\,(\sqrt{s}\,\bar E_\pm - 2\,\bar q_{\rm cm}^2)/\sqrt{s}
        \end{subarray}
        \right)
      $
      & &
      $\displaystyle
      +\frac{1}{2}\;\;2$ & &
      $\displaystyle
      -\sqrt{\frac{2}{3}}\,\frac{q_{\rm cm}^2}{2\,(2\,J)}\left(
        \begin{subarray}{l}\displaystyle
          \mp(2\,J+1)\,\bar E + (2\,J-1)\,\bar M
          \\ \\ \displaystyle
          \mp (2\,J-1)\,(\sqrt{s}\,\bar E_\pm - 2\,\bar q_{\rm cm}^2)/\sqrt{s}
        \end{subarray}
        \right)
      $
      & \\[2pt]
      $\displaystyle
      +\frac{1}{2}\;\;1$ & &
      $\displaystyle
      \mp \sqrt{\frac{2}{3}}\,\sqrt s\,\bar E_\pm\,E_\mp
      $
      & &
      $\displaystyle
      -\frac{1}{2}\;\;1$ & &
      $\displaystyle
      \pm \sqrt{\frac{2}{3}}\,s\sqrt s
      $
      & \\[2pt]
      $\displaystyle
      -\frac{1}{2}\;\;2$ & &
      $\displaystyle
      \sqrt{\frac{2}{3}}\,\frac{s\,E_\mp}{2\,(2\,J+2)\,\bar E_\mp}\left(
        \begin{subarray}{l}\displaystyle
          \mp(2\,J+1)\,\bar E - (2\,J-3)\,\bar M
          \\ \\ \displaystyle
          \pm (2\,J-1)\,(\sqrt{s}\,\bar E_\pm - 2\,\bar q_{\rm cm}^2)/\sqrt{s}
        \end{subarray}
        \right)
      $
      & &
      $\displaystyle
      -\frac{3}{2}\;\;2$ & &
      $\displaystyle
      -\sqrt{\frac{2}{3}}\,\frac{2\,J-1}{2\,J}\,\frac{s^2}{2\,q_{\rm cm}^2}\left(
        \begin{subarray}{l}\displaystyle
          \mp \bar E - \bar M
          \\ \\ \displaystyle
          \pm (\sqrt{s}\,\bar E_\pm - 2\,\bar q_{\rm cm}^2)/\sqrt{s}
        \end{subarray}
        \right)
      $
      & \\[2pt]
      $\displaystyle
      \phantom{+}k\;\;n$ & &
      $\displaystyle
      \Big[a^{J+k}_{\pm\,n}\Big]_{21}$
      & &
      $\displaystyle
      \phantom{+}k\;\;n$ & &
      $\displaystyle
      \Big[b^{J+k}_{\pm\,n}\Big]_{21}$
      & \\[5pt] \cline{3-3}\cline{7-7}
      $\displaystyle
      +\frac{3}{2}\;\;2$ & &
      $\displaystyle
      -\left(\frac{\bar q_{\rm cm}q_{\rm cm}}{s}\right)^2\,\frac{\sqrt{2\,J-1}\,\sqrt{2\,J+3}\,\bar E_\pm\,E_\mp }{2\sqrt{2}\,(2\,J+2)}
      $
      & &
      $\displaystyle
      +\frac{1}{2}\;\;2$ & &
      $\displaystyle
      \left(\frac{\bar q_{\rm cm}q_{\rm cm}}{s}\right)^2\frac{\sqrt{2\,J-1}\,\sqrt{2\,J+3}\,s }{2 \sqrt{2}\,(2\,J)}
      $
      & \\[2pt]
      $\displaystyle
      -\frac{1}{2}\;\;2$ & &
      $\displaystyle
      \frac{\sqrt{2\,J-1}\,\sqrt{2\,J+3}\,\bar E_\pm\,E_\mp }{2 \sqrt{2}\,(2\,J+2)}
      $
      & &
      $\displaystyle
      -\frac{3}{2}\;\;2$ & &
      $\displaystyle
      -\frac{\sqrt{2\,J-1}\,\sqrt{2\,J+3}\,s }{2\sqrt{2}\,(2\,J)}
      $
      & \\[9pt] \hline
    \end{tabular}
    \vskip0.6cm
    \caption{Non-vanishing coefficients $a^{J+k}_{\pm\,n}$ and $b^{J+k}_{\pm\,n}$ as introduced in (\ref{def-decomposition-TJ})
      for  the $0^-{\rm 1\over 2}^+\rightarrow 0^-{\rm 3\over 2}^+$ reaction.}
    \label{tab:abpm-01/2to03/2}
  \end{center}
\end{table*}

\begin{table*}[t]
  \renewcommand{\arraystretch}{2.5} 
  \setlength{\tabcolsep}{17pt} 
  \rescale
  \begin{center}
    \begin{tabular}{|l@{\quad:\qquad}c@{}c@{}c|l@{\quad:\qquad}c@{}c@{}c|}
      \hline
      $\displaystyle
      \phantom{+}k\;\;n$ & &
      $\displaystyle
      \Big[a^{J+k}_{\pm\,n}\Big]_{11}$
      & &
      $\displaystyle
      \phantom{+}k\;\;n$ & &
      $\displaystyle
      \Big[b^{J+k}_{\pm\,n}\Big]_{11}$
      & \\[5pt] \cline{3-3}\cline{7-7}
      $\displaystyle
      +\frac{3}{2}\;\;3$ & &
      $\displaystyle
      \pm\frac{q_{\rm cm}^2\,\bar E_\pm\,E_\mp\left(
          \sqrt{s}\,\bar E_\mp - \sqrt{s}\,{\bar\omega}\,(2\,J+1) + \bar q_{\rm cm}^2\,(2\,J-1)
        \right)}{(2\,J+2)\,s\sqrt s}
      $
      & &
      $\displaystyle
      +\frac{1}{2}\;\;3$ & &
      $\displaystyle
      \mp\frac{q_{\rm cm}^2\left((2\,J-1)\,\bar q_{\rm cm}^2 - (2\,J)\,{\bar\omega}\sqrt s\right)}{ (2\,J)\,\sqrt s}
      $
      & \\[2pt]
      $\displaystyle
      +\frac{1}{2}\;\;1$ & &
      $\displaystyle
      E_\mp\left(2\,{\bar\omega} - \bar E_\mp\right)
      $
      & &
      $\displaystyle
      -\frac{1}{2}\;\;1$ & &
      $\displaystyle
      s
      $
      & \\[2pt]
      $\displaystyle
      +\frac{1}{2}\;\;2$ & &
      $\displaystyle
      \pm \sqrt s\,\bar E_\pm\,E_\mp
      $
      & &
      $\displaystyle
      -\frac{1}{2}\;\;2$ & &
      $\displaystyle
      \mp s\sqrt s
      $
      & \\[2pt]
      $\displaystyle
      -\frac{1}{2}\;\;3$ & &
      $\displaystyle
      \mp\frac{\sqrt s\,E_\mp\left((2\,J-1)\,\bar E_\pm + \sqrt s\right)}{(2\,J+2)}
      $
      & &
      $\displaystyle
      -\frac{3}{2}\;\;3$ & &
      $\displaystyle
      \pm\frac{(2\,J-1)\,s\sqrt s }{(2\,J)}
      $
      & \\[2pt]
      $\displaystyle
      \phantom{+}k\;\;n$ & &
      $\displaystyle
      \Big[a^{J+k}_{\pm\,n}\Big]_{21}$
      & &
      $\displaystyle
      \phantom{+}k\;\;n$ & &
      $\displaystyle
      \Big[b^{J+k}_{\pm\,n}\Big]_{21}$
      & \\[5pt] \cline{3-3}\cline{7-7}
      $\displaystyle
      +\frac{3}{2}\;\;3$ & &
      $\displaystyle
      \left(\frac{\bar q_{\rm cm}q_{\rm cm}}{s}\right)^2\frac{\sqrt{2}\,\sqrt s\,E_\mp }{(2\,J+2)}
      $
      & &
      $\displaystyle
      -\frac{1}{2}\;\;1$ & &
      $\displaystyle
      \pm\sqrt{2}\,\sqrt s
      $
      & \\[2pt]
      $\displaystyle
      +\frac{1}{2}\;\;1$ & &
      $\displaystyle
      \mp\frac{\sqrt{2}\,\bar E_\mp\,E_\mp }{\sqrt s}
      $
      & & \multicolumn{4}{c|}{} \\[2pt]
      $\displaystyle
      -\frac{1}{2}\;\;3$ & &
      $\displaystyle
      -\frac{\sqrt{2}\,\sqrt s\,E_\mp }{(2\,J+2)}
      $
      & & \multicolumn{4}{c|}{} \\[2pt]
      $\displaystyle
      \phantom{+}k\;\;n$ & &
      $\displaystyle
      \Big[a^{J+k}_{\pm\,n}\Big]_{31}$
      & &
      $\displaystyle
      \phantom{+}k\;\;n$ & &
      $\displaystyle
      \Big[b^{J+k}_{\pm\,n}\Big]_{31}$
      & \\[5pt] \cline{3-3}\cline{7-7}
      $\displaystyle
      +\frac{3}{2}\;\;3$ & &
      $\displaystyle
      -\left(\frac{\bar q_{\rm cm}q_{\rm cm}}{s}\right)^2\frac{\sqrt{2\,J-1}\,\sqrt{2\,J+3}\,\bar E_\pm\,E_\mp }{2\sqrt{2}\,(2\,J+2)}
      $
      & &
      $\displaystyle
      +\frac{1}{2}\;\;3$ & &
      $\displaystyle
      \left(\frac{\bar q_{\rm cm}q_{\rm cm}}{s}\right)^2\frac{\sqrt{2\,J-1}\,\sqrt{2\,J+3}\,s }{2\sqrt{2}\,(2\,J)}
      $
      & \\[2pt]
      $\displaystyle
      -\frac{1}{2}\;\;3$ & &
      $\displaystyle
      \frac{\sqrt{2\,J-1}\,\sqrt{2\,J+3}\,\bar E_\pm\,E_\mp }{2\sqrt{2}\,(2\,J+2)}
      $
      & &
      $\displaystyle
      -\frac{3}{2}\;\;3$ & &
      $\displaystyle
      -\frac{\sqrt{2\,J-1}\,\sqrt{2\,J+3}\,s }{2\sqrt{2}\,(2\,J)}
      $
      & \\[9pt] \hline
    \end{tabular}
    \vskip0.6cm
    \caption{Non-vanishing coefficients $a^{J+k}_{\pm\,n}$ and $b^{J+k}_{\pm\,n}$ as introduced in (\ref{def-decomposition-TJ})
      for the $0^-{\rm 1\over 2}^+\rightarrow 1^-{\rm 1\over 2}^+$ reaction.}
    \label{tab:abpm-01/2to11/2}
  \end{center}
\end{table*}

Helicity partial-wave amplitudes are correlated at specific kinematical conditions, see for example the review \cite{CohenTannoudji1968239}.
This is seen once the amplitudes $t^{J}_{\pm,ab}(\sqrt{s})$ are expressed in terms of the invariant functions
$F^\pm_{n}(\sqrt{s},t)$. In contrast covariant partial-wave amplitudes $T^{J}_\pm(s) $ are free of kinematical constraints
and can therefore be used efficiently in partial-wave dispersion relations. They are associated with covariant states and
covariant projector polynomials which diagonalize the Bethe-Salpeter two-body scattering equation for local
interactions \cite{Lutz:2001yb,Lutz:2001mi,Lutz:2003fm,Stoica:2011cy,Lutz:2011xc}. We introduce
\begin{eqnarray}
T^{J}_\pm (\sqrt{s}\,) &=& \sqrt{\frac{4\,\bar M\,M\,s}{\bar E_\pm \,E_\pm }}\,\Bigg( \frac{s}{\bar{q}_{\rm cm} \,q_{\rm cm}} \Bigg)^{J-\frac{1}{2}}\,
\nonumber\\
& \times& \big[\bar{C}_{\pm}^{J}(\sqrt{s}\,) \big]^T\, t^J_\pm(\sqrt{s}\,) \, C^J_{\pm}(\sqrt{s}\,) \,,
  \label{defTJ}
\end{eqnarray}
with real triangular matrices $C^J_{\pm}(\sqrt{s}\,)$ and $\bar{C}^J_{\pm}(\sqrt{s}\,)$
characterizing the transformation for the initial and final states from the helicity basis to the new
covariant basis. We generate a convention in which the covariant partial-wave amplitudes $T^{J}_\pm (\sqrt{s}\,)$
satisfy the MacDowell relations
\begin{eqnarray}
T^{J}_- (\sqrt{s}) = T^{J}_+ (- \sqrt{s}) \,,
\label{res-MacDowell-general}
\end{eqnarray}
for all considered reactions. The transformation (\ref{defTJ}) implies a change in the phase-space distribution
\begin{eqnarray}
\rho_{\pm}^{J}(\sqrt{s}\, ) &=& - \Im\Big[T_{\pm}^{J}(\sqrt{s}\,)\Big]^{-1}  \label{def-rho} \\
  & =& \frac{E_\pm }{8 \,\pi\,\sqrt{s}} \,\left(\frac{q_{\rm cm}}{\sqrt{s}}\right)^{2\,J}
  \Big[ C^J_{\pm}(\sqrt{s}\,)\Big]^{-1}  \Big[C_{\pm}^{J}(\sqrt{s}\,)\Big]^{T,-1}\,,
 \nonumber
\end{eqnarray}
where we remind the reader that the first line in (\ref{def-rho}) holds for sufficiently large $s$ only.

We insist on transformation matrices that lead to an asymptotically bounded phase-space matrix,
i.e. we require
\begin{eqnarray}
&&  \lim_{\sqrt{s}\to \infty} \rho _{\pm}^{J}( \sqrt{s}\, ) < \infty \,,
\nonumber\\
&&  \lim_{\sqrt{s}\to \infty}\det \rho _{\pm}^{J}( \sqrt{s} ) =  {\rm const} \neq 0\,.
  \label{def-rho-constraint}
\end{eqnarray}
The condition (\ref{def-rho-constraint}) implies bounded partial-wave scattering amplitudes at asymptotically large $\sqrt{s}$.
In turn the dispersion-integral in the non-linear integral equation (\ref{def-U}) requires at most one subtraction to
render finite results. This can be seen from a rewrite
\begin{eqnarray}
&& \Im \,T_{ab}(\sqrt{s}\,) =\sum_{c,d}
T_{ac}(\sqrt{s}\,)
\,\rho_{cd}(\sqrt{s}\,)\,T^\dagger_{db}(\sqrt{s}\,)\,,
\label{unitarity-rewrite}
\end{eqnarray}
of the unitarity condition (\ref{def-rho}), where we suppress the reference to $J$ and $P$ for
notational simplicity. In order to gain insight into the non-linear integral equation
we derive the asymptotic behavior of the partial-wave amplitude. Assuming at first
that the phase-space matrix $\rho_{ab} \sim \delta_{ab} $ is diagonal the unitarity
condition (\ref{unitarity-rewrite}) implies the particular identity
\begin{eqnarray}
 [\Re\, T_{aa} ]^2 &=& \frac{\Im \,T_{aa}}{\rho_{aa}}-[\Im \,T_{aa} ]^2
\nonumber\\
 \qquad &-&\underbrace{
\sum_{c \neq a} \,\frac{[\Im\, T_{ac}]^2+[\Re \,T_{ac}]^2}{\rho_{aa}}\,\rho_{cc}
}_{>0} \,.
\label{asymp-a}
\end{eqnarray}
From (\ref{asymp-a}) it follows that
\begin{eqnarray}
  && \Big| \Im \,T_{aa} \Big| < \Big| \frac{1}{\rho_{aa}} \Big| \,, \qquad \quad
  \Big| \Re\, T_{aa} \Big| <  \Big| \frac{1}{\rho_{aa}} \Big|  \,,
\end{eqnarray}
the diagonal elements of the covariant partial-wave amplitudes are asymptotically bounded for large $\sqrt{s}$.
This holds if the phase-space matrix $\rho_{aa}$ is bounded asymptotically by a non-vanishing constant.
An identical conclusion may be drawn from (\ref{asymp-a}) for the off-diagonal elements. Our
conclusions can also be proven for a general triangular transformation matrix $C^J_{\pm}(\sqrt{s}\,)$ that is asymptotically
bounded and may lead to a non-diagonal phase-space matrix.

In a first step we reproduce the previous results of \cite{Lutz:2001yb}. The transformation matrix for the
$0\, {\tiny {1\over 2}} $ state is
\begin{eqnarray}
C^J_{\pm, \,0 \frac{1}{2}}(\sqrt{s}\,) = 1 \,.
\end{eqnarray}
For the partial-wave amplitudes of the $0^- {\textstyle {1\over 2}}^+ \to 0^- {\textstyle {1\over 2}}^+ $ reaction this leads to
\begin{eqnarray}
T^J_{\pm }(\sqrt{s}\,) =\pm\,\sqrt{s}\,\Big[  A^{J-{\rm 1\over 2}}_{\pm\, 1}(\sqrt{s}\,) - \frac{\bar E_\mp\,E_\mp}{s}\,A^{J+{\rm 1\over 2}}_{\mp\, 1}(\sqrt{s}\,)\Big]\,,
\label{res:01to01}
\end{eqnarray}
with
\begin{eqnarray}
 A^{J}_{\pm\, n}(\sqrt{s}\,) &=& \left( \frac{s}{\bar q_{\rm cm}\,q_{\rm cm}} \right)^{J}
 \nonumber\\
&\times&
 \int_{-1}^{1} \!
  \frac{\mathrm{d} \cos\theta}{2} \, F^\pm_n(\sqrt{s},t) \, P_{J}(\cos \theta) \,.
  \label{def-As}
\end{eqnarray}
The important merit of (\ref{res:01to01}, \ref{def-As}) is the absence of kinematical constraints, with the possible exception
at $s=0$. A singularity at $\bar q_{\rm cm}\,q_{\rm cm}=0$ in (\ref{def-As}) is not realized due to the properties of the
Legendre polynomials $P_{J}(\cos \theta)$. We recover the projector polynomials first derived
in \cite{Lutz:2001yb}. In a notation using (\ref{def-hatgamma}) and the building blocks introduced in \cite{Lutz:2001mi,Stoica:2011cy} we
write
\begin{eqnarray}
&& \pm \sqrt{s}\,{\mathcal Y}^J_\pm ( \bar r,\,r, w) = Y^{(1)}_{J +\frac{1}{2}}( \bar r,\,r, w) \,P_\pm
\nonumber\\
&& \qquad +\,  \frac{1}{s}\,Y^{(1)}_{J -\frac{1}{2}}( \bar r,\,r, w) \,(\bar r \cdot \hat \gamma)\,P_\mp  \,( \hat \gamma  \cdot r)\,,
\label{res-projectors}
\end{eqnarray}
with the generic polynomials
\begin{eqnarray}
&& Y^{(k)}_n(\bar r,\,r, w) =\left( \frac{\bar q_{\rm cm}\,q_{\rm cm}}{s} \right)^{n-k}  \left( \frac{\mathrm{d}}{\mathrm{d} \,\cos \theta } \right)^k\,P_n (\cos \theta)\,,
\nonumber\\
&& \qquad {\rm with}\qquad \cos \theta = - \frac{\bar r \cdot r}{\bar q_{\rm cm}\,q_{\rm cm}} =
-\frac{\bar r \cdot r}{\sqrt{\bar r^2\,r^2}}  \,,
\end{eqnarray}
regular at $\bar q_{\rm cm}\,q_{\rm cm} =0 $. A projector polynomial ${\mathcal Y}^J_\pm$ has the defining
property that if partial-wave expanded  via (\ref{def-Wigner}) it contributes exclusively to a single partial-wave
amplitude with $T^J_\pm(\sqrt{s}\,) = 1$.

\begin{table*}
  \renewcommand{\arraystretch}{2.5} 
  \setlength{\tabcolsep}{13pt} 
  \setlength{\arraycolsep}{23pt} 
  \rescale
  \begin{center}
    \begin{tabular}{|@{\;\;}r@{\;}l|@{\;\;}r@{\;}l|@{\;\;}r@{\;}l|@{\;\;}r@{\;}l|}
      \hline
      $\displaystyle
      (1,1):$ &
      \multicolumn{3}{@{\;}l|@{\;\;}}{
        $\displaystyle
        \frac{ m\,M\,s}{ E_\mp\,E_\pm}$
      }
      &
      $\displaystyle
      (2,1):$ &
      $\displaystyle
      \mp\frac{ M\,s\,({ \omega}- E_\pm)}{\sqrt{2}\,E_\mp\,E_\pm}$
      &
      $\displaystyle
      (2,2):$ &
      $\displaystyle
      \frac{\sqrt{2\,J+3}\,M\,\sqrt{s}}{2\,\sqrt{2}\,E_\mp}$
      \\[5pt] \hline
      $\displaystyle
      (3,1):$ &
      \multicolumn{3}{@{\;}l|@{\;\;}}{
        $\displaystyle
        \sqrt{\frac{2}{3}}\,{\omega}\,\Bigg(\frac{{\omega}\,s}{E_\mp\,E_\pm} + \frac{M\,(M \pm \sqrt{s})\,s}{2\,{\omega}\,E_\mp  E_\pm} \mp \frac{(M\pm \sqrt{s})\,s}{E_\mp\,E_\pm} - 2\,(\delta -1)\,\sqrt{s} \Bigg) $
      }
      &
      $\displaystyle
      (3,2):$ &
      $\displaystyle
      \frac{\sqrt{2\,J+3}\,M\,\sqrt{s}}{2\,\sqrt{6}\,E_\mp}$
      &
      $\displaystyle
      (3,3):$ &
      $\displaystyle
      \frac{\sqrt{2\,J+3}}{2\,\sqrt{2}}$
      \\[5pt] \hline
      $\displaystyle
      (4,1):$ &
      \multicolumn{3}{@{\;}l|@{\;\;}}{
        $\displaystyle
        \sqrt{\frac{2\,J-1}{2\,J+3}}\Bigg(
        2\,\sqrt{2}\,M\,\sqrt{s} \mp\frac{M\,s\,( M\pm \sqrt{s})}{\sqrt{2}\,E_\mp\,E_\pm}
        \Bigg)$
      }
      &
      $\displaystyle
      (4,2):$ &
      $\displaystyle
      \mp\frac{\sqrt{2\,J-1}\,M\,\sqrt{s}}{2\,\sqrt{2}\,E_\mp}$
      &
      $\displaystyle
      (4,4):$ &
      $\displaystyle
      \pm\frac{1}{2}$
      \\[5pt] \hline
      $\displaystyle
      (5,1):$ &
      \multicolumn{7}{@{\;}l|}{
        $\displaystyle
        \pm\sqrt{\frac{2\,J-1}{2\,J+3}}\,\frac{1}{\sqrt{3}}\,\Bigg(
        -\frac{m\,(\pm(\delta -1)\,\sqrt{s}- M)\,s}{E_\mp\,E_\pm}
        \pm 2\,m\,\sqrt{s}
        \mp 4\,m\,E_\mp
        \mp 8\,m\,E_\pm
        + 2\,M\,\Bigg[
        m \mp\frac{M\,({ \omega}- E_\pm)}{m}
        \Bigg]
        \Bigg)$
      }
      \\[5pt] \hline
      $\displaystyle
      (5,2):$ &
      $\displaystyle
      \pm\frac{\sqrt{2\,J-1}}{\sqrt{3}}\Bigg(
      m \mp\frac{M\,({ \omega}- E_\pm)}{m}
      \Bigg)$
      &
      $\displaystyle
      (5,3):$ &
      $\displaystyle
      -\frac{\sqrt{2\,J-1}\,({ \omega}- E_\pm)}{2\,m}$
      &
      $\displaystyle
      (5,4):$ &
      $\displaystyle
      \frac{({ \omega}- E_\pm)}{\sqrt{6}\,m}$
      &
      $\displaystyle
      (5,5):$ &
      $\displaystyle
      \pm\frac{ E_\pm}{ m\,\sqrt{s}}$
      \\[5pt] \hline
      $\displaystyle
      (6,1):$ &
      \multicolumn{7}{@{\;}l|}{
        \hspace{-0.8cm}$\displaystyle
        \begin{array}{l}\displaystyle
          \sqrt{\frac{2}{3}}\,\sqrt{\frac{(2\,J-3)\,(2\,J-1)}{(2\,J+3)\,(2\,J+5)}}\,\frac{ \omega\,\sqrt{s} }{ E_\mp\,E_\pm}\Bigg(
          -3\,s
          +2\,{ \omega}\,\sqrt{s}
          +\frac{ ( M\pm \sqrt{s})\,(\pm 2\,\sqrt{s}-3\, M)\,\sqrt{s}}{2\,{ \omega}}
          \Bigg)
          \\[1pt] \displaystyle
          \mp\sqrt{\frac{2}{3}}\,\sqrt{\frac{(2\,J-3)\,(2\,J-1)}{(2\,J+3)\,(2\,J+5)}}\,\Bigg(
          (2\,M\pm \sqrt{s})\,\sqrt{s}
          \mp 4\,( m^2+ M^2)
          +2\,{ \omega}\,\Bigg[
          \bigg(\frac{ M^4}{ m^2\,s}-\frac{ M^2}{s}-5 \bigg) ( M\pm \sqrt{s}) + \frac{2\,{ \omega}\,( M\pm 4\,\sqrt{s})}{\sqrt{s}}
          \Bigg]
          \Bigg)
        \end{array}$
      }
      \\[21pt] \hline
      $\displaystyle
      (6,2):$ &
      \multicolumn{7}{@{\;}l|}{
        $\displaystyle
        \pm\sqrt{\frac{(2\,J-3)\,(2\,J-1)}{2\,J+5}}\,\frac{1 }{\sqrt{6}\,\sqrt{s}}\,\Bigg(
        \mp\frac{ M\,s}{2\,E_\mp}
        +2\,{ \omega}\,\Bigg[
        -\frac{( M \pm \sqrt{s})\,M^3}{ m^2\,s} \pm \frac{  M}{\sqrt{s}}+\frac{ m^2}{s}
        \Bigg] \sqrt{s}
        -4\,( m^2+ M^2)
        \Bigg)$
      }
      \\[5pt] \hline
      $\displaystyle
      (6,3):$ &
      \multicolumn{7}{@{\;}l|}{
        $\displaystyle
        \pm\sqrt{\frac{(2\,J-3)\,(2\,J-1)}{2\,J+5}}\,\frac{{ \omega} }{\sqrt{2}\,\sqrt{s}}\,\Bigg(
        \frac{4\,M\,{ \omega}^2}{ m^2\,\sqrt{s}}
        \mp\frac{ M^2 }{ m^2}
        -\frac{4\,M\mp \sqrt{s}}{\sqrt{s}}
        -\frac{\sqrt{s}\,( M\pm \sqrt{s})\,({ \omega}- E_\mp)}{2\,m^2\,{ \omega}}
        \Bigg)$
      }
      \\[5pt] \hline
      $\displaystyle
      (6,4):$ &
      \multicolumn{7}{@{\;}l|}{
        $\displaystyle
        \sqrt{\frac{2\,J-3}{2\,J+5}}\,\frac{{ \omega} }{\sqrt{3}\,M\,\sqrt{s}}\,\Bigg(
        -\frac{( m^2- M^2)\,( M\pm \sqrt{s})}{ m^2}
        +\frac{\sqrt{s}\,( M\pm 2\,\sqrt{s})}{2\,{ \omega}}
        \mp\frac{4\,( m^2+ M^2)\,E_\mp\,E_\pm}{ m^2\,\sqrt{s}}
        \Bigg)$
      }
      \\[5pt] \hline
      $\displaystyle
      (6,5):$ &
      \multicolumn{3}{@{\;}l|@{\;\;}}{
        $\displaystyle
        \mp\sqrt{\frac{2\,J-3}{2\,J+5}}\,\frac{\sqrt{2}\,({ \omega}- E_\mp)\,E_\pm}{ m^2\,\sqrt{s}}$
      }
      &
      $\displaystyle
      (6,6):$ &
      \multicolumn{3}{@{\;}l|}{
        $\displaystyle
        \frac{  E_\mp\,E_\pm}{ m^2\,M\,s}$
      }
      \\[5pt] \hline
    \end{tabular}
    \vskip0.6cm
    \caption{Non-zero elements of the transformation matrices $ \Big[ C_{\pm,\,1\,{\rm 3\over 2}}^J(\sqrt s\,) \Big]_{ab}$}
    \label{tab:CJ-13/2}
  \end{center}
\end{table*}

A projector is defined for off-shell conditions with $q^2 \neq m^2, p^2 \neq M^2$
and $\bar q^2 \neq \bar m^2, \bar p^2 \neq \bar M^2$ and is independent on any mass parameter. This is because only then
a projector structure generates an analytic solution of its associated Bethe-Salpeter equation, where a particular
renormalization program built on dimensional regularization must be assumed \cite{Lutz:2001mi,Lutz:2003fm,Semke:2005sn}.
The projectors as given in (\ref{res-projectors}) are minimal in the sense that any possible alternative has necessarily
a higher mass dimension. A projector appropriately multiplied by positive powers of $q^2, \gamma \cdot p$ or
$\bar q^2, \gamma \cdot \bar p$ has all properties we insist on a projector to have. Negative powers are prohibited since
they would destroy the property that the projector has to solve the Bethe-Salpeter equation. A singular behaviour at for
instance $q^2=0$ or $\gamma \cdot p =0 $ renders the Bethe-Salpeter equation ill-defined. As emphasized before the
dimension of a projector must not be altered by a multiplication with any mass parameter. The only other available scale
$\sqrt{s}$ can not be used, since that would alter the asymptotic property (\ref{def-rho-constraint}).

We proceed with the more complicated $0\, {\tiny {3\over 2}} $, $1\,{\tiny {1\over 2}}$ and $1\,{\tiny {3\over 2}}$ states.
According to (\ref{def-helicity-states}) for a given $J$ and $P$ there are two, three and six possible partial-wave states respectively.
For the first two cases the transformation matrices take the form
\begin{eqnarray}
&& C^J_{\pm,\, 0 \frac{3}{2}}(\sqrt{s}\,) =
\left(
\begin{array}{cc}
  M\,\frac{\sqrt{s}}{E_\mp} & 0 \\
  \sqrt{n}\, \frac{ (2\,E \mp M)\,\sqrt{s}- 4\,E_\pm\,E_\mp}{\sqrt{3}\,\sqrt{n+2}\,E_\mp} & \pm \frac{E_\pm}{\sqrt{s}}
\end{array}
\right)\,,
\label{res-C032}
\\ \nonumber\\
&& C^J_{\pm,1\frac{1}{2}} (\sqrt{s}\,) =
  \left(
    \begin{array}{@{}ccc@{}}
      -\frac{ m \,\sqrt{s}}{  E_\mp} & 0 & 0 \\
      \frac{\pm\sqrt{s}  \left(1-\frac{{ \omega }}{ E_\mp}\right)}{\sqrt{2} } & 1 & 0 \\
      \frac{\sqrt{\frac{2\, J-1}{2\, J+3}} \left(\frac{\sqrt{s} { \omega}}{ E_\mp}-4  E_\pm+\sqrt{s}\right)}{\sqrt{2} }
      & \pm\sqrt{\frac{2\, J-1}{2\, J+3}}
      & \pm\frac{  E_\pm}{\sqrt{s}}
    \end{array}
  \right) \,, \nonumber
\end{eqnarray}
with $J=n+1/2$, $E = E_\pm \mp M$ and $\omega +E = \sqrt{s}$. The energies $E_\pm$ were introduced already in (\ref{def-Qpm}). For the most complicated case the
non-zero elements of the transformation matrix are given in Tab. \ref{tab:CJ-13/2}. The transformation
matrices are not unitary and therefore the associated phase-space matrices $\rho^J_\pm (\sqrt{s}\,)$ are not diagonal.
In contrast to the helicity states the phase-space matrix in the covariant states does have off-diagonal elements.
The quest for the elimination of kinematical constraints leads necessarily to off-diagonal elements.
Note that the condition of minimal projectors leads possibly to distinct dimensions of the different columns in a transformation
matrix. The asymptotic property of the phase-space matrix (\ref{def-rho-constraint}) is readily confirmed
using the explicit form (\ref{res-C032}) and Tab. \ref{tab:CJ-13/2}. To the best knowledge of the authors the transformation
matrices $C^J_{\pm}$ presented here are novel and were not derived in the literature before.

In order to verify the acclaimed properties of the transformation matrices it suffices to consider the
$0\, {\tiny {1\over 2}} \to 0 \,{\tiny {3\over 2}} $, $0\, {\tiny {1\over 2}} \to 1\, {\tiny {1\over 2}} $ and
$0\, {\tiny {1\over 2}} \to 1\, {\tiny {3\over 2}} $ reactions. Nevertheless, we cross checked our results against all considered two-body
reactions. Our derivations rely on extensive use of computer algebra programs that indeed verify our claims. While we obtained explicit expressions
for all of such  reactions, which are useful in numerical applications, the results are too
tedious to be shown fully here.

The decomposition of the covariant partial-wave amplitudes in terms
of the invariant amplitudes $F^\pm_n$ should not lead to a singular behavior. Following previous works \cite{Stoica:2011cy,Lutz:2011xc} we introduce
the convenient notation
\begin{eqnarray}
  T^J_\pm(\sqrt{s}\,) &=& \sum_{k,n}\,a_{\pm \,n}^{J+k}(\sqrt{s}\,)\,A_{\pm\, n}^{J+k}(\sqrt{s}\,)
  \nonumber\\
  &+& \sum_{k,n}\,b_{\pm \,n}^{J+k}(\sqrt{s}\,)\,A_{\mp\, n}^{J+k}(\sqrt{s}\,)\,,
  \label{def-decomposition-TJ}
\end{eqnarray}
where we introduced the functions $A_{\pm\, n}^{J+k}(\sqrt{s}\,)$ already in (\ref{def-As}). The coefficient functions
$a_{\pm \,n}^{J+k}(\sqrt{s}\,)$ and $b_{\pm \,n}^{J+k}(\sqrt{s}\,)$
for $0\, {\tiny {1\over 2}} \to 0 \,{\tiny {3\over 2}} $
and $0\, {\tiny {1\over 2}} \to 1\, {\tiny {1\over 2}} $ are shown in Tab.~\ref{tab:abpm-01/2to03/2}
and Tab.~\ref{tab:abpm-01/2to11/2}.

We turn to the associated projector polynomials. They do not depend on any mass parameter and are
regular in particular at $r^2 =0$, $\bar r^2=0$ and $\bar r \cdot r= 0$. Since they should
have minimal dimension they may come with different dimensions. Their derivation involves repeated use of the
on-shell identities
\begin{eqnarray}
  && \bar u(\bar p)\,( \bar r\cdot \hat\gamma  )\,P_\pm = \pm\,\bar E_\mp \, \bar u(\bar p)\,P_\pm\,,
  \quad
  \nonumber\\
  && P_\pm\,(\hat\gamma\cdot r)\,u(p) = \pm\, E_\mp\,P_\pm \,  u(p) \,.
  \label{Vr-property}
\end{eqnarray}
We recall that if a projector ${\mathcal Y}^J_{\pm,ab}( \bar r,\,r, w)$ is partial-wave expanded
via (\ref{def-Wigner}) it contributes exclusively to a single
partial-wave amplitude with $T^J_{\pm,ab}(\sqrt{s}\,) = 1$.
Since the explicit expressions are rather lengthy not all of them are shown here.
We provide explicit results for the two production cases $0\, {\tiny {1\over 2}} \to 0 \,{\tiny {3\over 2}} , \,1\, {\tiny {1\over 2}} $.
The projector polynomials are expressed in terms of the
tensor structures introduced in (\ref{def-01/2to03/2}, \ref{def-01/2to11/2}), where
the two sectors are discriminated by the use of specific Lorentz indices. We find  \hfill
\begin{widetext}
  \begin{eqnarray}
    \label{eq-projectors-YJ-01/2to03/2}
    && {\mathcal Y}^{n+\frac{1}{2}}_{\pm,11, \bar \nu} ( \bar r,\,r, w)
    =
    \pm\sqrt{\frac{3}{2}}\,\frac{ 1}{s^{3/2}}\, Y^{(1)}_{n+1}\,T^{(1)}_{\mp,\bar\nu}
    \mp\sqrt{\frac{3}{2}}\,\frac{1 }{s^{5/2}}\,Y^{(1)}_{n} \,(\bar r \cdot \hat \gamma)\,T^{(1)}_{\pm,\bar\nu}\,(\hat \gamma \cdot r)
    \,,\nonumber\\
    && {\mathcal Y}^{n+\frac{1}{2}}_{\pm,21, \bar \nu} ( \bar r,\,r, w)
    =
    \sqrt{\frac{n}{n+2}}\,\frac{2\,\sqrt{2} }{s}\,Y^{(1)}_{n+1}\,T^{(1)}_{\mp,\bar\nu}
    -\frac{\sqrt{2} }{\sqrt{n\,(n+2)} s}\,Y^{(2)}_{n+1}\,T^{(2)}_{\mp,\bar\nu}
    \nonumber\\ && \qquad
    -\,\frac{r^2\,(\bar q^2 -\bar p^2 -s) }{\sqrt{2}\,\sqrt{n\,(n+2)}\,s^4}\,Y^{(2)}_{n-1}\,(\bar r \cdot \hat \gamma)\,T^{(1)}_{\pm,\bar\nu} \,(\hat \gamma \cdot r)
    \nonumber\\ && \qquad
    +\,Y^{(2)}_{n}\,\left(
      \frac{ r^2\,(\bar q^2 -\bar p^2 -s)}{\sqrt{2}\,\sqrt{n\,(n+2)}\,s^3}\,T^{(1)}_{\mp,\bar\nu}
      +\frac{\sqrt{2} }{\sqrt{n\,(n+2)}\,s^2}\,(\bar r \cdot \hat \gamma)\,T^{(2)}_{\pm,\bar\nu}\,(\hat \gamma \cdot r)
    \right)
    \nonumber\\ && \qquad
    \pm\,\sqrt{2}\,Y^{(1)}_{n}\,\left(
      \sqrt{\frac{n}{n+2}}\,\frac{(\pm 2\,\Slash{\bar p}\,\sqrt{s} + \bar q^2 - \bar p^2 -s)}{s^{5/2}}
      + \frac{1}{\sqrt{n\,(n+2)}\,s^{3/2}}
    \right)\,T^{(1)}_{\pm,\bar\nu}\,(\hat \gamma \cdot r)
    \,,
 \nonumber\\ \nonumber\\ \nonumber\\
    \label{eq-projectors-YJ-01/2to11/2}
    && {\mathcal Y}^{n+\frac{1}{2}}_{\pm,11, \bar \mu} ( \bar r,\,r, w)
    =
    \mp \frac{1}{s^{3/2}}\,Y^{(1)}_{n+1}\,T^{(2)}_{\mp,\bar\mu}
    \pm \frac{1 }{s^{5/2}}\,Y^{(1)}_{n}\,(\bar r\cdot \hat\gamma)\,T^{(2)}_{\pm,\bar\mu}\,(\hat\gamma\cdot r)
    \,,\nonumber\\
    && {\mathcal Y}^{n+\frac{1}{2}}_{\pm,21, \bar \mu}\,( \bar r,\,r, w)
    =
    Y^{(1)}_{n+1}\,\left(\pm\frac{1}{\sqrt{2}\,\sqrt{s}}\,T^{(1)}_{\mp,\bar\mu} + \frac{1}{\sqrt{2}\,s}\,T^{(2)}_{\mp,\bar\mu}\right)
    \nonumber\\ && \qquad
    \pm\,Y^{(1)}_{n}\,\left(
      \frac{ \left(\pm \,2\, \slash{\!\!\!\bar p}\,\sqrt{s} + 3 \,\bar q^2 -3\, \bar p^2 + s\right)}{2\,\sqrt{2}\,s^{5/2}}\,T^{(2)}_{\pm,\bar\mu}\,(\hat\gamma\cdot r)
      +\frac{1 }{\sqrt{2} s^{3/2}}\,(\bar r\cdot \hat \gamma)\,T^{(1)}_{\pm,\bar\mu} (\hat \gamma \cdot r)
    \right)
    \,,\nonumber\\
    && {\mathcal Y}^{n+\frac{1}{2}}_{\pm,31, \bar \mu} ( \bar r,\,r, w)
    =
    -\sqrt{\frac{n}{n+2}}\,\frac{2\,\sqrt{2} }{s}\,Y^{(1)}_{n+1}\,T^{(2)}_{\mp,\bar\mu}
    -\frac{\sqrt{2}  }{\sqrt{n\,(n+2)}\,s}\,Y^{(2)}_{n+1}\,T^{(3)}_{\mp,\bar\mu}
    \nonumber\\ && \qquad
    -\, \frac{r^2 \,(s -\bar p^2 +\bar q^2) }{\sqrt{2} \sqrt{n \,(n+2)} s^4}\,Y^{(2)}_{n-1}\,(\bar r\cdot \hat\gamma)\,T^{(2)}_{\pm,\bar\mu}\,(\hat\gamma\cdot r)
    \nonumber\\ && \qquad
    +\,Y^{(2)}_{n} \left(
      \frac{\sqrt{2}  }{\sqrt{n\,(n+2)}\,s^2}\,(\bar r\cdot \hat\gamma)\,T^{(3)}_{\pm,\bar\mu}\,(\hat\gamma\cdot r)
      +\frac{ r^2\,(s -\bar p^2 +\bar q^2)}{\sqrt{2}\,\sqrt{n\,(n+2)}\,s^3}\,T^{(2)}_{\mp,\bar\mu}
    \right)
    \nonumber\\ && \qquad
    +\,Y^{(1)}_{n} \left(
      \sqrt{\frac{n}{n+2}}\,\frac{2\,\sqrt{2} }{ s^2}\,(\bar r\cdot \hat\gamma)\,T^{(2)}_{\pm,\bar\mu}\,(\hat \gamma \cdot r)
      -\frac{\sqrt{2}  }{\sqrt{n\,(n+2)} s}\,T^{(1)}_{\pm,\bar\mu}\,(\hat \gamma \cdot r)
    \right)    \,.
  \end{eqnarray}
\end{widetext}

\section{Summary}
\label{sec:summary}

In this work we studied the generic properties of two-body reactions that are of central importance for the
computation of the resonance spectrum of baryons in the hadrogenesis conjecture.
We have constructed covariant partial-wave amplitudes for two-body reactions
with $J^P=0^-,1^-$ and $J^P=\frac{1}{2}^+,\frac{3}{2}^+$ particles which
are free from kinematical constraints. Those covariant partial-wave amplitudes are conveniently
used in partial-wave dispersion relations. Explicit transformations from the conventional
helicity states to the covariant states were derived and presented in this work.
It was illustrated that the covariant partial-wave amplitudes are associated with projector polynomials that
generate analytic solutions of the Bethe-Salpeter equation in the presence of short-range interactions.

In an initial step we identified complete sets of invariant functions that parameterize the considered reaction
amplitudes and are expected to satisfy Mandelstam's dispersion-integral representation. A convenient projection algebra
was constructed that is instrumental in a derivation of the invariant amplitudes by means of computer algebra codes.


\begin{acknowledgement}
M.F.M.L. thanks E.E. Kolomeitsev for collaboration at an early stage of the project.
\end{acknowledgement}



\appendix
\phantomsection
\addcontentsline{toc}{section}{Appendix A}
\section*{Appendix A}
\label{app-A}
\setcounter{section}{1}
\setcounter{equation}{0}
\numberwithin{equation}{section}

In this Appendix we specify the projectors introduced in
(\ref{def-11/2to03/2}- \ref{def-03/2to13/2}). The various sectors are
discriminated by the use of specific Lorentz indices. While the presence of the index $\mu$ and $\bar \mu $
signals a spin-one particle, the indices $\nu$ and $\bar \nu$
a spin-three-half particle in the initial and final state respectively. For the (\ref{def-11/2to03/2}) system we find
\begin{eqnarray}
  &&
  Q^{\bar \nu \mu}_{\pm, 1}
  =
  - \frac{\sqrt s}{\bar E_\pm}\,\Big[(\bar r\cdot r)\,P^{\bar \nu \mu}_{\pm ,1} - \bar E_\pm\,E_\mp\,P^{\bar \nu \mu}_{\mp ,1}\Big]
  \nonumber\\ && \qquad
  \mp\,\frac{1}{\bar E_\pm}\,\Big[ (\bar r\cdot r)\,Q^{\bar \nu \mu}_{\mp, 3} + \bar E_\pm\,E_\mp\,Q^{\bar \nu \mu}_{\pm, 3}\Big]
  \,, \nonumber\\ &&
  Q^{\bar \nu \mu}_{\pm, 2}
  =
  \pm\,w\,P^{\bar \nu \mu}_{\mp ,4}
  \pm\,\frac{1}{\sqrt s}\,Q^{\bar \nu \mu}_{\mp, 1}
  \nonumber\\ && \qquad
  \mp\,{\textstyle{1\over 2}}\,(\bar \delta - 1)\,\frac{s}{v^2}\,E_\pm\,\Big[ (\bar r\cdot r)\,Q^{\bar \nu \mu}_{\pm, 1} + \bar E_\mp\,E_\mp\,Q^{\bar \nu \mu}_{\mp, 1}\Big]
  \,, \nonumber\\ &&
  Q^{\bar \nu \mu}_{\pm, 3}
  =
  \pm\,\frac{s}{v^2}\,\sqrt{s}\,(\bar r\cdot r)\,\Big[ (\bar r\cdot r)\,P^{\bar \nu \mu}_{\mp ,1} - \bar E_\mp\,E_\pm\,P^{\bar \nu \mu}_{\pm ,1}\Big]
  \nonumber\\ && \qquad
  \pm\,\frac{s}{v^2}\,\bar E_\mp\,\Big[ (\bar r\cdot r)\,P^{\bar \nu \mu}_{\pm ,2} - \bar E_\pm\,E_\pm\,P^{\bar \nu \mu}_{\mp ,2}\Big]
  \,, \nonumber\\ &&
  Q^{\bar \nu \mu}_{\pm, 4}
  =
  \frac{s}{v^2}\,\Big[ (\bar r\cdot r)\,P^{\bar \nu \mu}_{\mp ,5} - \bar E_\mp\,E_\mp\,P^{\bar \nu \mu}_{\pm ,5}\Big]
  \nonumber\\ && \qquad
  \mp\,\frac{s}{v^2}\,E_\mp\,\Big[ (\bar r\cdot r)\,Q^{\bar \nu \mu}_{\pm, 2} + \bar E_\mp\,E_\pm\,Q^{\bar \nu \mu}_{\mp, 2}\Big]
  \nonumber\\ && \qquad
  -\, {\textstyle{1\over 2}}\,(\bar \delta - 1)\,\frac{s}{v^2}\,(r\cdot r)\,Q^{\bar \nu \mu}_{\pm, 1}
  \,, \nonumber\\ &&
  Q^{\bar \nu \mu}_{\pm, 5}
  =
  \frac{s}{v^2}\,\Big[ (\bar r\cdot r)\,P^{\bar \nu \mu}_{\mp ,3} - \bar E_\mp\,E_\mp\,P^{\bar \nu \mu}_{\pm ,3}\Big]
  \nonumber\\ && \qquad
  \pm\,{\textstyle{1\over 2}}\,(\delta + 1)\,\frac{s}{v^2}\,\bar E_\mp\,\Big[ (\bar r\cdot r)\,Q^{\bar \nu \mu}_{\mp, 3} + \bar E_\pm\,E_\mp\,Q^{\bar \nu \mu}_{\pm, 3}\Big]
  \nonumber\\ && \qquad
  -\, {\textstyle{1\over 2}}\,(\delta + 1)\,\frac{s}{v^2}\,(\bar r\cdot\bar r)\,Q^{\bar \nu \mu}_{\pm, 1}
  \,, \nonumber\\ &&
  Q^{\bar \nu \mu}_{\pm, 6}
  =
  \frac{s}{v^2}\,\Big[ (\bar r\cdot r)\,P^{\bar \nu \mu}_{\mp ,6} - \bar E_\mp\,E_\mp\,P^{\bar \nu \mu}_{\pm ,6}\Big]
  \nonumber\\ && \qquad
  \pm\,{\textstyle{1\over 2}}\,(\delta + 1)\,\frac{s}{v^2}\,\bar E_\mp\,\Big[ (\bar r\cdot r)\,Q^{\bar \nu \mu}_{\mp, 2} + \bar E_\pm\,E_\mp\,Q^{\bar \nu \mu}_{\pm, 2}\Big]
  \nonumber\\ && \qquad
  -\, {\textstyle{1\over 4}}\,(\bar \delta - 1)\,(\delta + 1)\,\frac{s}{v^2}\,(\bar r\cdot r)\,Q^{\bar \nu \mu}_{\pm, 1}
  \,,\nonumber
\end{eqnarray}
where
\begin{eqnarray}
  &&\bar p_{\bar \nu}\, P^{\bar \nu \mu }_{\pm, k} = 0 =\Lambda\,P^{\bar \nu \mu }_{\pm, k}\,\bar\Lambda\,\gamma_{\bar \nu}\,,
  \qquad   q_{ \mu}\, P^{\bar \nu \mu }_{\pm, k} = 0\,,
  \nonumber\\ \nonumber\\
  &&
  P^{\bar \nu \mu}_{\pm, 1}
  = \big[
  \rbot^{\bar\nu}\,i\,\gamma_5\,P_\pm
  - v^{\bar\nu}\,(\sqrt s/v^2)\,\bar E_\pm\,P_\mp
  \big]\,v^{\mu}/v^2
  \,,\nonumber\\ &&
  P^{\bar \nu \mu}_{\pm, 2}
  = \big[
  \rbot^{\bar\nu}\,P_\pm
  + v^{\bar\nu}\,(\sqrt s/v^2)\,\bar E_\pm\,i\,\gamma_5\,P_\mp
  \big]\,\rbarbot^{\mu}
  \,,\nonumber\\ &&
  P^{\bar \nu \mu}_{\pm, 3}
  = \big[
  \rbot^{\bar\nu}\,P_\pm
  + v^{\bar\nu}\,(\sqrt s/v^2)\,\bar E_\pm\,i\,\gamma_5\,P_\mp
  \big]\,\wLbarbot^{\mu}
  \,,\nonumber\\ &&
  P^{\bar \nu \mu}_{\pm, 4}
  = \big[
  \wRbot^{\bar\nu}\,i\,\gamma_5\,P_\pm
  - v^{\bar\nu}\,((\bar r\cdot r)\,P_\mp \pm\,\bar M\,E_\mp\,P_\pm)/v^2
  \big]\,v^{\mu}/v^2
  \,,\nonumber\\ &&
  P^{\bar \nu \mu}_{\pm, 5}
  = \big[
  \wRbot^{\bar\nu}\,P_\pm
  + v^{\bar\nu}\,i\,\gamma_5\,((\bar r\cdot r)\,P_\mp \pm\,\bar M\,E_\pm\,P_\pm)/v^2
  \big]\,\rbarbot^{\mu}
  \,,\nonumber\\ &&
  P^{\bar \nu \mu}_{\pm, 6}
  = \big[
  \wRbot^{\bar\nu}\,P_\pm
  + v^{\bar\nu}\,i\,\gamma_5\,((\bar r\cdot r)\,P_\mp \pm\,\bar M\,E_\pm\,P_\pm)/v^2
  \big]\,\wLbarbot^{\mu}
  \,.\nonumber
\end{eqnarray}
For the (\ref{def-01/2to13/2}) case
\begin{eqnarray}
  &&
  Q^{\bar \mu \bar \nu}_{\pm , 1}
  =
  \pm\,\frac{\sqrt{s}}{\bar M}\,\Big[ P^{\bar \mu \bar \nu}_{\mp ,5} + \sqrt{s}\,\bar E_\mp\,P^{\bar \mu \bar \nu}_{\pm ,4}\Big]
  \,,\nonumber\\ &&
  Q^{\bar \mu \bar \nu}_{\pm , 2}
  =
  \pm\,(1/\bar E_\mp)\,P^{\bar \mu \bar \nu}_{\mp ,5}
  \nonumber\\ && \qquad
  \pm\,{\textstyle{1\over 2}}\,(\bar\delta - 1)\,\frac{s}{v^2}\,\frac{(\bar r\cdot r)}{\bar E_\mp}\,\Big[ (\bar r\cdot r)\,Q^{\bar \mu \bar \nu}_{\pm,1}
  + \bar E_\mp\,E_\mp\,Q^{\bar \mu \bar \nu}_{\mp,1}\Big]
  \,,\nonumber\\ &&
  Q^{\bar \mu \bar \nu}_{\pm , 3}
  =
  \mp\,\sqrt{s}\,P^{\bar \mu \bar \nu}_{\pm ,1}
  \mp\,\frac{s}{v^2}\,\bar E_\pm\,\Big[ (\bar r\cdot r)\,Q^{\bar \mu \bar \nu}_{\pm,1}
  + \bar E_\mp\,E_\mp\,Q^{\bar \mu \bar \nu}_{\mp,1}\Big]
  \,,\nonumber\\ &&
  Q^{\bar \mu \bar \nu}_{\pm , 4}
  =
  \frac{s}{v^2}\,\Big[ (\bar r\cdot r)\,P^{\bar \mu \bar \nu}_{\mp ,6} - \bar E_\mp\,E_\mp\,P^{\bar \mu \bar \nu}_{\pm ,6}\Big]
  \nonumber\\ && \qquad
  \pm\,{\textstyle{1\over 2}}\,(\bar\delta + 1)\,\frac{s}{v^2}\,E_\mp\,\Big[ (\bar r\cdot r)\,Q^{\bar \mu \bar \nu}_{\mp,2}
  + \bar E_\mp\,E_\pm\,Q^{\bar \mu \bar \nu}_{\pm,2}\Big]
  \nonumber\\ && \qquad
  +\, {\textstyle{1\over 4}}\,(\bar\delta + 1)\,(\bar\delta - 1)\,\frac{s}{v^2}\,(r\cdot r)\,Q^{\bar \mu \bar \nu}_{\pm,1}
  \,,\nonumber\\ &&
  Q^{\bar \mu \bar \nu}_{\pm , 5}
  =
  \frac{s}{v^2}\,\Big[ (\bar r\cdot r)\,P^{\bar \mu \bar \nu}_{\mp ,3} - \bar E_\mp\,E_\mp\,P^{\bar \mu \bar \nu}_{\pm ,3}\Big]
  \nonumber\\ && \qquad
  \pm\,{\textstyle{1\over 2}}\,(\bar\delta + 1)\,\frac{s}{v^2}\,E_\mp\,\Big[ (\bar r\cdot r)\,Q^{\bar \mu \bar \nu}_{\mp,3}
  + \bar E_\mp\,E_\pm\,Q^{\bar \mu \bar \nu}_{\pm,3}\Big]
  \nonumber\\ && \qquad
  +\, {\textstyle{1\over 2}}\,(\bar\delta + 1)\,\frac{s}{v^2}\,(\bar r\cdot r)\,Q^{\bar \mu \bar \nu}_{\pm,1}
  \,,\nonumber\\ &&
  Q^{\bar \mu \bar \nu}_{\pm , 6}
  =
  \frac{s}{v^2}\,\Big[ (\bar r\cdot r)\,P^{\bar \mu \bar \nu}_{\mp ,2} - \bar E_\mp\,E_\mp\,P^{\bar \mu \bar \nu}_{\pm ,2}\Big]
  \nonumber\\ && \qquad
  \mp\,\frac{s}{v^2}\,\bar E_\mp\,\Big[ (\bar r\cdot r)\,Q^{\bar \mu \bar \nu}_{\pm,3}
  + \bar E_\pm\,E_\mp\,Q^{\bar \mu \bar \nu}_{\mp,3}\Big]
  \nonumber\\ && \qquad
  +\, \frac{s}{v^2}\,(\bar r\cdot\bar r)\,Q^{\bar \mu \bar \nu}_{\pm,1}
  \,,\nonumber
\end{eqnarray}
where
\begin{eqnarray}
  &&
  \bar p_{ \bar\nu}\, P^{\bar \mu \bar \nu }_{\pm, k} = 0= \Lambda\,P^{\bar \mu \bar \nu }_{\pm, k}\,\bar\Lambda\,\gamma_{\bar \nu}\,,\qquad
  \bar q_{\bar \mu}\, P^{\bar \mu \bar \nu }_{\pm, k} =0\,,
  \nonumber\\ \nonumber\\
  &&
  P^{\bar \mu \bar \nu}_{\pm, 1}
  = \big[
  \rbot^{\bar \nu}\,i\,\gamma_5\,P_\pm
  - v^{\bar \nu}\,(\sqrt{s}/v^2)\,\bar E_\pm\,P_\mp
  \big]\,v^{\bar \mu}/v^2
  \,,\nonumber\\ &&
  P^{\bar \mu \bar \nu}_{\pm, 2}
  = \big[
  \rbot^{\bar \nu}\,P_\pm
  + v^{\bar \nu}\,(\sqrt{s}/v^2)\,\bar E_\pm\,i\,\gamma_5\,P_\mp
  \big]\,\rbot^{\bar \mu}
  \,,\nonumber\\ &&
  P^{\bar \mu \bar \nu}_{\pm, 3}
  = \big[
  \rbot^{\bar \nu}\,P_\pm
  + v^{\bar \nu}\,(\sqrt{s}/v^2)\,\bar E_\pm\,i\,\gamma_5\,P_\mp
  \big]\,\wLbot^{\bar \mu}
  \,,\nonumber\\ &&
  P^{\bar \mu \bar \nu}_{\pm, 4}
  = \big[
  \wRbot^{\bar \nu}\,i\,\gamma_5\,P_\pm
  - v^{\bar \nu}\,((\bar r\cdot r)\,P_\mp \pm\,\bar M\,E_\mp\,P_\pm)/v^2
  \big]\,v^{\bar \mu}/v^2
  \,,\nonumber\\ &&
  P^{\bar \mu \bar \nu}_{\pm, 5}
  = \big[
  \wRbot^{\bar \nu}\,P_\pm
  + v^{\bar \nu}\,i \,\gamma_5\,((\bar r\cdot r)\,P_\mp \pm\,\bar M\,E_\pm\,P_\pm)/v^2
  \big]\,\rbot^{\bar \mu}
  \,,\nonumber\\ &&
  P^{\bar \mu \bar \nu}_{\pm, 6}
  = \big[
  \wRbot^{\bar \nu}\,P_\pm
  + v^{\bar \nu}\,i \,\gamma_5\,((\bar r\cdot r)\,P_\mp \pm\,\bar M\,E_\pm\,P_\pm)/v^2
  \big]\,\wLbot^{\bar \mu}
  \,.\nonumber
\end{eqnarray}
For the (\ref{def-03/2to03/2}) case
\begin{eqnarray}
  &&
  Q^{\bar \nu \nu}_{\pm, 1}
  =
  P^{\bar \nu \nu}_{\mp,1}
  - \frac{s}{v^2}\,2\,\bar E_\mp\,E_\mp\,\Big[ (\bar r\cdot r)\,P^{\bar \nu \nu}_{\pm,1} -
  \bar E_\pm\,E_\pm\,P^{\bar \nu \nu}_{\mp,1}\Big]
  \,,\nonumber\\ &&
  Q^{\bar \nu \nu}_{\pm, 2}
  =
  \frac{s}{v^2}\,\Big[ (\bar r\cdot r)\,P^{\bar \nu \nu}_{\mp,3} - \bar E_\mp\,E_\mp\,P^{\bar \nu \nu}_{\pm,3}\Big]
  \nonumber\\ && \qquad
  -\, \frac{\sqrt s}{v^2}\,E_\mp\,\Big[ (\bar r\cdot r)\,Q^{\bar \nu \nu}_{\mp,1}
  + \bar E_\pm\,E_\pm\,Q^{\bar \nu \nu}_{\pm,1}\Big]
  \,,\nonumber\\ &&
  Q^{\bar \nu \nu}_{\pm, 3}
  =
  \frac{s}{v^2}\,\Big[ (\bar r\cdot r)\,P^{\bar \nu \nu}_{\mp,2} - \bar E_\mp\,E_\mp\,P^{\bar \nu \nu}_{\pm,2}\Big]
  \nonumber\\ && \qquad
  -\, \frac{\sqrt s}{v^2}\,\bar E_\mp\,\Big[ (\bar r\cdot r)\,Q^{\bar \nu \nu}_{\mp,1}
  + \bar E_\pm\,E_\pm\,Q^{\bar \nu \nu}_{\pm,1}\Big]
  \,,\nonumber\\ &&
  Q^{\bar \nu \nu}_{\pm, 4}
  =
  \frac{s}{v^2}\,\Big[ (\bar r\cdot r)\,P^{\bar \nu \nu}_{\mp,4} - \bar E_\mp\,E_\mp\,P^{\bar \nu \nu}_{\pm,4}\Big]
  - (1/s)\,Q^{\bar \nu \nu}_{\mp,1}
  \nonumber\\ && \qquad
  +\, {\textstyle{1\over 2}}\,(\bar\delta - 1)\,\frac{\sqrt s}{v^2}\,E_\pm\,\Big[ (\bar r\cdot r)\,Q^{\bar \nu \nu}_{\pm,1}
  + \bar E_\mp\,E_\mp\,Q^{\bar \nu \nu}_{\mp,1}\Big]
  \nonumber\\ && \qquad
  \pm\,{\textstyle{1\over 2}}\,(\delta - 1)\,\frac{\sqrt s}{v^2}\,\bar M\,\Big[ (\bar r\cdot r)\,Q^{\bar \nu \nu}_{\pm,1}
  + \bar E_\mp\,E_\mp\,Q^{\bar \nu \nu}_{\mp,1}\Big]
  \,,\nonumber
\end{eqnarray}
where
\begin{eqnarray}
  && \bar p_{\bar \nu}\, P^{\bar \nu \nu }_{\pm, k} =0
  = \Lambda\,P^{\bar \nu \nu }_{\pm, k}\,\bar\Lambda\,\gamma_{\bar \nu}\,, \qquad
  p_{ \nu}\, P^{\bar \nu \nu }_{\pm, k} =0=\gamma_\nu\,\Lambda\,P^{\bar \nu \nu }_{\pm, k}\,\bar\Lambda\,,
  \nonumber\\ \nonumber\\
  &&
  P^{\bar \nu \nu}_{\pm, 1}
  =
  \rbot^{\bar \nu}\,\rbarbot^{\nu}\,P_\pm
  + v^{\bar \nu}\,v^{\nu}\,(s/v^2)\,\bar E_\pm\,E_\pm\,P_\mp/v^2
  \nonumber\\ && \qquad
  -\, \rbot^{\bar \nu}\,v^{\nu}\,(\sqrt{s}/v^2)\,E_\pm\,i\,\gamma_5\,P_\pm
  \nonumber\\ && \qquad
  +\, v^{\bar \nu}\,\rbarbot^{\nu}\,(\sqrt{s}/v^2)\,\bar E_\pm\,i\,\gamma_5\,P_\mp
  \,,\nonumber\\ &&
  P^{\bar \nu \nu}_{\pm, 2}
  =
  \rbot^{\bar \nu}\,\wRbarbot^{\nu}\,P_\pm
  + v^{\bar \nu}\,\wRbarbot^{\nu}\,(\sqrt{s}/v^2)\,\bar E_\pm\,i\,\gamma_5\,P_\mp
  \nonumber\\ && \qquad
  -\, \rbot^{\bar \nu}\,v^{\nu}\,i \,\gamma_5\,((\bar r\cdot r)\,P_\pm \pm\,M\,\bar E_\pm\,P_\mp)/v^2
  \nonumber\\ && \qquad
  +\, v^{\bar \nu}\,v^{\nu}\,(\sqrt{s}/v^2)\,\bar E_\pm\,((\bar r\cdot r)\,P_\mp \pm\,M\,\bar E_\mp\,P_\pm)/v^2
  \,,\nonumber\\ &&
  P^{\bar \nu \nu}_{\pm, 3}
  =
  \wRbot^{\bar \nu}\,\rbarbot^{\nu}\,P_\pm
  - \wRbot^{\bar \nu}\,v^{\nu}\,(\sqrt{s}/v^2)\,E_\pm\,i\,\gamma_5\,P_\pm
  \nonumber\\ && \qquad
  +\, v^{\bar \nu}\,\rbarbot^{\nu}\,i\,\gamma_5\,((\bar r\cdot r)\,P_\mp \pm\,\bar M\,E_\pm\,P_\pm)/v^2
  \nonumber\\ && \qquad
  +\, v^{\bar \nu}\,v^{\nu}\,(\sqrt{s}/v^2)\,E_\pm\,((\bar r\cdot r)\,P_\mp \pm\,\bar M\,E_\mp\,P_\pm)/v^2
  \,,\nonumber\\ &&
  P^{\bar \nu \nu}_{\pm, 4}
  =
  \wRbot^{\bar \nu}\,\wRbarbot^{\nu}\,P_\pm
  + v^{\bar \nu}\,v^{\nu}\,\big[
  (1/s)\,P_\mp
  \nonumber\\ && \qquad \quad
  \mp\,{\textstyle{1\over 2}}\,(\bar\delta - 1)\,(\sqrt{s}/v^2)\,M\,((\bar r\cdot r)\,P_\pm - \bar E_\pm\,E_\pm\,P_\mp)
  \nonumber\\ && \qquad \quad
  \mp\,{\textstyle{1\over 2}}\,(\delta - 1)\,(\sqrt{s}/v^2)\,\bar M\,((\bar r\cdot r)\,P_\pm - \bar E_\pm\,E_\pm\,P_\mp)
  \nonumber\\ && \qquad \quad
  +\, {\textstyle{1\over 2}}\,(\bar\delta - 1)\,{\textstyle{1\over 2}}\,(\delta - 1)\,(s/v^2)\,\bar E_\pm\,E_\pm\,P_\mp
  \big]/v^2
  \nonumber\\ && \qquad
  -\, \wRbot^{\bar \nu}\,v^{\nu}\,i \,\gamma_5\,((\bar r\cdot r)\,P_\pm \pm\,M\,\bar E_\pm\,P_\mp)/v^2
  \nonumber\\ && \qquad
  +\, v^{\bar \nu}\,\wRbarbot^{\nu}\,i \,\gamma_5\,((\bar r\cdot r)\,P_\mp \pm\,\bar M\,E_\pm\,P_\pm)/v^2
  \,.\nonumber
\end{eqnarray}
For the (\ref{def-11/2to13/2}) case
\begin{eqnarray}
  &&
  Q_{\pm, 1}^{\bar \mu \bar \nu \mu}
  =
  {\textstyle{3}}\,(\bar\delta - 1)\,\frac{s}{v^2}\,\frac{\sqrt s}{\bar M}\,(r\cdot r)\,\bar E_\mp\,\Big[
  \nonumber\\ && \qquad \qquad
  s\,\bar E_\pm\,E_\mp\,\big[ (\bar r\cdot r)\,P_{\mp,1}^{\bar \mu \bar \nu \mu}
  - \bar E_\mp\,E_\pm\,P_{\pm,1}^{\bar \mu \bar \nu \mu}\big]
  \nonumber\\ && \qquad \qquad
  +\, \sqrt{s}\,E_\mp\,\big[ (\bar r\cdot r)\,P_{\pm,2}^{\bar \mu \bar \nu \mu}
  - \bar E_\pm\,E_\pm\,P_{\mp,2}^{\bar \mu \bar \nu \mu}\big]
  \nonumber\\ && \qquad \qquad
  -\, \sqrt{s}\,\bar E_\pm\,\big[ (\bar r\cdot r)\,P_{\mp,7}^{\bar \mu \bar \nu \mu}
  - \bar E_\mp\,E_\mp\,P_{\pm,7}^{\bar \mu \bar \nu \mu}\big]
  \nonumber\\ && \qquad \qquad
  +\, \big[ (\bar r\cdot r)\,P_{\pm,8}^{\bar \mu \bar \nu \mu}
  - \bar E_\pm\,E_\mp\,P_{\mp,8}^{\bar \mu \bar \nu \mu}\big]
  \Big]
  \nonumber\\ && \qquad
  +\, 4\,\frac{s}{v^2}\,\frac{\sqrt s}{\bar M}\,(\bar r\cdot\bar r)\,E_\mp\,\Big[
  \nonumber\\ && \qquad \qquad
  s\,\bar E_\mp\,E_\pm\,\big[ (\bar r\cdot r)\,P_{\pm,4}^{\bar \mu \bar \nu \mu}
  - \bar E_\pm\,E_\mp\,P_{\mp,4}^{\bar \mu \bar \nu \mu}\big]
  \nonumber\\ && \qquad \qquad
  +\, \sqrt{s}\,E_\pm\,\big[ (\bar r\cdot r)\,P_{\mp,5}^{\bar \mu \bar \nu \mu}
  - \bar E_\mp\,E_\mp\,P_{\pm,5}^{\bar \mu \bar \nu \mu}\big]
  \nonumber\\ && \qquad \qquad
  -\, \sqrt{s}\,\bar E_\mp\,\big[ (\bar r\cdot r)\,P_{\pm,10}^{\bar \mu \bar \nu \mu}
  - \bar E_\pm\,E_\pm\,P_{\mp,10}^{\bar \mu \bar \nu \mu}\big]
  \nonumber\\ && \qquad \qquad
  +\, \big[ (\bar r\cdot r)\,P_{\mp,11}^{\bar \mu \bar \nu \mu}
  - \bar E_\mp\,E_\pm\,P_{\pm,11}^{\bar \mu \bar \nu \mu}\big]
  \Big]
  \nonumber\\ && \qquad
  +\, {\textstyle{2}}\,(\bar\delta - 1)\,\frac{\sqrt s}{\bar M}\,E_\mp\,\Big[
  s\,\bar E_\mp\,E_\pm\,P_{\pm,1}^{\bar \mu \bar \nu \mu}
  + \sqrt{s}\,E_\pm\,P_{\mp,2}^{\bar \mu \bar \nu \mu}
  \nonumber\\ && \qquad \qquad \qquad
  -\, \sqrt{s}\,\bar E_\mp\,P_{\pm,7}^{\bar \mu \bar \nu \mu}
  + P_{\mp,8}^{\bar \mu \bar \nu \mu}
  \Big]
  \nonumber\\ && \qquad
  \pm\,E_\mp\,\Big[
  - s\,\bar E_\mp\,E_\pm\,P_{\pm,1}^{\bar \mu \bar \nu \mu}
  + \sqrt{s}\,E_\pm\,P_{\mp,2}^{\bar \mu \bar \nu \mu}
  \nonumber\\ && \qquad \qquad \qquad
  +\, \sqrt{s}\,\bar E_\mp\,P_{\pm,7}^{\bar \mu \bar \nu \mu}
  + P_{\mp,8}^{\bar \mu \bar \nu \mu}
  \Big]
  \nonumber\\ && \qquad
  +\, (\bar r\cdot r)\,\frac{1}{\bar M}\,\Big[
  s\,\bar E_\mp\,E_\mp\,P_{\mp,1}^{\bar \mu \bar \nu \mu}
  + \sqrt{s}\,E_\mp\,P_{\pm,2}^{\bar \mu \bar \nu \mu}
  \nonumber\\ && \qquad \qquad \qquad
  -\, \sqrt{s}\,\bar E_\mp\,P_{\mp,7}^{\bar \mu \bar \nu \mu}
  + P_{\pm,8}^{\bar \mu \bar \nu \mu}
  \Big]
  \nonumber\\ && \qquad
  -\, \frac{s}{\bar M}\,\sqrt{s}\,\bar E_\mp\,\Big[ (\bar r\cdot r)\,P_{\pm,4}^{\bar \mu \bar \nu \mu}
  - 3\,\bar E_\pm\,E_\mp\,P_{\mp,4}^{\bar \mu \bar \nu \mu}\Big]
  \nonumber\\ && \qquad
  -\, \frac{s}{\bar M}\,\Big[ (\bar r\cdot r)\,P_{\mp,5}^{\bar \mu \bar \nu \mu}
  - 3\,\bar E_\mp\,E_\mp\,P_{\pm,5}^{\bar \mu \bar \nu \mu}\Big]
  \nonumber\\ && \qquad
  +\, 2\,\frac{\sqrt s}{\bar M}\,\bar E_\mp\,\Big[ P_{\pm,11}^{\bar \mu \bar \nu \mu}
  - \sqrt{s}\,\bar E_\pm\,P_{\mp,10}^{\bar \mu \bar \nu \mu}\Big]
  \,,\nonumber\\ &&
  Q_{\pm, 2}^{\bar \mu \bar \nu \mu}
  =
  -\, (\bar\delta - 1)\,\frac{s}{v^2}\,\frac{\sqrt s}{\bar M}\,(r\cdot r)\,\bar E_\pm\,\Big[
  \nonumber\\ && \qquad \qquad
  s\,\bar E_\mp\,E_\pm\,\big[ (\bar r\cdot r)\,P_{\pm,1}^{\bar \mu \bar \nu \mu}
  - \bar E_\pm\,E_\mp\,P_{\mp,1}^{\bar \mu \bar \nu \mu}\big]
  \nonumber\\ && \qquad \qquad
  +\, \sqrt{s}\,E_\pm\,\big[ (\bar r\cdot r)\,P_{\mp,2}^{\bar \mu \bar \nu \mu}
  - \bar E_\mp\,E_\mp\,P_{\pm,2}^{\bar \mu \bar \nu \mu}\big]
  \nonumber\\ && \qquad \qquad
  -\, \sqrt{s}\,\bar E_\mp\,\big[ (\bar r\cdot r)\,P_{\pm,7}^{\bar \mu \bar \nu \mu}
  - \bar E_\pm\,E_\pm\,P_{\mp,7}^{\bar \mu \bar \nu \mu}\big]
  \nonumber\\ && \qquad \qquad
  +\, \big[ (\bar r\cdot r)\,P_{\mp,8}^{\bar \mu \bar \nu \mu}
  - \bar E_\mp\,E_\pm\,P_{\pm,8}^{\bar \mu \bar \nu \mu}\big]
  \Big]
  \nonumber\\ && \qquad
  -\, 2\,\frac{s}{v^2}\,\frac{\sqrt s}{\bar M}\,(\bar r\cdot\bar r)\,E_\pm\,\Big[
  \nonumber\\ && \qquad \qquad
  s\,\bar E_\pm\,E_\mp\,\big[ (\bar r\cdot r)\,P_{\mp,4}^{\bar \mu \bar \nu \mu}
  - \bar E_\mp\,E_\pm\,P_{\pm,4}^{\bar \mu \bar \nu \mu}\big]
  \nonumber\\ && \qquad \qquad
  +\, \sqrt{s}\,E_\mp\,\big[ (\bar r\cdot r)\,P_{\pm,5}^{\bar \mu \bar \nu \mu}
  - \bar E_\pm\,E_\pm\,P_{\mp,5}^{\bar \mu \bar \nu \mu}\big]
  \nonumber\\ && \qquad \qquad
  -\, \sqrt{s}\,\bar E_\pm\,\big[ (\bar r\cdot r)\,P_{\mp,10}^{\bar \mu \bar \nu \mu}
  - \bar E_\mp\,E_\mp\,P_{\pm,10}^{\bar \mu \bar \nu \mu}\big]
  \nonumber\\ && \qquad \qquad
  +\, \big[ (\bar r\cdot r)\,P_{\pm,11}^{\bar \mu \bar \nu \mu}
  - \bar E_\pm\,E_\mp\,P_{\mp,11}^{\bar \mu \bar \nu \mu}\big]
  \Big]
  \nonumber\\ && \qquad
  +\, {\textstyle{1\over 2}}\,(\bar\delta - 1)\,\frac{s}{\bar M}\,\bar E_\pm\,E_\pm\,\Big[ P_{\mp,7}^{\bar \mu \bar \nu \mu}
  - \sqrt{s}\,E_\mp\,P_{\mp,1}^{\bar \mu \bar \nu \mu}\Big]
  \nonumber\\ && \qquad
  \pm\,\sqrt{s}\,(\bar r\cdot r)\,\Big[ P_{\pm,7}^{\bar \mu \bar \nu \mu}
  - \sqrt{s}\,E_\pm\,P_{\pm,1}^{\bar \mu \bar \nu \mu}\Big]
  \nonumber\\ && \qquad
  +\, \frac{1}{\bar M}\,\bar E_\pm\,E_\pm\,\Big[ P_{\pm,8}^{\bar \mu \bar \nu \mu}
  + \sqrt{s}\,E_\mp\,P_{\pm,2}^{\bar \mu \bar \nu \mu}\Big]
  \nonumber\\ && \qquad
  -\, \frac{\sqrt s}{\bar M}\,\bar E_\pm\,\Big[ s\,\bar E_\mp\,E_\pm\,P_{\pm,4}^{\bar \mu \bar \nu \mu}
  + \sqrt{s}\,E_\pm\,P_{\mp,5}^{\bar \mu \bar \nu \mu}
  \nonumber\\ && \qquad \qquad \qquad
  -\, \sqrt{s}\,\bar E_\mp\,P_{\pm,10}^{\bar \mu \bar \nu \mu}
  + P_{\mp,11}^{\bar \mu \bar \nu \mu}
  \Big]
  \,,\nonumber\\ &&	
  Q_{\pm, 3}^{\bar \mu \bar \nu \mu}
  =
  \pm\,\frac{\sqrt s}{\bar M}\,\Big[ P_{\mp,17}^{\bar \mu \bar \nu \mu}
  - \sqrt{s}\,\bar E_\mp\,P_{\pm,16}^{\bar \mu \bar \nu \mu}\Big]
  \nonumber\\ && \qquad
  \mp\,{\textstyle{1\over 4}}\,(\bar\delta - 1)\,(\delta + 1)\,\frac{s}{v^2}\,\frac{\sqrt s}{\bar M}\,(\bar r\cdot r)\,\Big[
  \nonumber\\ && \qquad \qquad \qquad
  (\bar r\cdot r)\,Q_{\pm,6}^{\bar \mu \bar \nu \mu}
  + \bar E_\mp\,E_\pm\,Q_{\mp,6}^{\bar \mu \bar \nu \mu}\Big]
  \nonumber\\ && \qquad
  \pm\,{\textstyle{1\over 2}}\,(\delta + 1)\,\frac{s}{v^2}\,\bar E_\mp\,\Big[ (\bar r\cdot r)\,Q_{\mp,1}^{\bar \mu \bar \nu \mu}
  + \bar E_\pm\,E_\pm\,Q_{\pm,1}^{\bar \mu \bar \nu \mu}\Big]
  \,,\nonumber\\ &&
  Q_{\pm, 4}^{\bar \mu \bar \nu \mu}
  =
  \frac{\sqrt s}{\bar E_\pm}\,\Big[ (\bar r\cdot r)\,P_{\pm,3}^{\bar \mu \bar \nu \mu}
  - \bar E_\pm\,E_\pm\,P_{\mp,3}^{\bar \mu \bar \nu \mu}\Big]
  \nonumber\\ && \qquad
  +\, {\textstyle{1\over 2}}\,(\bar\delta + 1)\,\Big[
  - \frac{E_\pm}{\bar E_\pm}\,Q_{\pm,8}^{\bar \mu \bar \nu \mu}
  \nonumber\\ && \qquad \qquad
  \pm\,\frac{s}{v^2}\,\frac{(\bar r\cdot r)}{\bar E_\pm}\,\big[
  (\bar r\cdot r)\,Q_{\mp,1}^{\bar \mu \bar \nu \mu}
  + \bar E_\pm\,E_\pm\,Q_{\pm,1}^{\bar \mu \bar \nu \mu}\big]
  \nonumber\\ && \qquad \qquad
  \pm\,\frac{s}{v^2}\,E_\pm\,\big[ (\bar r\cdot r)\,Q_{\mp,2}^{\bar \mu \bar \nu \mu}
  + \bar E_\mp\,E_\mp\,Q_{\pm,2}^{\bar \mu \bar \nu \mu}\big]
  \Big]
  \nonumber\\ && \qquad
  \mp\,\frac{1}{\bar E_\pm}\,\Big[ (\bar r\cdot r)\,Q_{\mp,12}^{\bar \mu \bar \nu \mu}
  + \bar E_\pm\,E_\pm\,Q_{\pm,12}^{\bar \mu \bar \nu \mu}\Big]
  \,,\nonumber\\ &&
  Q_{\pm, 5}^{\bar \mu \bar \nu \mu}
  =
  \frac{\sqrt s}{\bar E_\pm}\,\Big[ (\bar r\cdot r)\,P_{\pm,7}^{\bar \mu \bar \nu \mu}
  - \bar E_\pm\,E_\pm\,P_{\mp,7}^{\bar \mu \bar \nu \mu}\Big]
  \nonumber\\ && \qquad
  \mp\,\frac{s}{v^2}\,E_\pm\,\Big[
  (\bar r\cdot r)\,(Q_{\pm,1}^{\bar \mu \bar \nu \mu} + Q_{\mp,2}^{\bar \mu \bar \nu \mu})
  \nonumber\\ && \qquad \qquad \qquad
  +\, \bar E_\mp\,E_\mp\,(Q_{\mp,1}^{\bar \mu \bar \nu \mu} + Q_{\pm,2}^{\bar \mu \bar \nu \mu})
  \Big]
  \nonumber\\ && \qquad
  \mp\,\frac{1}{\bar E_\pm}\,Q_{\pm,2}^{\bar \mu \bar \nu \mu}
  - \frac{E_\pm}{\bar E_\pm}\,Q_{\pm,8}^{\bar \mu \bar \nu \mu}
  \,,\nonumber\\ &&
  Q_{\pm, 6}^{\bar \mu \bar \nu \mu}
  =
  \frac{2}{(\bar\delta - 1)}\,\frac{1}{(r\cdot r)}\,\Big[ (\bar r\cdot r)\,P_{\mp,11}^{\bar \mu \bar \nu \mu}
  - \bar E_\mp\,E_\pm\,P_{\pm,11}^{\bar \mu \bar \nu \mu}\Big]
  \nonumber\\ && \qquad
  +\, \frac{2}{(\bar\delta - 1)}\,\frac{\bar E_\mp}{E_\mp}\,\Big[
  Q_{\pm,7}^{\bar \mu \bar \nu \mu}
  \nonumber\\ && \qquad \qquad \qquad
  \pm\,\frac{1}{E_\pm}\,\big[ (\bar r\cdot r)\,Q_{\pm,9}^{\bar \mu \bar \nu \mu}
  + \bar E_\pm\,E_\pm\,Q_{\mp,9}^{\bar \mu \bar \nu \mu}\big]
  \Big]
  \nonumber\\ && \qquad
  \mp\,\frac{s}{v^2}\,\bar E_\mp\,\Big[ (\bar r\cdot r)\,(Q_{\mp,1}^{\bar \mu \bar \nu \mu} + Q_{\pm,2}^{\bar \mu \bar \nu \mu})
  \nonumber\\ && \qquad \qquad \qquad
  +\, \bar E_\pm\,E_\pm\,(Q_{\pm,1}^{\bar \mu \bar \nu \mu} + Q_{\mp,2}^{\bar \mu \bar \nu \mu})
  \Big]
  \nonumber\\ && \qquad
  \mp\,\frac{1}{E_\mp}\,Q_{\pm,1}^{\bar \mu \bar \nu \mu}
  + \frac{(\bar r\cdot r)}{(r\cdot r)}\,Q_{\pm,5}^{\bar \mu \bar \nu \mu}
  \,,\nonumber\\ &&
  Q_{\pm, 7}^{\bar \mu \bar \nu \mu}
  =
  s\,\Big[ (\bar r\cdot r)\,P_{\pm,4}^{\bar \mu \bar \nu \mu}
  - \bar E_\pm\,E_\mp\,P_{\mp,4}^{\bar \mu \bar \nu \mu}\Big]
  \nonumber\\ && \qquad
  \pm\,{\textstyle{1\over 2}}\,(\bar\delta - 1)\,\frac{s}{v^2}\,E_\mp\,\Big[
  (\bar r\cdot r)\,(Q_{\mp,1}^{\bar \mu \bar \nu \mu}
  + Q_{\pm,2}^{\bar \mu \bar \nu \mu})
  \nonumber\\ && \qquad \qquad \qquad
  +\, \bar E_\pm\,E_\pm\,(Q_{\pm,1}^{\bar \mu \bar \nu \mu}
  + Q_{\mp,2}^{\bar \mu \bar \nu \mu})
  \Big]
  \nonumber\\ && \qquad
  \mp\,\frac{1}{\sqrt s}\,(Q_{\pm,1}^{\bar \mu \bar \nu \mu} + Q_{\mp,2}^{\bar \mu \bar \nu \mu})
  \,,\nonumber\\ &&
  Q_{\pm, 8}^{\bar \mu \bar \nu \mu}
  =
  s\,\Big[ (\bar r\cdot r)\,P_{\pm,1}^{\bar \mu \bar \nu \mu}
  - \bar E_\pm\,E_\mp\,P_{\mp,1}^{\bar \mu \bar \nu \mu}\Big]
  \nonumber\\ && \qquad
  \mp\,\frac{s}{v^2}\,\bar E_\pm\,\Big[ (\bar r\cdot r)\,(Q_{\pm,1}^{\bar \mu \bar \nu \mu} + Q_{\mp,2}^{\bar \mu \bar \nu \mu})
  \nonumber\\ && \qquad \qquad \qquad
  +\, \bar E_\mp\,E_\mp\,(Q_{\mp,1}^{\bar \mu \bar \nu \mu} + Q_{\pm,2}^{\bar \mu \bar \nu \mu})
  \Big]
  \,,\nonumber\\ &&
  Q_{\pm, 9}^{\bar \mu \bar \nu \mu}
  =
  \pm\,\sqrt{s}\,P_{\pm,10}^{\bar \mu \bar \nu \mu}
  \mp\,\frac{1}{\sqrt s}\,Q_{\pm,5}^{\bar \mu \bar \nu \mu}
  \nonumber\\ && \qquad
  +\, {\textstyle{1\over 2}}\,(\bar\delta - 1)\,\frac{s}{v^2}\,E_\pm\,\Big[
  E_\mp\,Q_{\pm,2}^{\bar \mu \bar \nu \mu}
  \nonumber\\ && \qquad \qquad \qquad
  \pm\,\big[ (\bar r\cdot r)\,Q_{\mp,5}^{\bar \mu \bar \nu \mu}
  + \bar E_\pm\,E_\mp\,Q_{\pm,5}^{\bar \mu \bar \nu \mu}\big]
  \Big]
  \nonumber\\ && \qquad
  -\, \frac{\sqrt s}{v^2}\,E_\pm\,\Big[ (\bar r\cdot r)\,Q_{\pm,1}^{\bar \mu \bar \nu \mu} \mp\,\bar M\,E_\mp\,Q_{\mp,1}^{\bar \mu \bar \nu \mu}\Big]
  \nonumber\\ && \qquad
  \mp\,\frac{s}{v^2}\,E_\pm\,\Big[ (\bar r\cdot r)\,Q_{\pm,7}^{\bar \mu \bar \nu \mu} + \bar E_\pm\,E_\mp\,Q_{\mp,7}^{\bar \mu \bar \nu \mu}\Big]
  \,,\nonumber\\ &&
  Q_{\pm,10}^{\bar \mu \bar \nu \mu}
  =
  \mp\,\sqrt{s}\,P_{\mp,6}^{\bar \mu \bar \nu \mu}
  \pm\,\frac{1}{\sqrt s}\,Q_{\mp,4}^{\bar \mu \bar \nu \mu}
  \nonumber\\ && \qquad
  +\, {\textstyle{1\over 4}}\,(\bar\delta + 1)\,(\bar\delta - 1)\,\frac{s}{v^2}\,(r\cdot r)\,Q_{\pm,1}^{\bar \mu \bar \nu \mu}
  \nonumber\\ && \qquad
  \mp\,{\textstyle{1\over 2}}\,(\bar\delta - 1)\,\frac{s}{v^2}\,E_\mp\,\Big[ (\bar r\cdot r)\,Q_{\pm,4}^{\bar \mu \bar \nu \mu}
  + \bar E_\mp\,E_\pm\,Q_{\mp,4}^{\bar \mu \bar \nu \mu}
  \Big]
  \nonumber\\ && \qquad
  \pm\,{\textstyle{1\over 2}}\,(\bar\delta + 1)\,\frac{s}{v^2}\,E_\mp\,\Big[
  \big[ (\bar r\cdot r)\,Q_{\mp,7}^{\bar \mu \bar \nu \mu}
  + \bar E_\mp\,E_\pm\,Q_{\pm,7}^{\bar \mu \bar \nu \mu}\big]
  \nonumber\\ && \qquad \quad
  \pm\,\frac{1}{\sqrt s}\,\big[ (\bar r\cdot r)\,Q_{\pm,2}^{\bar \mu \bar \nu \mu}
  \pm\,\bar M\,E_\pm\,Q_{\mp,2}^{\bar \mu \bar \nu \mu}\big]
  \Big]
  \,,\nonumber\\ &&
  Q_{\pm,11}^{\bar \mu \bar \nu \mu}
  =
  \pm\,\sqrt{s}\,P_{\pm,13}^{\bar \mu \bar \nu \mu}
  \nonumber\\ && \qquad
  -\, {\textstyle{1\over 2}}\,(\delta + 1)\,\frac{s}{v^2}\,\Big[
  (\bar r\cdot\bar r)\,(Q_{\mp,1}^{\bar \mu \bar \nu \mu}
  + Q_{\pm,2}^{\bar \mu \bar \nu \mu})
  \nonumber\\ && \qquad \qquad \qquad
  \mp\,\bar E_\pm\,\big[ (\bar r\cdot r)\,Q_{\mp,8}^{\bar \mu \bar \nu \mu}
  + \bar E_\mp\,E_\pm\,Q_{\pm,8}^{\bar \mu \bar \nu \mu}\big]
  \Big]
  \nonumber\\ && \qquad
  \mp\,\frac{s}{v^2}\,\bar E_\pm\,\Big[ (\bar r\cdot r)\,Q_{\pm,3}^{\bar \mu \bar \nu \mu}
  + \bar E_\mp\,E_\pm\,Q_{\mp,3}^{\bar \mu \bar \nu \mu}\Big]
  \,,\nonumber\\ &&
  Q_{\pm,12}^{\bar \mu \bar \nu \mu}
  =
  \mp\,\frac{s}{v^2}\,\sqrt{s}\,(\bar r\cdot r)\,\Big[ (\bar r\cdot r)\,P_{\mp,3}^{\bar \mu \bar \nu \mu}
  - \bar E_\mp\,E_\mp\,P_{\pm,3}^{\bar \mu \bar \nu \mu}\Big]
  \nonumber\\ && \qquad
  \pm\,\frac{s}{v^2}\,\bar E_\mp\,\Big[ (\bar r\cdot r)\,P_{\pm,9}^{\bar \mu \bar \nu \mu}
  - \bar E_\pm\,E_\mp\,P_{\mp,9}^{\bar \mu \bar \nu \mu}\Big]
  \nonumber\\ && \qquad
  +\, {\textstyle{1\over 2}}\,(\bar\delta + 1)\,\frac{s}{v^2}\,\Big[
  (\bar r\cdot r)\,Q_{\pm,1}^{\bar \mu \bar \nu \mu}
  \pm\,(\bar r\cdot r)\,\bar E_\mp\,Q_{\mp,5}^{\bar \mu \bar \nu \mu}
  \nonumber\\ && \qquad \qquad \qquad
  \pm\,E_\mp\,\big[ (\bar r\cdot r)\,Q_{\mp,8}^{\bar \mu \bar \nu \mu}
  + \bar E_\mp\,E_\pm\,Q_{\pm,8}^{\bar \mu \bar \nu \mu}\big]
  \Big]
  \,,\nonumber\\ &&
  Q_{\pm,13}^{\bar \mu \bar \nu \mu}
  =
  \pm\,\sqrt{s}\,P_{\pm,16}^{\bar \mu \bar \nu \mu}
  \mp\,\frac{1}{\sqrt s}\,Q_{\pm,3}^{\bar \mu \bar \nu \mu}
  \nonumber\\ && \qquad
  -\, {\textstyle{1\over 4}}\,(\bar\delta - 1)\,(\delta + 1)\,\frac{s}{v^2}\,(\bar r\cdot r)\,(Q_{\mp,1}^{\bar \mu \bar \nu \mu}
  + Q_{\pm,2}^{\bar \mu \bar \nu \mu})
  \nonumber\\ && \qquad
  \pm\,{\textstyle{1\over 2}}\,(\bar\delta - 1)\,\frac{s}{v^2}\,E_\pm\,\Big[ (\bar r\cdot r)\,Q_{\mp,3}^{\bar \mu \bar \nu \mu}
  + \bar E_\pm\,E_\mp\,Q_{\pm,3}^{\bar \mu \bar \nu \mu}
  \Big]
  \nonumber\\ && \qquad
  \pm\,{\textstyle{1\over 2}}\,(\delta + 1)\,\frac{s}{v^2}\,\Big[
  \bar E_\pm\,\big[ (\bar r\cdot r)\,Q_{\mp,7}^{\bar \mu \bar \nu \mu}
  + \bar E_\mp\,E_\pm\,Q_{\pm,7}^{\bar \mu \bar \nu \mu}\big]
  \nonumber\\ && \qquad
  -\, \frac{\bar M}{\sqrt s}\,\big[ (\bar r\cdot r)\,Q_{\mp,1}^{\bar \mu \bar \nu \mu}
  + \bar E_\pm\,E_\pm\,Q_{\pm,1}^{\bar \mu \bar \nu \mu}\big]
  \Big]
  \,,\nonumber\\ &&
  Q_{\pm,14}^{\bar \mu \bar \nu \mu}
  =
  \pm\,\frac{1}{E_\pm}\,P_{\mp,8}^{\bar \mu \bar \nu \mu}
  \mp\,\frac{1}{E_\pm}\,Q_{\pm,6}^{\bar \mu \bar \nu \mu}
  \nonumber\\ && \qquad
  \mp\,\frac{s}{v^2}\,\bar E_\mp\,\Big[
  (\bar r\cdot r)\,(Q_{\mp,6}^{\bar \mu \bar \nu \mu} + Q_{\pm,8}^{\bar \mu \bar \nu \mu})
  \nonumber\\ && \qquad \qquad \qquad
  +\, \bar E_\pm\,E_\mp\,(Q_{\pm,6}^{\bar \mu \bar \nu \mu} + Q_{\mp,8}^{\bar \mu \bar \nu \mu})
  \Big]
  \nonumber\\ && \qquad
  -\, \frac{s}{v^2}\,\frac{\bar E_\mp}{E_\pm}\,\Big[ (\bar r\cdot r)\,Q_{\pm,2}^{\bar \mu \bar \nu \mu}
  + \bar E_\pm\,E_\pm\,Q_{\pm,1}^{\bar \mu \bar \nu \mu}\Big]
  \nonumber\\ && \qquad
  \pm\,\frac{s}{v^2}\,\frac{(\bar r\cdot\bar r)}{E_\pm}\,\Big[ (\bar r\cdot r)\,Q_{\pm,5}^{\bar \mu \bar \nu \mu}
  + \bar E_\mp\,E_\pm\,Q_{\mp,5}^{\bar \mu \bar \nu \mu}\Big]
  \,,\nonumber\\ &&
  Q_{\pm,15}^{\bar \mu \bar \nu \mu}
  =
  \frac{s}{v^2}\,((\bar r\cdot r)\,P_{\mp,18}^{\bar \mu \bar \nu \mu} - \bar E_\mp\,E_\pm\,P_{\pm,18}^{\bar \mu \bar \nu \mu})
  \nonumber\\ && \qquad
  \mp\,{\textstyle{1\over 8}}\,(\bar\delta + 1)\,(\bar\delta - 1)\,(\delta + 1)\,\left(\frac{s}{v^2}\right)^2\,\Big[
  \nonumber\\ && \qquad \qquad
  (r\cdot r)\,\bar E_\mp\,\big[ (\bar r\cdot r)\,Q_{\mp,1}^{\bar \mu \bar \nu \mu}
  + \bar E_\pm\,E_\pm\,Q_{\pm,1}^{\bar \mu \bar \nu \mu}\big]
  \nonumber\\ && \qquad \qquad
  - \,(\bar r\cdot r)\,E_\pm\,\big[ (\bar r\cdot r)\,Q_{\mp,2}^{\bar \mu \bar \nu \mu}
  + \bar E_\mp\,E_\mp\,Q_{\pm,2}^{\bar \mu \bar \nu \mu}\big]
  \Big]
  \nonumber\\ && \qquad
  + \,{\textstyle{1\over 4}}\,(\bar\delta + 1)\,(\bar\delta - 1)\,\frac{s}{v^2}\,(r\cdot r)\,Q_{\pm,3}^{\bar \mu \bar \nu \mu}
  \nonumber\\ && \qquad
  - \,{\textstyle{1\over 4}}\,(\bar\delta - 1)\,(\delta + 1)\,\frac{s}{v^2}\,(\bar r\cdot r)\,Q_{\pm,4}^{\bar \mu \bar \nu \mu}
  \nonumber\\ && \qquad
  - \,{\textstyle{1\over 4}}\,(\bar\delta + 1)\,(\delta + 1)\,\frac{s}{v^2}\,\Big[
  2\,(\bar r\cdot r)\,Q_{\mp,7}^{\bar \mu \bar \nu \mu}
  - \bar E_\mp\,E_\pm\,Q_{\pm,7}^{\bar \mu \bar \nu \mu}
  \Big]
  \nonumber\\ && \qquad
  \pm\,{\textstyle{1\over 2}}\,(\bar\delta + 1)\,\frac{s}{v^2}\,E_\pm\,\Big[ (\bar r\cdot r)\,Q_{\mp,13}^{\bar \mu \bar \nu \mu}
  + \bar E_\mp\,E_\mp\,Q_{\pm,13}^{\bar \mu \bar \nu \mu}
  \Big]
  \nonumber\\ && \qquad
  \pm\,{\textstyle{1\over 2}}\,(\delta + 1)\,\frac{s}{v^2}\,\bar E_\mp\,\Big[ (\bar r\cdot r)\,Q_{\mp,10}^{\bar \mu \bar \nu \mu}
  + \bar E_\pm\,E_\pm\,Q_{\pm,10}^{\bar \mu \bar \nu \mu}
  \Big]
  \,,\nonumber\\ &&
  Q_{\pm,16}^{\bar \mu \bar \nu \mu}
  =
  \frac{s}{v^2}\,\Big[ (\bar r\cdot r)\,P_{\mp,12}^{\bar \mu \bar \nu \mu} - \bar E_\mp\,E_\pm\,P_{\pm,12}^{\bar \mu \bar \nu \mu}\Big]
  \nonumber\\ && \qquad
  +\, {\textstyle{1\over 4}}\,(\bar\delta + 1)\,(\bar\delta - 1)\,\frac{s}{v^2}\,(r\cdot r)\,\Big[
  Q_{\pm,5}^{\bar \mu \bar \nu \mu}
  \nonumber\\ && \qquad \qquad
  \pm\,\frac{s}{v^2}\,E_\pm\,\big[ (\bar r\cdot r)\,(Q_{\pm,1}^{\bar \mu \bar \nu \mu}
  + Q_{\mp,2}^{\bar \mu \bar \nu \mu})
  \nonumber\\ && \qquad \qquad \qquad
  +\, \bar E_\mp\,E_\mp\,(Q_{\mp,1}^{\bar \mu \bar \nu \mu}
  + Q_{\pm,2}^{\bar \mu \bar \nu \mu})
  \big]
  \Big]
  \nonumber\\ && \qquad
  -\, {\textstyle{1\over 2}}\,(\bar\delta - 1)\,\frac{s}{v^2}\,(r\cdot r)\,Q_{\pm,4}^{\bar \mu \bar \nu \mu}
  - {\textstyle{1\over 2}}\,(\bar\delta + 1)\,\frac{s}{v^2}\,\Big[
  (r\cdot r)\,Q_{\mp,7}^{\bar \mu \bar \nu \mu}
  \nonumber\\ && \qquad \qquad \qquad
  \mp\,E_\pm\,\big[ (\bar r\cdot r)\,Q_{\mp,9}^{\bar \mu \bar \nu \mu}
  + \bar E_\mp\,E_\mp\,Q_{\pm,9}^{\bar \mu \bar \nu \mu}\big]
  \Big]
  \nonumber\\ && \qquad
  \mp\,\frac{s}{v^2}\,E_\pm\,\Big[ (\bar r\cdot r)\,Q_{\pm,10}^{\bar \mu \bar \nu \mu}
  + \bar E_\mp\,E_\mp\,Q_{\mp,10}^{\bar \mu \bar \nu \mu}\Big]
  \,,\nonumber\\ &&
  Q_{\pm,17}^{\bar \mu \bar \nu \mu}
  =
  \frac{s}{v^2}\,\Big[ (\bar r\cdot r)\,P_{\mp,15}^{\bar \mu \bar \nu \mu} - \bar E_\mp\,E_\pm\,P_{\pm,15}^{\bar \mu \bar \nu \mu}\Big]
  \nonumber\\ && \qquad
  -\, {\textstyle{1\over 4}}\,(\bar\delta + 1)\,(\delta + 1)\,\frac{s}{v^2}\,\Big[
  \nonumber\\ && \qquad \qquad
  \big[ 2\,(\bar r\cdot r)\,Q_{\mp,8}^{\bar \mu \bar \nu \mu}
  - \bar E_\mp\,E_\pm\,Q_{\pm,8}^{\bar \mu \bar \nu \mu}\big]
  \nonumber\\ && \qquad \qquad
  \pm\,\frac{s}{v^2}\,\bar E_\mp\,\big[ (\bar r\cdot r)\,((\bar r\cdot r)\,Q_{\mp,1}^{\bar \mu \bar \nu \mu}
  + \bar E_\pm\,E_\pm\,Q_{\pm,1}^{\bar \mu \bar \nu \mu})
  \nonumber\\ && \qquad \qquad \quad
  +\, \bar E_\pm\,E_\pm\,((\bar r\cdot r)\,Q_{\mp,2}^{\bar \mu \bar \nu \mu}
  + \bar E_\mp\,E_\mp\,Q_{\pm,2}^{\bar \mu \bar \nu \mu})
  \big]
  \Big]
  \nonumber\\ && \qquad
  +\, {\textstyle{1\over 2}}\,(\bar\delta + 1)\,\frac{s}{v^2}\,\Big[
  (\bar r\cdot r)\,Q_{\pm,3}^{\bar \mu \bar \nu \mu}
  \nonumber\\ && \qquad \qquad \qquad
  \pm\,E_\pm\,\big[ (\bar r\cdot r)\,Q_{\mp,11}^{\bar \mu \bar \nu \mu}
  + \bar E_\mp\,E_\mp\,Q_{\pm,11}^{\bar \mu \bar \nu \mu}\big]
  \Big]
  \nonumber\\ && \qquad
  -\, {\textstyle{1\over 2}}\,(\delta + 1)\,\frac{s}{v^2}\,\Big[
  (\bar r\cdot\bar r)\,Q_{\pm,4}^{\bar \mu \bar \nu \mu}
  \nonumber\\ && \qquad \qquad \qquad
  \mp\,\bar E_\mp\,\big[ (\bar r\cdot r)\,Q_{\mp,12}^{\bar \mu \bar \nu \mu}
  + \bar E_\pm\,E_\pm\,Q_{\pm,12}^{\bar \mu \bar \nu \mu}\big]
  \Big]
  \,,\nonumber\\ &&
  Q_{\pm,18}^{\bar \mu \bar \nu \mu}
  =
  \frac{s}{v^2}\,\Big[ (\bar r\cdot r)\,P_{\mp,14}^{\bar \mu \bar \nu \mu} - \bar E_\mp\,E_\pm\,P_{\pm,14}^{\bar \mu \bar \nu \mu}\Big]
  \nonumber\\ && \qquad
  -\, {\textstyle{1\over 2}}\,(\delta + 1)\,\frac{s}{v^2}\,\Big[
  (\bar r\cdot\bar r)\,(Q_{\mp,8}^{\bar \mu \bar \nu \mu}
  + Q_{\pm,6}^{\bar \mu \bar \nu \mu})
  \nonumber\\ && \qquad \qquad
  \mp\,\bar E_\mp\,\big[ (\bar r\cdot r)\,Q_{\mp,14}^{\bar \mu \bar \nu \mu}
  + \bar E_\pm\,E_\pm\,Q_{\pm,14}^{\bar \mu \bar \nu \mu}\big]
  \nonumber\\ && \qquad \qquad
  \pm\,\frac{s}{v^2}\,(\bar r\cdot\bar r)\,\bar E_\mp\,\big[ (\bar r\cdot r)\,(Q_{\mp,1}^{\bar \mu \bar \nu \mu}
  + Q_{\pm,2}^{\bar \mu \bar \nu \mu})
  \nonumber\\ && \qquad \qquad \qquad
  +\, \bar E_\pm\,E_\pm\,(Q_{\pm,1}^{\bar \mu \bar \nu \mu}
  + Q_{\mp,2}^{\bar \mu \bar \nu \mu})
  \big]
  \Big]
  + \frac{s}{v^2}\,(\bar r\cdot\bar r)\,Q_{\pm,3}^{\bar \mu \bar \nu \mu}
  \nonumber\\ && \qquad
  \mp\,\frac{s}{v^2}\,\bar E_\mp\,\Big[ (\bar r\cdot r)\,Q_{\pm,11}^{\bar \mu \bar \nu \mu}
  + \bar E_\pm\,E_\pm\,Q_{\mp,11}^{\bar \mu \bar \nu \mu}\Big]
  \,,\nonumber
\end{eqnarray}
where
\begin{eqnarray}
  && \bar q_{\bar \mu}\, P^{\bar \mu \bar \nu \mu}_{\pm, k} =0\,, \quad
  q_{\mu}\, P^{\bar \mu \bar \nu \mu}_{\pm, k} =0\, \quad
  \bar p_{ \bar\nu}\, P^{\bar \mu \bar \nu \mu}_{\pm, k} =0=
  \Lambda\,P^{\bar \mu \bar \nu \mu}_{\pm, k}\,\bar\Lambda\,\gamma_{\bar \nu}\,,
  \nonumber\\ \nonumber\\
  &&
  P_{\pm, 1}^{\bar \mu \bar \nu \mu}
  = \big[
  \rbot^{\bar \nu}\,i\,\gamma_5\,P_\pm
  - v^{\bar \nu}\,(\sqrt{s}/v^2)\,\bar E_\pm\,P_\mp
  \big]\,v^{\bar \mu}\,v^{\mu}/v^2/v^2
  \,,\nonumber\\ &&
  P_{\pm, 2}^{\bar \mu \bar \nu \mu}
  = \big[
  \rbot^{\bar \nu}\,P_\pm
  + v^{\bar \nu}\,(\sqrt{s}/v^2)\,\bar E_\pm\,i\,\gamma_5\,P_\mp
  \big]\,\rbot^{\bar \mu}\,v^{\mu}/v^2
  \,,\nonumber\\ &&
  P_{\pm, 3}^{\bar \mu \bar \nu \mu}
  = \big[
  \rbot^{\bar \nu}\,P_\pm
  + v^{\bar \nu}\,(\sqrt{s}/v^2)\,\bar E_\pm\,i\,\gamma_5\,P_\mp
  \big]\,\wLbot^{\bar \mu}\,v^{\mu}/v^2
  \,,\nonumber\\ &&
  P_{\pm, 4}^{\bar \mu \bar \nu \mu}
  = \big[
  -\, v^{\bar \nu}\,((\bar r\cdot r)\,P_\mp \pm\,\bar M\,E_\mp\,P_\pm)/v^2
  \nonumber\\ && \qquad \qquad
  +\, \wRbot^{\bar \nu}\,i\,\gamma_5\,P_\pm
  \big]\,v^{\bar \mu}\,v^{\mu}/v^2/v^2
  \,,\nonumber\\ &&
  P_{\pm, 5}^{\bar \mu \bar \nu \mu}
  = \big[
  v^{\bar \nu}\,i\,\gamma_5\,((\bar r\cdot r)\,P_\mp \pm\,\bar M\,E_\pm\,P_\pm)/v^2
  \nonumber\\ && \qquad \qquad
  +\, \wRbot^{\bar \nu}\,P_\pm
  \big]\,\rbot^{\bar \mu}\,v^{\mu}/v^2
  \,,\nonumber\\ &&
  P_{\pm, 6}^{\bar \mu \bar \nu \mu}
  = \big[
  v^{\bar \nu}\,i\,\gamma_5\,((\bar r\cdot r)\,P_\mp \pm\,\bar M\,E_\pm\,P_\pm)/v^2
  \nonumber\\ && \qquad \qquad
  +\, \wRbot^{\bar \nu}\,P_\pm
  \big]\,\wLbot^{\bar \mu}\,v^{\mu}/v^2
  \,,\nonumber\\ &&
  P_{\pm, 7}^{\bar \mu \bar \nu \mu}
  = \big[
  \rbot^{\bar \nu}\,P_\pm
  + v^{\bar \nu}\,(\sqrt{s}/v^2)\,\bar E_\pm\,i\,\gamma_5\,P_\mp
  \big]\,v^{\bar \mu}\,\rbarbot^{\mu}/v^2
  \,,\nonumber\\ &&
  P_{\pm, 8}^{\bar \mu \bar \nu \mu}
  = \big[
  \rbot^{\bar \nu}\,i\,\gamma_5\,P_\pm
  - v^{\bar \nu}\,(\sqrt{s}/v^2)\,\bar E_\pm\,P_\mp
  \big]\,\rbot^{\bar \mu}\,\rbarbot^{\mu}
  \,,\nonumber\\ &&
  P_{\pm, 9}^{\bar \mu \bar \nu \mu}
  = \big[
  \rbot^{\bar \nu}\,i\,\gamma_5\,P_\pm
  - v^{\bar \nu}\,(\sqrt{s}/v^2)\,\bar E_\pm\,P_\mp
  \big]\,\wLbot^{\bar \mu}\,\rbarbot^{\mu}
  \,,\nonumber\\ &&
  P_{\pm,10}^{\bar \mu \bar \nu \mu}
  = \big[
  v^{\bar \nu}\,i\,\gamma_5\,((\bar r\cdot r)\,P_\mp \pm\,\bar M\,E_\pm\,P_\pm)/v^2
  \nonumber\\ && \qquad \qquad
  +\, \wRbot^{\bar \nu}\,P_\pm
  \big]\,v^{\bar \mu}\,\rbarbot^{\mu}/v^2
  \,,\nonumber\\ &&
  P_{\pm,11}^{\bar \mu \bar \nu \mu}
  = \big[
  -\, v^{\bar \nu}\,((\bar r\cdot r)\,P_\mp \pm\,\bar M\,E_\mp\,P_\pm)/v^2
  \nonumber\\ && \qquad \qquad
  +\, \wRbot^{\bar \nu}\,i\,\gamma_5\,P_\pm
  \big]\,\rbot^{\bar \mu}\,\rbarbot^{\mu}
  \,,\nonumber\\ &&
  P_{\pm,12}^{\bar \mu \bar \nu \mu}
  = \big[
  -\, v^{\bar \nu}\,((\bar r\cdot r)\,P_\mp \pm\,\bar M\,E_\mp\,P_\pm)/v^2
  \nonumber\\ && \qquad \qquad
  +\, \wRbot^{\bar \nu}\,i\,\gamma_5\,P_\pm
  \big]\,\wLbot^{\bar \mu}\,\rbarbot^{\mu}
  \,,\nonumber\\ &&
  P_{\pm,13}^{\bar \mu \bar \nu \mu}
  = \big[
  \rbot^{\bar \nu}\,P_\pm
  + v^{\bar \nu}\,(\sqrt{s}/v^2)\,\bar E_\pm\,i\,\gamma_5\,P_\mp
  \big]\,v^{\bar \mu}\,\wLbarbot^{\mu}/v^2
  \,,\nonumber\\ &&
  P_{\pm,14}^{\bar \mu \bar \nu \mu}
  = \big[
  \rbot^{\bar \nu}\,i\,\gamma_5\,P_\pm
  - v^{\bar \nu}\,(\sqrt{s}/v^2)\,\bar E_\pm\,P_\mp
  \big]\,\rbot^{\bar \mu}\,\wLbarbot^{\mu}
  \,,\nonumber\\ &&
  P_{\pm,15}^{\bar \mu \bar \nu \mu}
  = \big[
  \rbot^{\bar \nu}\,i\,\gamma_5\,P_\pm
  - v^{\bar \nu}\,(\sqrt{s}/v^2)\,\bar E_\pm\,P_\mp
  \big]\,\wLbot^{\bar \mu}\,\wLbarbot^{\mu}
  \,,\nonumber\\ &&
  P_{\pm,16}^{\bar \mu \bar \nu \mu}
  = \big[
  v^{\bar \nu}\,i\,\gamma_5\,((\bar r\cdot r)\,P_\mp \pm\,\bar M\,E_\pm\,P_\pm)/v^2
  \nonumber\\ && \qquad \qquad
  +\, \wRbot^{\bar \nu}\,P_\pm
  \big]\,v^{\bar \mu}\,\wLbarbot^{\mu}/v^2
  \,,\nonumber\\ &&
  P_{\pm,17}^{\bar \mu \bar \nu \mu}
  = \big[
  -\, v^{\bar \nu}\,((\bar r\cdot r)\,P_\mp \pm\,\bar M\,E_\mp\,P_\pm)/v^2
  \nonumber\\ && \qquad \qquad
  +\, \wRbot^{\bar \nu}\,i\,\gamma_5\,P_\pm
  \big]\,\rbot^{\bar \mu}\,\wLbarbot^{\mu}
  \,,\nonumber\\ &&
  P_{\pm,18}^{\bar \mu \bar \nu \mu}
  = \big[
  -\, v^{\bar \nu}\,((\bar r\cdot r)\,P_\mp \pm\,\bar M\,E_\mp\,P_\pm)/v^2
  \nonumber\\ && \qquad \qquad
  +\, \wRbot^{\bar \nu}\,i\,\gamma_5\,P_\pm
  \big]\,\wLbot^{\bar \mu}\,\wLbarbot^{\mu}
  \,.\nonumber
\end{eqnarray}
For the (\ref{def-03/2to13/2}) case
\begin{eqnarray}
  &&
  Q_{\pm, 1}^{\bar \mu \bar \nu \nu}
  =
  -\, (\bar\delta - 1)\,\frac{s}{v^2}\,\frac{\sqrt s}{\bar M}\,(r\cdot r)\,\bar E_\pm\,\Big[
  \nonumber\\ && \qquad \qquad
  \big[ (\bar r\cdot r)\,P_{\mp,5}^{\bar \mu \bar \nu \nu}
  - \bar E_\mp\,E_\pm\,P_{\pm,5}^{\bar \mu \bar \nu \nu}\big]
  \nonumber\\ && \qquad \qquad
  -\, \sqrt{s}\,\bar E_\mp\,\big[ (\bar r\cdot r)\,P_{\pm,1}^{\bar \mu \bar \nu \nu}
  - \bar E_\pm\,E_\pm\,P_{\mp,1}^{\bar \mu \bar \nu \nu}\big]
  \Big]
  \nonumber\\ && \qquad
  -\, 2\,\frac{s}{v^2}\,\frac{\sqrt s}{\bar M}\,(\bar r\cdot\bar r)\,E_\pm\,\Big[
  \big[ (\bar r\cdot r)\,P_{\pm,7}^{\bar \mu \bar \nu \nu}
  - \bar E_\pm\,E_\mp\,P_{\mp,7}^{\bar \mu \bar \nu \nu}\big]
  \nonumber\\ && \qquad \qquad
  -\, \sqrt{s}\,\bar E_\pm\,\big[ (\bar r\cdot r)\,P_{\mp,3}^{\bar \mu \bar \nu \nu}
  - \bar E_\mp\,E_\mp\,P_{\pm,3}^{\bar \mu \bar \nu \nu}\big]
  \Big]
  \nonumber\\ && \qquad
  +\, {\textstyle{1\over 2}}\,(\bar\delta - 1)\,\frac{s}{\bar M}\,\bar E_\pm\,E_\pm\,P_{\mp,1}^{\bar \mu \bar \nu \nu}
  \pm\,\sqrt{s}\,(\bar r\cdot r)\,P_{\pm,1}^{\bar \mu \bar \nu \nu}
  \nonumber\\ && \qquad
  +\, \frac{1}{\bar M}\,\bar E_\pm\,\Big[
  s\,\bar E_\mp\,P_{\pm,3}^{\bar \mu \bar \nu \nu}
  - \sqrt{s}\,P_{\mp,7}^{\bar \mu \bar \nu \nu}
  + E_\pm\,P_{\pm,5}^{\bar \mu \bar \nu \nu}
  \Big]
  \,,\nonumber\\ &&
  Q_{\pm, 2}^{\bar \mu \bar \nu \nu}
  =
  -\, 2\,\frac{s}{v^2}\,\bar E_\mp\,E_\pm\,\Big[ (\bar r\cdot r)\,P_{\pm,9}^{\bar \mu \bar \nu \nu}
  - \bar E_\pm\,E_\mp\,P_{\mp,9}^{\bar \mu \bar \nu \nu}\Big]
  + P_{\mp,9}^{\bar \mu \bar \nu \nu}
  \nonumber\\ && \qquad
  \pm\,{\textstyle{1\over 2}}\,(\bar\delta + 1)\,\frac{s}{v^2}\,E_\pm\,\Big[
  \big[ (\bar r\cdot r)\,Q_{\mp,1}^{\bar \mu \bar \nu \nu}
  + \bar E_\mp\,E_\mp\,Q_{\pm,1}^{\bar \mu \bar \nu \nu}\big]
  \nonumber\\ && \qquad \qquad
  \mp\,\bar E_\mp\,\big[ (\bar r\cdot r)\,Q_{\mp,5}^{\bar \mu \bar \nu \nu}
  - \bar E_\pm\,E_\mp\,Q_{\pm,5}^{\bar \mu \bar \nu \nu}\big]
  \Big]
  \nonumber\\ && \qquad
  +\, {\textstyle{1\over 2}}\,(\bar\delta + 1)\,Q_{\pm,5}^{\bar \mu \bar \nu \nu}
  \,,\nonumber\\ &&
  Q_{\pm, 3}^{\bar \mu \bar \nu \nu}
  =
  -\, 2\,\frac{s}{v^2}\,\bar E_\mp\,E_\pm\,\Big[ (\bar r\cdot r)\,P_{\pm,5}^{\bar \mu \bar \nu \nu}
  - \bar E_\pm\,E_\mp\,P_{\mp,5}^{\bar \mu \bar \nu \nu}\Big]
  + P_{\mp,5}^{\bar \mu \bar \nu \nu}
  \nonumber\\ && \qquad
  \mp\,\frac{s}{v^2}\,\bar E_\mp\,\Big[
  \big[ (\bar r\cdot r)\,Q_{\pm,1}^{\bar \mu \bar \nu \nu}
  + \bar E_\pm\,E_\pm\,Q_{\mp,1}^{\bar \mu \bar \nu \nu}\big]
  \nonumber\\ && \qquad \qquad
  \pm\,\bar E_\pm\,\big[ (\bar r\cdot r)\,Q_{\pm,5}^{\bar \mu \bar \nu \nu}
  - \bar E_\mp\,E_\pm\,Q_{\mp,5}^{\bar \mu \bar \nu \nu}\big]
  \Big]
  \,,\nonumber\\ &&
  Q_{\pm, 4}^{\bar \mu \bar \nu \nu}
  =
  \pm\,\frac{\sqrt s}{\bar M}\,\Big[ P_{\mp,8}^{\bar \mu \bar \nu \nu}
  - \sqrt{s}\,\bar E_\mp\,P_{\pm,4}^{\bar \mu \bar \nu \nu}\Big]
  \nonumber\\ && \qquad
  \mp\,{\textstyle{1\over 2}}\,(\bar\delta - 1)\,(\delta - 1)\,\frac{\sqrt s}{\bar M}\,\Big[
  Q_{\pm,3}^{\bar \mu \bar \nu \nu}
  \nonumber\\ && \qquad \qquad
  +\, \frac{s}{v^2}\,\bar E_\mp\,E_\pm\,\big[ (\bar r\cdot r)\,Q_{\mp,3}^{\bar \mu \bar \nu \nu}
  + \bar E_\pm\,E_\mp\,Q_{\pm,3}^{\bar \mu \bar \nu \nu}\big]
  \Big]
  \nonumber\\ && \qquad
  \mp\,(\delta - 1)\,\frac{s}{v^2}\,\bar E_\mp\,\Big[ (\bar r\cdot r)\,Q_{\pm,1}^{\bar \mu \bar \nu \nu}
  + \bar E_\pm\,E_\pm\,Q_{\mp,1}^{\bar \mu \bar \nu \nu}
  \Big]
  \nonumber\\ && \qquad
  \mp\,\frac{(\bar r\cdot r)}{\bar M\,w}\,Q_{\mp,3}^{\bar \mu \bar \nu \nu}
  \pm\,\frac{M}{\sqrt s}\,Q_{\pm,3}^{\bar \mu \bar \nu \nu}
  \pm\,\frac{2}{\sqrt s}\,Q_{\mp,1}^{\bar \mu \bar \nu \nu}
  \,,\nonumber\\ &&
  Q_{\pm, 5}^{\bar \mu \bar \nu \nu}
  =
  \frac{\sqrt s}{\bar E_\pm}\,\Big[ (\bar r\cdot r)\,P_{\pm,1}^{\bar \mu \bar \nu \nu}
  - \bar E_\pm\,E_\pm\,P_{\mp,1}^{\bar \mu \bar \nu \nu}\Big]
  \mp\,\frac{1}{\bar E_\pm}\,Q_{\pm,1}^{\bar \mu \bar \nu \nu}
  \,,\nonumber\\ &&
  Q_{\pm, 6}^{\bar \mu \bar \nu \nu}
  =
  \pm\,\sqrt{s}\,P_{\pm,4}^{\bar \mu \bar \nu \nu}
  + \frac{1}{s}\,Q_{\mp,1}^{\bar \mu \bar \nu \nu}
  \mp\,\frac{1}{\sqrt s}\,Q_{\pm,4}^{\bar \mu \bar \nu \nu}
  \nonumber\\ && \qquad
  \pm\,{\textstyle{1\over 2}}\,(\bar\delta + 1)\,\frac{\sqrt s}{v^2}\,M\,\Big[ (\bar r\cdot r)\,Q_{\pm,1}^{\bar \mu \bar \nu \nu}
  + \bar E_\pm\,E_\pm\,Q_{\mp,1}^{\bar \mu \bar \nu \nu}\Big]
  \nonumber\\ && \qquad
  \pm\,{\textstyle{1\over 2}}\,(\bar\delta - 1)\,\frac{s}{v^2}\,E_\pm\,\Big[ (\bar r\cdot r)\,Q_{\mp,4}^{\bar \mu \bar \nu \nu}
  + \bar E_\pm\,E_\mp\,Q_{\pm,4}^{\bar \mu \bar \nu \nu}\Big]
  \nonumber\\ && \qquad
  +\, {\textstyle{1\over 2}}\,(\delta - 1)\,\frac{\sqrt s}{v^2}\,\bar E_\pm\,\Big[ (\bar r\cdot r)\,Q_{\pm,1}^{\bar \mu \bar \nu \nu}
  - \bar E_\mp\,E_\pm\,Q_{\mp,1}^{\bar \mu \bar \nu \nu}\Big]
  \nonumber\\ && \qquad
  \mp\,M\,\frac{\sqrt s}{v^2}\,\Big[ (\bar r\cdot r)\,Q_{\pm,1}^{\bar \mu \bar \nu \nu}
  + \bar E_\pm\,E_\pm\,Q_{\mp,1}^{\bar \mu \bar \nu \nu}\Big]
  \,,\nonumber\\ &&
  Q_{\pm, 7}^{\bar \mu \bar \nu \nu}
  =
  \pm\,\sqrt{s}\,P_{\pm,3}^{\bar \mu \bar \nu \nu}
  \mp\,\frac{1}{\sqrt s}\,Q_{\pm,5}^{\bar \mu \bar \nu \nu}
  \nonumber\\ && \qquad
  \pm\,{\textstyle{1\over 2}}\,(\bar\delta - 1)\,\frac{s}{v^2}\,E_\pm\,\Big[ (\bar r\cdot r)\,Q_{\mp,5}^{\bar \mu \bar \nu \nu}
  + \bar E_\pm\,E_\mp\,Q_{\pm,5}^{\bar \mu \bar \nu \nu}\Big]
  \nonumber\\ && \qquad
  +\, \frac{\sqrt s}{v^2}\,E_\pm\,\Big[ (\bar r\cdot r)\,Q_{\mp,1}^{\bar \mu \bar \nu \nu}
  - \bar E_\pm\,E_\mp\,Q_{\pm,1}^{\bar \mu \bar \nu \nu}\Big]
  \,,\nonumber\\ &&
  Q_{\pm, 8}^{\bar \mu \bar \nu \nu}
  =
  \pm\,\sqrt{s}\,P_{\pm,2}^{\bar \mu \bar \nu \nu}
  \nonumber\\ && \qquad
  \mp\,\frac{s}{v^2}\,\bar E_\pm\,\Big[ (\bar r\cdot r)\,Q_{\pm,4}^{\bar \mu \bar \nu \nu}
  + \bar E_\mp\,E_\pm\,Q_{\mp,4}^{\bar \mu \bar \nu \nu}\Big]
  \nonumber\\ && \qquad
  +\, \frac{\sqrt s}{v^2}\,\bar E_\pm\,\Big[ (\bar r\cdot r)\,Q_{\mp,1}^{\bar \mu \bar \nu \nu}
  - \bar E_\mp\,E_\pm\,Q_{\pm,1}^{\bar \mu \bar \nu \nu}\Big]
  \,,\nonumber\\ &&
  Q_{\pm, 9}^{\bar \mu \bar \nu \nu}
  =
  \frac{s}{v^2}\,\Big[ (\bar r\cdot r)\,P_{\mp,12}^{\bar \mu \bar \nu \nu}
  - \bar E_\mp\,E_\pm\,P_{\pm,12}^{\bar \mu \bar \nu \nu}\Big]
  - \frac{1}{s}\,Q_{\mp,2}^{\bar \mu \bar \nu \nu}
  \nonumber\\ && \qquad
  +\, {\textstyle{1\over 4}}\,(\bar\delta - 1)\,(\bar\delta + 1)\,\frac{s}{v^2}\,(r\cdot r)\,Q_{\pm,4}^{\bar \mu \bar \nu \nu}
  \nonumber\\ && \qquad
  +\, {\textstyle{1\over 2}}\,(\delta - 1)\,\frac{\sqrt s}{v^2}\,\bar E_\pm\,\Big[ (\bar r\cdot r)\,Q_{\pm,2}^{\bar \mu \bar \nu \nu}
  + \bar E_\mp\,E_\pm\,Q_{\mp,2}^{\bar \mu \bar \nu \nu}\Big]
  \nonumber\\ && \qquad
  \pm\,{\textstyle{1\over 2}}\,(\bar\delta + 1)\,\frac{1}{v^2}\,\Big[
  s\,E_\pm\,\big[ (\bar r\cdot r)\,Q_{\mp,6}^{\bar \mu \bar \nu \nu}
  + \bar E_\mp\,E_\mp\,Q_{\pm,6}^{\bar \mu \bar \nu \nu}\big]
  \nonumber\\ && \qquad \qquad
  -\, \sqrt{s}\,M\,\big[ (\bar r\cdot r)\,Q_{\pm,2}^{\bar \mu \bar \nu \nu}
  + \bar E_\mp\,E_\pm\,Q_{\mp,2}^{\bar \mu \bar \nu \nu}\big]
  \nonumber\\ && \qquad \qquad
  -\, E_\mp\,\big[ (\bar r\cdot r)\,Q_{\pm,1}^{\bar \mu \bar \nu \nu}
  - \bar M\,M\,\frac{E_\pm}{E_\mp}\,Q_{\mp,1}^{\bar \mu \bar \nu \nu}\big]
  \Big]
  \nonumber\\ && \qquad
  \pm\,\frac{\sqrt s}{v^2}\,M\,\Big[ (\bar r\cdot r)\,Q_{\pm,2}^{\bar \mu \bar \nu \nu}
  + \bar E_\mp\,E_\pm\,Q_{\mp,2}^{\bar \mu \bar \nu \nu}\Big]
  \,,\nonumber\\ &&
  Q_{\pm,10}^{\bar \mu \bar \nu \nu}
  =
  \frac{s}{v^2}\,\Big[ (\bar r\cdot r)\,P_{\mp,11}^{\bar \mu \bar \nu \nu}
  - \bar E_\mp\,E_\pm\,P_{\pm,11}^{\bar \mu \bar \nu \nu}\Big]
  \nonumber\\ && \qquad
  +\, {\textstyle{1\over 4}}\,(\bar\delta - 1)\,(\bar\delta + 1)\,\frac{s}{v^2}\,(r\cdot r)\,Q_{\pm,5}^{\bar \mu \bar \nu \nu}
  \nonumber\\ && \qquad
  \pm\,{\textstyle{1\over 2}}\,(\bar\delta + 1)\,\frac{s}{v^2}\,E_\pm\,\Big[
  - \frac{E_\mp}{\sqrt s}\,Q_{\pm,1}^{\bar \mu \bar \nu \nu}
  \nonumber\\ && \qquad \qquad
  +\, \big[ (\bar r\cdot r)\,Q_{\mp,7}^{\bar \mu \bar \nu \nu}
  + \bar E_\mp\,E_\mp\,Q_{\pm,7}^{\bar \mu \bar \nu \nu}\big]
  \Big]
  \nonumber\\ && \qquad
  - \frac{\sqrt s}{v^2}\,E_\pm\,\Big[ (\bar r\cdot r)\,Q_{\mp,2}^{\bar \mu \bar \nu \nu}
  + \bar E_\pm\,E_\mp\,Q_{\pm,2}^{\bar \mu \bar \nu \nu}\Big]
  \,,\nonumber\\ &&
  Q_{\pm,11}^{\bar \mu \bar \nu \nu}
  =
  \frac{s}{v^2}\,\Big[ (\bar r\cdot r)\,P_{\mp,10}^{\bar \mu \bar \nu \nu}
  - \bar E_\mp\,E_\pm\,P_{\pm,10}^{\bar \mu \bar \nu \nu}\Big]
  \nonumber\\ && \qquad
  +\, {\textstyle{1\over 2}}\,(\bar\delta + 1)\,\frac{s}{v^2}\,\Big[
  \big[ (\bar r\cdot r)\,Q_{\pm,4}^{\bar \mu \bar \nu \nu}
  \mp\,\frac{1}{\sqrt s}\,\bar E_\mp\,E_\mp\,Q_{\pm,1}^{\bar \mu \bar \nu \nu}\big]
  \nonumber\\ && \qquad \qquad
  \pm\,E_\pm\,\big[ (\bar r\cdot r)\,Q_{\mp,8}^{\bar \mu \bar \nu \nu}
  + \bar E_\mp\,E_\mp\,Q_{\pm,8}^{\bar \mu \bar \nu \nu}\big]
  \Big]
  \nonumber\\ && \qquad
  -\, \frac{\sqrt s}{v^2}\,\bar E_\mp\,\Big[ (\bar r\cdot r)\,Q_{\mp,2}^{\bar \mu \bar \nu \nu}
  + \bar E_\pm\,E_\mp\,Q_{\pm,2}^{\bar \mu \bar \nu \nu}\Big]
  \,,\nonumber\\ &&
  Q_{\pm,12}^{\bar \mu \bar \nu \nu}
  =
  \frac{s}{v^2}\,\Big[ (\bar r\cdot r)\,P_{\mp,6}^{\bar \mu \bar \nu \nu}
  - \bar E_\mp\,E_\pm\,P_{\pm,6}^{\bar \mu \bar \nu \nu}\Big]
  \nonumber\\ && \qquad
  \mp\,\frac{s}{v^2}\,\bar E_\mp\,\Big[ (\bar r\cdot r)\,Q_{\pm,8}^{\bar \mu \bar \nu \nu}
  + \bar E_\pm\,E_\pm\,Q_{\mp,8}^{\bar \mu \bar \nu \nu}\Big]
  \nonumber\\ && \qquad
  -\, \frac{\sqrt s}{v^2}\,\bar E_\mp\,\Big[ (\bar r\cdot r)\,Q_{\mp,3}^{\bar \mu \bar \nu \nu}
  + \bar E_\pm\,E_\mp\,Q_{\pm,3}^{\bar \mu \bar \nu \nu}\Big]
  \nonumber\\ && \qquad
  +\, \frac{s}{v^2}\,(\bar r\cdot\bar r)\,Q_{\pm,4}^{\bar \mu \bar \nu \nu}
  \mp\,\frac{\sqrt s}{v^2}\,(\bar r\cdot\bar r)\,Q_{\mp,1}^{\bar \mu \bar \nu \nu}
  \,,\nonumber
\end{eqnarray}
where
\begin{eqnarray}
  &&
  \bar q_{\bar \mu}\, P^{\bar \mu \bar \nu \nu}_{\pm, k} =0\,,\quad
  \bar p_{\bar\nu}\, P^{\bar \mu \bar \nu \nu}_{\pm, k} =0=
  \Lambda\,P^{\bar \mu \bar \nu \nu}_{\pm, k}\,\bar\Lambda\,\gamma_{\bar \nu}\,,\quad
  \nonumber\\
  && p_{\nu}\, P^{\bar \mu \bar \nu \nu}_{\pm, k} =0=
  \gamma_{\nu}\,\Lambda\,P^{\bar \mu \bar \nu \nu}_{\pm, k}\,\bar\Lambda\,,
  \nonumber\\ \nonumber\\
  &&
  P_{\pm, 1}^{\bar \mu \bar \nu \nu}
  =
  v^{\bar \mu}\,\big[\rbot^{\bar \nu}\,\rbarbot^{\nu}\,P_\pm
  + v^{\bar \nu}\,v^{\nu}\,(s/v^2)\,\bar E_\pm\,E_\pm\,P_\mp/v^2
  \nonumber\\ && \quad
  -\, \rbot^{\bar \nu}\,v^{\nu}\,(\sqrt{s}/v^2)\,E_\pm\,i\,\gamma_5\,P_\pm
  \nonumber\\ && \quad
  +\, v^{\bar \nu}\,\rbarbot^{\nu}\,(\sqrt{s}/v^2)\,\bar E_\pm\,i\,\gamma_5\,P_\mp
  \big]/v^2
  \,,\nonumber\\ &&
  P_{\pm, 2}^{\bar \mu \bar \nu \nu}
  =
  v^{\bar \mu}\,\big[\rbot^{\bar \nu}\,\wRbarbot^{\nu}\,P_\pm
  + v^{\bar \nu}\,\wRbarbot^{\nu}\,(\sqrt{s}/v^2)\,\bar E_\pm\,i\,\gamma_5\,P_\mp
  \nonumber\\ && \quad
  -\, \rbot^{\bar \nu}\,v^{\nu}\,i\,\gamma_5\,((\bar r\cdot r)\,P_\pm \pm\,M\,\bar E_\pm\,P_\mp)/v^2
  \nonumber\\ && \quad
  +\, v^{\bar \nu}\,v^{\nu}\,(\sqrt{s}/v^2)\,\bar E_\pm\,((\bar r\cdot r)\,P_\mp \pm\,M\,\bar E_\mp\,P_\pm)/v^2
  \big]/v^2
  \,,\nonumber\\ &&
  P_{\pm, 3}^{\bar \mu \bar \nu \nu}
  =
  v^{\bar \mu}\,\big[\wRbot^{\bar \nu}\,\rbarbot^{\nu}\,P_\pm
  - \wRbot^{\bar \nu}\,v^{\nu}\,(\sqrt{s}/v^2)\,E_\pm\,i\,\gamma_5\,P_\pm
  \nonumber\\ && \quad
  +\, v^{\bar \nu}\,\rbarbot^{\nu}\,i\,\gamma_5\,((\bar r\cdot r)\,P_\mp \pm\,\bar M\,E_\pm\,P_\pm)/v^2
  \nonumber\\ && \quad
  +\, v^{\bar \nu}\,v^{\nu}\,(\sqrt{s}/v^2)\,E_\pm\,((\bar r\cdot r)\,P_\mp \pm\,\bar M\,E_\mp\,P_\pm)/v^2
  \big]/v^2
  \,,\nonumber\\ &&
  P_{\pm, 4}^{\bar \mu \bar \nu \nu}
  =
  v^{\bar \mu}\,\big[\wRbot^{\bar \nu}\,\wRbarbot^{\nu}\,P_\pm
  + v^{\bar \nu}\,v^{\nu}\,\big\{
  (1/s)\,P_\mp
  \nonumber\\ && \quad \quad
  +\, {\textstyle{1\over 2}}\,(\bar\delta - 1)\,{\textstyle{1\over 2}}\,(\delta - 1)\,(s/v^2)\,\bar E_\pm\,E_\pm\,P_\mp
  \nonumber\\ && \quad \quad
  \mp\,{\textstyle{1\over 2}}\,(\bar\delta - 1)\,(\sqrt{s}/v^2)\,M\,((\bar r\cdot r)\,P_\pm - \bar E_\pm\,E_\pm\,P_\mp)
  \nonumber\\ && \quad \quad
  \mp\,{\textstyle{1\over 2}}\,(\delta - 1)\,(\sqrt{s}/v^2)\,\bar M\,((\bar r\cdot r)\,P_\pm - \bar E_\pm\,E_\pm\,P_\mp)
  \big\}/v^2
  \nonumber\\ && \quad
  -\, \wRbot^{\bar \nu}\,v^{\nu}\,i\,\gamma_5\, ((\bar r\cdot r)\,P_\pm \pm\,M\,\bar E_\pm\,P_\mp)/v^2
  \nonumber\\ && \quad
  +\, v^{\bar \nu}\,\wRbarbot^{\nu}\,i\,\gamma_5\, ((\bar r\cdot r)\,P_\mp \pm\,\bar M\,E_\pm\,P_\pm)/v^2
  \big]/v^2
  \,,\nonumber\\ &&
  P_{\pm, 5}^{\bar \mu \bar \nu \nu}
  =
  \rbot^{\bar \mu}\,\big[
  \rbot^{\bar \nu}\,v^{\nu}\,(\sqrt{s}/v^2)\,E_\mp\,P_\pm
  - v^{\bar \nu}\,\rbarbot^{\nu}\,(\sqrt{s}/v^2)\,\bar E_\pm\,P_\mp
  \nonumber\\ && \quad
  +\, \rbot^{\bar \nu}\,\rbarbot^{\nu}\,i\,\gamma_5\,P_\pm
  + v^{\bar \nu}\,v^{\nu}\,(s/v^2)\,\bar E_\pm\,E_\mp\,i\,\gamma_5\,P_\mp/v^2
  \big]
  \,,\nonumber\\ &&
  P_{\pm, 6}^{\bar \mu \bar \nu \nu}
  =
  \rbot^{\bar \mu}\,\big[\rbot^{\bar \nu}\,\wRbarbot^{\nu}\,i\,\gamma_5\,P_\pm
  - v^{\bar \nu}\,\wRbarbot^{\nu}\,(\sqrt{s}/v^2)\,\bar E_\pm\,P_\mp
  \nonumber\\ && \quad
  +\, \rbot^{\bar \nu}\,v^{\nu}\,((\bar r\cdot r)\,P_\pm \mp\,M\,\bar E_\pm\,P_\mp)/v^2
  \nonumber\\ && \quad
  +\, v^{\bar \nu}\,v^{\nu}\,(\sqrt{s}/v^2)\,\bar E_\pm\,i\,\gamma_5\, ((\bar r\cdot r)\,P_\mp \mp\,M\,\bar E_\mp\,P_\pm)/v^2
  \big]
  \,,\nonumber\\ &&
  P_{\pm, 7}^{\bar \mu \bar \nu \nu}
  =
  \rbot^{\bar \mu}\,\big[\wRbot^{\bar \nu}\,\rbarbot^{\nu}\,i\,\gamma_5\,P_\pm
  + \wRbot^{\bar \nu}\,v^{\nu}\,(\sqrt{s}/v^2)\,E_\mp\,P_\pm
  \nonumber\\ && \quad
  -\, v^{\bar \nu}\,\rbarbot^{\nu}\,((\bar r\cdot r)\,P_\mp \pm\,\bar M\,E_\mp\,P_\pm)/v^2
  \nonumber\\ && \quad
  +\, v^{\bar \nu}\,v^{\nu}\,(\sqrt{s}/v^2)\,E_\mp\,i\,\gamma_5\, ((\bar r\cdot r)\,P_\mp \pm\,\bar M\,E_\pm\,P_\pm)/v^2
  \big]
  \,,\nonumber\\ &&
  P_{\pm, 8}^{\bar \mu \bar \nu \nu}
  =
  \rbot^{\bar \mu}\,\big[\wRbot^{\bar \nu}\,\wRbarbot^{\nu}\,i\,\gamma_5\,P_\pm
  + v^{\bar \nu}\,v^{\nu}\,\big\{
  (1/s)\,i\,\gamma_5\,P_\mp
  \nonumber\\ && \quad \quad
  \pm\,{\textstyle{1\over 2}}\,(\bar\delta - 1)\,(\sqrt{s}/v^2)\,M\,i\,\gamma_5\, ((\bar r\cdot r)\,P_\pm - \bar E_\pm\,E_\mp\,P_\mp)
  \nonumber\\ && \quad \quad
  \mp\,{\textstyle{1\over 2}}\,(\delta - 1)\,(\sqrt{s}/v^2)\,\bar M\,i\,\gamma_5\, ((\bar r\cdot r)\,P_\pm - \bar E_\pm\,E_\mp\,P_\mp)
  \nonumber\\ && \quad \quad
  +\, {\textstyle{1\over 2}}\,(\bar\delta - 1)\,{\textstyle{1\over 2}}\,(\delta - 1)\,(s/v^2)\,\bar E_\pm\,E_\mp\,i\,\gamma_5\,P_\mp
  \big\}/v^2
  \nonumber\\ && \quad
  +\, \wRbot^{\bar \nu}\,v^{\nu}\,((\bar r\cdot r)\,P_\pm \mp\,M\,\bar E_\pm\,P_\mp)/v^2
  \nonumber\\ && \quad
  -\, v^{\bar \nu}\,\wRbarbot^{\nu}\,((\bar r\cdot r)\,P_\mp \pm\,\bar M\,E_\mp\,P_\pm)/v^2
  \big]
  \,,\nonumber\\ &&
  P_{\pm, 9}^{\bar \mu \bar \nu \nu}
  =
  \wLbot^{\bar \mu}\,\big[
  \rbot^{\bar \nu}\,v^{\nu}\,(\sqrt{s}/v^2)\,E_\mp\,P_\pm
  - v^{\bar \nu}\,\rbarbot^{\nu}\,(\sqrt{s}/v^2)\,\bar E_\pm\,P_\mp
  \nonumber\\ && \quad
  +\, \rbot^{\bar \nu}\,\rbarbot^{\nu}\,i\,\gamma_5\,P_\pm
  + v^{\bar \nu}\,v^{\nu}\,(s/v^2)\,\bar E_\pm\,E_\mp\,i\,\gamma_5\,P_\mp/v^2
  \big]
  \,,\nonumber\\ &&
  P_{\pm,10}^{\bar \mu \bar \nu \nu}
  =
  \wLbot^{\bar \mu}\,\big[\rbot^{\bar \nu}\,\wRbarbot^{\nu}\,i\,\gamma_5\,P_\pm
  - v^{\bar \nu}\,\wRbarbot^{\nu}\,(\sqrt{s}/v^2)\,\bar E_\pm\,P_\mp
  \nonumber\\ && \quad
  +\, \rbot^{\bar \nu}\,v^{\nu}\,((\bar r\cdot r)\,P_\pm \mp\,M\,\bar E_\pm\,P_\mp)/v^2
  \nonumber\\ && \quad
  +\, v^{\bar \nu}\,v^{\nu}\,(\sqrt{s}/v^2)\,\bar E_\pm\,i\,\gamma_5\, ((\bar r\cdot r)\,P_\mp \mp\,M\,\bar E_\mp\,P_\pm)/v^2
  \big]
  \,,\nonumber\\ &&
  P_{\pm,11}^{\bar \mu \bar \nu \nu}
  =
  \wLbot^{\bar \mu}\,\big[\wRbot^{\bar \nu}\,\rbarbot^{\nu}\,i\,\gamma_5\,P_\pm
  + \wRbot^{\bar \nu}\,v^{\nu}\,(\sqrt{s}/v^2)\,E_\mp\,P_\pm
  \nonumber\\ && \quad
  -\, v^{\bar \nu}\,\rbarbot^{\nu}\,((\bar r\cdot r)\,P_\mp \pm\,\bar M\,E_\mp\,P_\pm)/v^2
  \nonumber\\ && \quad
  +\, v^{\bar \nu}\,v^{\nu}\,(\sqrt{s}/v^2)\,E_\mp\,i\,\gamma_5\, ((\bar r\cdot r)\,P_\mp \pm\,\bar M\,E_\pm\,P_\pm)/v^2
  \big]
  \,,\nonumber\\ &&
  P_{\pm,12}^{\bar \mu \bar \nu \nu}
  =
  \wLbot^{\bar \mu}\,\big[\wRbot^{\bar \nu}\,\wRbarbot^{\nu}\,i\,\gamma_5\,P_\pm
  + v^{\bar \nu}\,v^{\nu}\,\big\{
  (1/s)\,i\,\gamma_5\,P_\mp
  \nonumber\\ && \quad \quad
  \pm\,{\textstyle{1\over 2}}\,(\bar\delta - 1)\,(\sqrt{s}/v^2)\,M\,i\,\gamma_5\, ((\bar r\cdot r)\,P_\pm - \bar E_\pm\,E_\mp\,P_\mp)
  \nonumber\\ && \quad \quad
  \mp\,{\textstyle{1\over 2}}\,(\delta - 1)\,(\sqrt{s}/v^2)\,\bar M\,i\,\gamma_5\, ((\bar r\cdot r)\,P_\pm - \bar E_\pm\,E_\mp\,P_\mp)
  \nonumber\\ && \quad \quad
  +\, {\textstyle{1\over 2}}\,(\bar\delta - 1)\,{\textstyle{1\over 2}}\,(\delta - 1)\,(s/v^2)\,\bar E_\pm\,E_\mp\,i\,\gamma_5\,P_\mp
  \big\}/v^2
  \nonumber\\ && \quad
  +\, \wRbot^{\bar \nu}\,v^{\nu}\,((\bar r\cdot r)\,P_\pm \mp\,M\,\bar E_\pm\,P_\mp)/v^2
  \nonumber\\ && \quad
  -\, v^{\bar \nu}\,\wRbarbot^{\nu}\,((\bar r\cdot r)\,P_\mp \pm\,\bar M\,E_\mp\,P_\pm)/v^2
  \big]
  \,.\nonumber
\end{eqnarray}

%
\bibliographystyle{unsrt}
\bibliography{bibliography}

%

\end{document}